\documentclass{article}
\usepackage[utf8]{inputenc}

\usepackage{amsmath}
\usepackage{amssymb}

\usepackage{graphicx}
\usepackage[dvipsnames]{xcolor}
\usepackage{tikz}
\usetikzlibrary{decorations.pathreplacing}
\definecolor{col1}{RGB}{255,250,245}
\definecolor{col2}{RGB}{245,250,250}
\definecolor{colP}{RGB}{136,34,85}
\definecolor{colI}{RGB}{51,34,136}
\definecolor{colD}{RGB}{17,119,51}
\definecolor{colN}{RGB}{0,0,0}

\usepackage{comment}
\usepackage{multirow}
\usepackage{pdflscape}
\usepackage{afterpage}
\usepackage{adjustbox}
\usepackage{lscape}

\usepackage{natbib}

\usepackage{todonotes}

\title{Accounting for multiple imputation-induced variability for differential analysis in mass spectrometry-based  label-free quantitative proteomics}
\author{Marie Chion\,$^{\text{1,2}*}$, Christine Carapito\,$^{\text{2}}$ and Frédéric Bertrand\,$^{\text{1,3}}$}
\date{ \small $^{\text{1}}$Institut de Recherche Mathématique Avancée, UMR 7501, 7 rue René Descartes, 67084 Strasbourg Cedex, France.\\
$^{\text{2}}$Laboratoire de Spectrométrie de Masse Bio-Organique, Institut Pluridisciplinaire Hubert Curien, UMR 7178, 25 rue Becquerel, 67087 Strasbourg Cedex, France.\\
$^{\text{3}}$Laboratoire de Modélisation et Sûreté des Systèmes, Institut Charles Delaunay, UMR CNRS 6281, Université de Technologie de Troyes, 12 Rue Marie Curie, 42060 Troyes Cedex, France.\\
$^*$ \textbf{Contact:} marie.chion@protonmail.com}

\begin{document}

\maketitle

\section*{Abstract}
\textbf{Motivation:} Imputing missing values is common practice in label-free quantitative proteomics. Imputation aims at replacing a missing value with a user-defined one. However, the imputation itself may not be optimally considered downstream of the imputation process, as imputed datasets are often considered as if they had always been complete. Hence, the uncertainty due to the imputation is not adequately taken into account. We provide a rigorous multiple imputation strategy, leading to a less biased estimation of the parameters' variability thanks to Rubin’s rules. The imputation-based peptide’s intensities’ variance estimator is then moderated using Bayesian hierarchical models. This estimator is finally included in moderated $t$-test statistics to provide differential analyses results. This workflow can be used both at peptide and protein-level in quantification datasets. For protein-level results based on peptide-level quantification data, an aggregation step is also included.\\
\textbf{Results:} Our methodology, named \texttt{mi4p}, was compared to the state-of-the-art \texttt{limma} workflow implemented in the \texttt{DAPAR} \texttt{R} package, both on simulated and real datasets. We observed a trade-off between sensitivity and specificity, while the overall performance of \texttt{mi4p} outperforms \texttt{DAPAR} in terms of $F$-Score.\\
\textbf{Availability:} The methodology here described is implemented under the \texttt{R} environment and can be found on GitHub: \texttt{https://github.com/mariechion/mi4p}. The \texttt{R} scripts which led to the results presented here can also be found on this repository. The real datasets are available on ProteomeXchange under the dataset identifiers PXD003841 and PXD027800.

\section{Introduction}

Dealing with incomplete data is one of the main challenges as far as statistical analysis is concerned. Different strategies can be used to tackle this issue. The simplest way consists of deleting from the dataset the observations for which there are too many missing values, leading to a complete-case dataset. However, it causes information loss, might create bias and could ultimately result in poorly informative datasets.

Another way to cope with missing data is to use methods that account for the missing information. For the last decades, researchers advocated the use of a single technique called imputation. Imputing missing values consists of replacing a missing value with a value derived using a user-defined formula (such as the mean, the median or a value provided by an expert, thus considering the user's knowledge). Hence it makes it possible to perform the analysis as if the data were complete. More particularly, the vector of parameters of interest can be then estimated.

\begin{figure}[!ht]
    \centering
    
\begin{tikzpicture}[scale=1.5]

\draw[dotted,fill=col1] (-1,0.4) rectangle (-0.8,0.6);
\draw[dotted,fill=col2] (-0.8,0.4) rectangle (-0.6,0.6);
\draw[dotted,fill=col1] (-0.6,0.4) rectangle (-0.4,0.6);
\draw[dotted,fill=col2] (-0.4,0.4) rectangle (-0.2,0.6);
\draw[dotted,fill=col1] (-0.2,0.4) rectangle (0,0.6);

\draw[thick] (-1,0.4)--(0,0.4)--(0,0.6)--(-1,0.6)--(-1,0.4);

\draw[dotted,fill=col1] (1.7,0.4) rectangle (1.9,0.6);
\draw[dotted,fill=col2] (1.9,0.4) rectangle (2.1,0.6);
\draw[dotted,fill=col1] (2.1,0.4) rectangle (2.3,0.6);
\draw[dotted,fill=col2] (2.3,0.4) rectangle (2.5,0.6);
\draw[dotted,fill=col1] (2.5,0.4) rectangle (2.7,0.6);

\draw[thick] (1.7,0.4)--(2.7,0.4)--(2.7,0.6)--(1.7,0.6)--(1.7,0.4);

\draw[dotted,fill=col1] (4,0.4) rectangle (4.2,0.6);
\draw[dotted,fill=col2] (4.2,0.4) rectangle (4.4,0.6);
\draw[dotted,fill=col1] (4.4,0.4) rectangle (4.6,0.6);
\draw[dotted,fill=col2] (4.6,0.4) rectangle (4.8,0.6);
\draw[dotted,fill=col1] (4.8,0.4) rectangle (5,0.6);

\draw[thick] (4,0.4)--(5,0.4)--(5,0.6)--(4,0.6)--(4,0.4);

\draw[thick,colN,<->] (-1,0.75)--(0,0.75);
\draw[decorate,decoration={brace,amplitude=5pt},thick] (0,0.35)--(-1,0.35); 

\draw (-0.5,0.925) node{\color{colN}\bf \small N observations};
\draw (-0.5,0) node{\color{colI}\bf \small I groups};

\draw[thick,colN,<->] (1.7,0.75)--(2.7,0.75);
\draw[decorate,decoration={brace,amplitude=5pt},thick] (2.7,0.35)--(1.7,0.35); 

\draw (2.2,0.925) node{\color{colN}\bf \small N observations};
\draw (2.2,0) node{\color{colI}\bf \small I groups};

\draw[thick,<->] (4.025,0.25)--(4.975,0.25);
\draw[thick,colN,<->] (4.025,0.75)--(4.975,0.75);
\draw (4.5,0.925) node{\color{colI} \bf \small I parameters};
\draw (4.5,0) node{\color{colI}\bf \small I groups};

\draw (0.7,0.5) node[above]{\scriptsize Single} node[below]{\scriptsize imputation};
\draw (3.4,0.5) node[above]{\scriptsize Estimation};

\draw[thick,->] (0.25,0.5)--(1.25,0.5);
\draw[thick,->] (3,0.5)--(3.8,0.5);

\draw (-0.5,-0.5) node{\small \bf (1)} (2.2,-0.5) node{\small \bf (2)}
(4.5,-0.5) node{\small \bf (3)};

\end{tikzpicture}
    
    \caption{{\bf Single imputation.} {\bf(1)} Initial dataset with missing values. It is supposed to be made of \textcolor{colN}{\bf N} observations that are split into {\bf I} groups. {\bf(2)} Single imputation provides an imputed dataset. {\bf(3)} The vector of parameters of interest is estimated based on the single imputed dataset.}
    \label{fig:SI}
\end{figure}
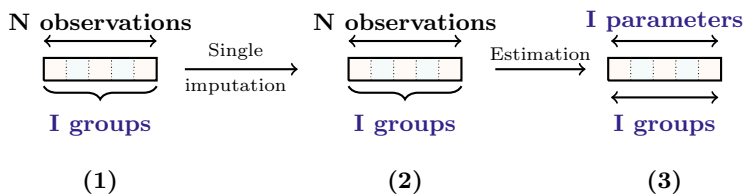

Single imputation means completing the dataset once and considering the imputed dataset as if it was never incomplete, see Figure~\ref{fig:SI}. However, single imputation has the major disadvantage of discarding the variability from the missing data and the imputation process. It may also lead to a biased estimator of the vector of parameters of interest.

\begin{figure}[!htb]
\centering
\begin{tikzpicture}[scale=1.5]

\draw[dotted,fill=col1] (-1,0.4) rectangle (-0.8,0.6);
\draw[dotted,fill=col2] (-0.8,0.4) rectangle (-0.6,0.6);
\draw[dotted,fill=col1] (-0.6,0.4) rectangle (-0.4,0.6);
\draw[dotted,fill=col2] (-0.4,0.4) rectangle (-0.2,0.6);
\draw[dotted,fill=col1] (-0.2,0.4) rectangle (0,0.6);

\draw[thick] (-1,0.4)--(0,0.4)--(0,0.6)--(-1,0.6)--(-1,0.4);

\draw[dotted,fill=col1] (1.8,0.7) rectangle (2.0,0.9);
\draw[dotted,fill=col2] (2.0,0.7) rectangle (2.2,0.9);
\draw[dotted,fill=col1] (2.2,0.7) rectangle (2.4,0.9);
\draw[dotted,fill=col2] (2.4,0.7) rectangle (2.6,0.9);
\draw[dotted,fill=col1] (2.6,0.7) rectangle (2.8,0.9);

\draw[thick] (1.8,0.7)--(2.8,0.7)--(2.8,0.9)--(1.8,0.9)--(1.8,0.7);

\draw[dotted,fill=col1] (1.7,0.5) rectangle (1.9,0.7);
\draw[dotted,fill=col2] (1.9,0.5) rectangle (2.1,0.7);
\draw[dotted,fill=col1] (2.1,0.5) rectangle (2.3,0.7);
\draw[dotted,fill=col2] (2.3,0.5) rectangle (2.5,0.7);
\draw[dotted,fill=col1] (2.5,0.5) rectangle (2.7,0.7);

\draw[dashed,thick] (1.7,0.5)--(2.7,0.5)--(2.7,0.7)--(1.7,0.7)--(1.7,0.5);

\draw[dotted,fill=col1] (1.6,0.3) rectangle (1.8,0.5);
\draw[dotted,fill=col2] (1.8,0.3) rectangle (2.0,0.5);
\draw[dotted,fill=col1] (2.0,0.3) rectangle (2.2,0.5);
\draw[dotted,fill=col2] (2.2,0.3) rectangle (2.4,0.5);
\draw[dotted,fill=col1] (2.4,0.3) rectangle (2.6,0.5);

\draw[dashed,thick] (1.6,0.3)--(2.6,0.3)--(2.6,0.5)--(1.6,0.5)--(1.6,0.3);

\draw[dotted,fill=col1] (1.5,0.1) rectangle (1.7,0.3);
\draw[dotted,fill=col2] (1.7,0.1) rectangle (1.9,0.3);
\draw[dotted,fill=col1] (1.9,0.1) rectangle (2.1,0.3);
\draw[dotted,fill=col2] (2.1,0.1) rectangle (2.3,0.3);
\draw[dotted,fill=col1] (2.3,0.1) rectangle (2.5,0.3);

\draw[thick] (1.5,0.1)--(2.5,0.1)--(2.5,0.3)--(1.5,0.3)--(1.5,0.1);

\draw[dotted,fill=col1] (4.1,1) rectangle (4.3,1.2);
\draw[dotted,fill=col2] (4.3,1) rectangle (4.5,1.2);
\draw[dotted,fill=col1] (4.5,1) rectangle (4.7,1.2);
\draw[dotted,fill=col2] (4.7,1) rectangle (4.9,1.2);
\draw[dotted,fill=col1] (4.9,1) rectangle (5.1,1.2);

\draw[thick] (4.1,1)--(5.1,1)--(5.1,1.2)--(4.1,1.2)--(4.1,1);

\draw[dotted,fill=col1] (4.1,-0.7) rectangle (4.3,-0.5);
\draw[dotted,fill=col2] (4.3,-0.7) rectangle (4.5,-0.5);
\draw[dotted,fill=col1] (4.5,-0.7) rectangle (4.7,-0.5);
\draw[dotted,fill=col2] (4.7,-0.7) rectangle (4.9,-0.5);
\draw[dotted,fill=col1] (4.9,-0.7) rectangle (5.1,-0.5);

\draw[dotted,fill=col2] (4.1,-0.5) rectangle (4.3,-0.3);
\draw[dotted,fill=col1] (4.3,-0.5) rectangle (4.5,-0.3);
\draw[dotted,fill=col2] (4.5,-0.5) rectangle (4.7,-0.3);
\draw[dotted,fill=col1] (4.7,-0.5) rectangle (4.9,-0.3);
\draw[dotted,fill=col2] (4.9,-0.5) rectangle (5.1,-0.3);

\draw[dotted,fill=col1] (4.1,-0.3) rectangle (4.3,-0.1);
\draw[dotted,fill=col2] (4.3,-0.3) rectangle (4.5,-0.1);
\draw[dotted,fill=col1] (4.5,-0.3) rectangle (4.7,-0.1);
\draw[dotted,fill=col2] (4.7,-0.3) rectangle (4.9,-0.1);
\draw[dotted,fill=col1] (4.9,-0.3) rectangle (5.1,-0.1);

\draw[dotted,fill=col2] (4.1,-0.1) rectangle (4.3,0.1);
\draw[dotted,fill=col1] (4.3,-0.1) rectangle (4.5,0.1);
\draw[dotted,fill=col2] (4.5,-0.1) rectangle (4.7,0.1);
\draw[dotted,fill=col1] (4.7,-0.1) rectangle (4.9,0.1);
\draw[dotted,fill=col2] (4.9,-0.1) rectangle (5.1,0.1);

\draw[dotted,fill=col1] (4.1,0.1) rectangle (4.3,0.3);
\draw[dotted,fill=col2] (4.3,0.1) rectangle (4.5,0.3);
\draw[dotted,fill=col1] (4.5,0.1) rectangle (4.7,0.3);
\draw[dotted,fill=col2] (4.7,0.1) rectangle (4.9,0.3);
\draw[dotted,fill=col1] (4.9,0.1) rectangle (5.1,0.3);

\draw[thick] (4.1,-0.7)--(5.1,-0.7)--(5.1,0.3)--(4.1,0.3)--(4.1,-0.7);

\draw[thick,colN,<->] (-1,0.75)--(0,0.75);
\draw[decorate,decoration={brace,amplitude=5pt},thick] (0,0.35)--(-1,0.35); 

\draw (-0.5,1) node{\color{colN}\bf \small N observations};
\draw (-0.5,0) node{\color{colI}\bf \small I groups};

\draw[thick,colN,<->] (1.8,1)--(2.8,1);
\draw[decorate,decoration={brace,amplitude=5pt},thick] (2.5,0.05)--(1.5,0.05);

\draw (2.3,1.2) node{\color{colN}\bf \small  N observations};
\draw (2,-0.3) node{\color{colI}\bf \small I groups};

\draw[thick,colD,<->] (2.7,0.1)--(3,0.7);
\draw (3.1,0.4) node{\color{colD}\bf \small D};
\draw (3.2,0.2) node{\color{colD}\bf \small draws};

\draw (4.6,0.65) node{\color{colI}\bf \small I groups};
\draw (5.5,-0.15) node{\color{colI}\bf \small I };

\draw[thick,<->] (5.1,0.9)--(4.1,0.9); 
\draw[thick,<->] (4.1,0.45)--(5.1,0.45);
\draw[thick,<->] (5.3,0.3)--(5.3,-0.7);

\draw (0.725,0.7) node{\scriptsize Multiple} 
(0.75,0.3) node{\scriptsize imputation} ;

\draw (-0.5,-0.7) node{\small \bf (1)} (2.2,-0.7) node{\small \bf (2)}
(5.9,1.1) node{\small \bf (3a)} (5.9,-0.1) node{\small \bf (3b)};

\draw[thick,->] (0.25,0.5)--(1.25,0.5);
\draw[thick,->] (3.4,0.55)--(4,1.1);
\draw[thick,->] (3.4,0.45)--(4,-0.1);

\end{tikzpicture}    

\caption{{\bf Multiple imputation strategy.} {\bf(1)} Initial dataset with missing values. It is supposed to have \textcolor{colN}{\bf N} observations that are split into \textcolor{colI}{\bf I} groups. {\bf(2)} Multiple imputation provides \textcolor{colD}{\bf D}~estimators for the vector of parameters of interest. {\bf(3a)} The \textcolor{colD}{\bf D}~estimators are combined using the first Rubin's rule to get the combined estimator. {\bf(3a)} The estimator of the variance-covariance matrix of the combined estimator is provided by the second Rubin's rule.}
\label{figMI}
\end{figure}
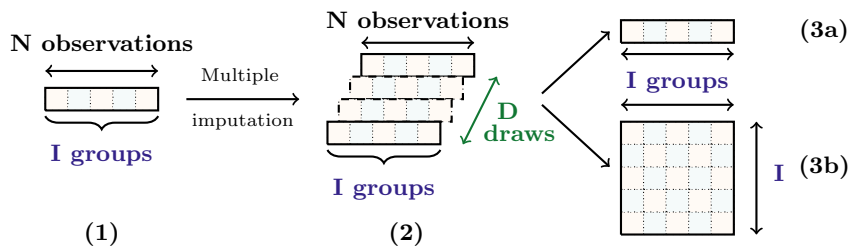

Multiple imputation described by \cite{LittleRubin} closes this loophole by generating several imputed datasets. These datasets are then used to build a combined estimator of the vector of parameters of interest, by usually using the mean of the estimates among all the imputed datasets, see Figure~\ref{figMI}. This combined estimator is known as the first Rubin's rule. The second Rubin's rule states a formula to estimate the variance of the combined estimator, decomposing it as the sum of the intra-imputation variance component and the between-imputation component. The rule of thumb suggested by \cite{White} takes the number of imputed datasets as the percentage of missing values in the original dataset. Recent work focused on better estimating the Fraction of Missing Information \citep{Pan2018} or improving that rule \citep{vonHippel2018}. Note that Rubin's rules cannot be used in order to get a combined imputed dataset but instead provide an estimator of the vector of parameters of interest and an estimator of its covariance matrix both based on multiple imputation, see Figure~\ref{figMI}.

Dealing with missing values is also one of the main struggles in label-free quantitative proteomics. Intensities of thousands of peptides are obtained by liquid chromatography-tandem mass spectrometry, using extracted ion chromatograms. Several reasons may cause missing peptides' intensities. Either the considered peptide is missing in the given biological sample, and the intensity is then missing not at random (MNAR), or it was missed during the acquisition process and, the intensity is then missing at random (MAR).

In state-of-the-art software for statistical analysis in label-free quantitative proteomics, single imputation is the most commonly used method to deal with missing values. In the \texttt{MSstats} \texttt{R} package (available on Bioconductor), \cite{MSstats} distinguish missing completely at random values and missing values due to low intensities. The user can then choose to impute the censored value using a threshold value or an Accelerated Failure Time model. The Perseus software by \cite{Perseus} offers three methods for single imputation: either imputing by "NaN", impute by a user-defined constant or impute according to a Gaussian distribution in order to simulate intensities, which are lower than the limit of detection. Recently, \cite{Goeminne2020} implemented a single imputation method based on a hurdle model in their \texttt{MSqRob} \texttt{R} package \citep{Goeminne2018}. As far as machine learning is concerned, \cite{XGboost} suggested a method for imputing missing values in label-free mass spectrometry-based proteomics datasets, called \texttt{XGboost}.

The ProStaR software based on the \verb+DAPAR+ \texttt{R} package and developed by \cite{DAPAR} splits missing values into two categories, whether they are Missing in an Entire Condition (MEC) or Partially Observed Values (POV) \citep{Lazar}. The software allows single imputation, using either a small quantile from the distribution of the considered biological sample, the $k$-Nearest Neighbours (kNN) algorithm or the Structured Least Squares Adaptative algorithm or by choosing a fixed value. The PANDA-view software developed by \cite{PANDA} also enables the use of the kNN algorithm or a fixed value. Moreover, both software programs give the possibility to impute the dataset several times before combining the imputed datasets in order to get a final dataset without any missing values. PANDA-view relies on the \texttt{mice} \texttt{R} package by \cite{mice}, whereas ProStaR accounts for the nature of missing values and imputes them with the \texttt{imp4p} \texttt{R} package implemented by \cite{imp4p}. However, both software programs consider the final dataset as if it had always been complete. The uncertainty due to multiple imputation is not properly taken into account downstream of the imputation step.

In the following, we will conduct the multiple imputation process to its end, as described by \cite{LittleRubin} and use the imputed datasets to provide a combined estimator of the vector of parameters of interest as well as a combined estimator of its variance-covariance matrix estimator. We will then project this matrix to get a unidimensional variance estimator before moderating it using the empirical Bayes procedure defined in the seminal paper of \cite{Smyth} and later developed by \cite{Phipson}. It is well known that such a moderating step highly improves the following statistical analyses such as significance testing of confidence interval estimation, both at the peptide level \citep{Suomi2015, Goeminne2015} or the protein level \citep{Goeminne2015, Goeminne2016}. 

\section{Materials}
\subsection{Simulated datasets}
We evaluated our methodology on three types of simulated datasets. First, we considered an experimental design where the distributions of the two groups to be compared scarcely overlap. This design led to a fixed effect one-way analysis of variance model (ANOVA), which can be written as:
\begin{equation}
    y_{ij} = \mu + \delta_{ij} + \epsilon_{ij}
\end{equation}
with $\mu = 100$, $\delta_{ij} = 100$ if  $1 \leq i \leq 10$ and $j=2$ and $\delta_{ij} = 0$ otherwise and $\epsilon_{ijk} \sim \mathcal{N}(0,1)$. Here, $y_{ij}$ represents the log-transformed abundance of peptide $i$ in the $j$-th sample.
Thus, we generated $100$ datasets by considering $200$ individuals and $10$ variables, divided into $2$ groups of $5$ variables, using the following steps:
\begin{enumerate}
    \item For the first 10 rows of the data frame, set as differentially expressed, draw the first 5 observations (first group) from a Gaussian distribution with a mean of 100 and a standard deviation of 1. Then draw the remaining 5 observations (second group) from a Gaussian distribution with a mean of 200 and a standard deviation of 1.
    \item For the remaining 190 rows, set as non-differentially expressed, draw the first 5 observations as well as the last 5 observations from a Gaussian distribution with a mean of 100 and a standard deviation of 1.
\end{enumerate}

Secondly, we considered an experimental design, where the distributions of the two groups to be compared might highly overlap. Hence, we based it on the random hierarchical ANOVA model by \cite{Lazar}, derived from \cite{Karpievitch2012}. The simulation design follows the following model:
\begin{equation}
    y_{ij} = P_{i} + G_{ik} + \epsilon_{ij\textcolor{Purple}{\bf k}}
\end{equation}
where $y_{ij}$ is the log-transformed abundance of peptide $i$ in the $j$-th sample, $P_{i}$ is the mean value of peptide $i$, $G_{ik}$ is the mean differences between the condition groups, and $\epsilon_{ij}$ is the random error terms, which stands for the peptide-wise variance.
We generated 100 datasets by considering 1000 individuals and 20 variables, divided into 2 groups of 10 variables, using the following steps: 
\begin{enumerate}
    \item Generate the peptide-wise effect $P_{i}$ by drawing 1000 observations from a Gaussian distribution with a mean of 1.5 and a standard deviation of 0.5.
    \item Generate the group effect $G_{ik}$ by drawing 200 observations (for the 200 individuals set as differentially expressed) from a Gaussian distribution with a mean of 1.5 and a standard deviation of 0.5 and 800 observations fixed to 0.
    \item Build the first group dataset by replicating 10 times the sum of $P_{i}$ and the random error term, drawn from a Gaussian distribution of mean 0 and standard deviation 0.5.
    \item Build the second group dataset by replicating 10 times the sum of $P_{i}$, $G_{ik}$ and the random error term drawn from a Gaussian distribution of mean 0 and standard deviation 0.5.
    \item Bind both datasets to get the complete dataset.
\end{enumerate}

Finally, we considered an experimental design similar to the second one, but with random effects $P_{i}$ and $G_{ik}$. The 100 datasets were generated as follows. 
\begin{enumerate}
    \item For the first group, replicate 10 times (for the 10 variables in this group) a draw from a mixture of 2 Gaussian distributions. The first one has the following parameters: a mean of 1.5 and a standard deviation of 0.5 (corresponds to $P_{i}$). The second one has the following parameters: a mean of 0 and a standard deviation of 0.5 (corresponds to $\epsilon_{ij}$).
    \item For the second group replicate 10 times (for the 10 variables in this group) a draw from a mixture of the following 3 distributions.
\begin{enumerate}
    \item The first one is a Gaussian distribution with the following parameters: a mean of 1.5 and a standard deviation of 0.5 (corresponds to $P_{i}$).
    \item The second one is the mixture of a Gaussian distribution with a mean of 1.5 and a standard deviation of 0.5 for the 200 first rows (set as differentially expressed) and a zero vector for the remaining 800 rows (set as not differentially expressed). This mixture illustrates the $G_{ik}$ term in the previous model.
    \item The third distribution has the following parameters: a mean of 0 and a standard deviation of 0.5 (corresponds to $\epsilon_{ij}$).
\end{enumerate}
\end{enumerate}
All simulated datasets were then amputed to produce MCAR missing values in the following proportions: 1\%, 5\%, 10\%, 15\%, 20\% and 25\%.

\subsection{Real datasets}
We challenged our methodology on several real datasets coming from two different experiments described hereafter.

We consider a first real dataset from \cite{Muller}. The experiment involved six peptide mixtures, composed of a constant yeast (\textit{Saccharomyces cerevisiae}) background, into which increasing amounts of UPS1 standard proteins mixtures (Sigma) were spiked at 0.5, 1, 2.5, 5, 10 and 25 fmol, respectively. 
In a second well-calibrated dataset, yeast was replaced by a more complex total lysate of \textit{Arabidopsis thaliana} in which UPS1 was spiked in 7 different amounts, namely 0.05, 0.25, 0.5, 1.25, 2.5, 5 and 10 fmol. For each mixture, technical triplicates were constituted.
The \textit{Saccharomyces cerevisiae} dataset was acquired on a nanoLC-MS/MS coupling composed of nanoAcquity UPLC device (Waters) coupled to a Q-Exactive Plus mass spectrometer (Thermo Scientific, Bremen, Germany) as extensively described in \cite{Muller}. The \textit{Arabidopsis thaliana} dataset was acquired on a nanoLC-MS/MS coupling composed of nanoAcquity UPLC device (Waters) coupled to a Q-Exactive HF-X mass spectrometer (Thermo Scientific, Bremen, Germany) as described in Supplementary data, Section S7.3.

For the \textit{Saccharomyces cerevisiae} and \textit{Arabidopsis thaliana} datasets, Maxquant software was used to identify peptides and derive extracted ion chromatograms. Peaks were assigned with the Andromeda search engine with full trypsin specificity. The database used for the searches was concatenated in-house with the \textit{Saccharomyces cerevisiae} entries extracted from the UniProtKB-SwissProt database (16 April 2015, 7806 entries) or the \textit{Arabidopsis thaliana} entries (09 April 2019, 15 818 entries) and those of the  UPS1 proteins (48 entries). The maximum false discovery rate was 1\% at peptide and protein levels using a target-decoy strategy.
For the \textit{Arabidopsis thaliana} + UPS1 experiment, data were extracted both with and without Match Between Runs and 2 pre-filtering criteria were applied before statistical analysis: only peptides with, on the one hand, at least 1 out of 3 quantified values in each condition and, on the other hand at least 2 out of 3, were kept. Thus, 4 datasets derived from the \textit{Arabidopsis thaliana} + UPS1 were considered. 
The same filtering criteria were applied for the \textit{Saccharomyces cerevisiae} + UPS1 experiment, but only on data extracted with Match Between Runs, leading to 2 datasets being considered.

\section{Methods}

\subsection{Normalization}
Normalising peptides' or proteins' intensities aims at reducing batch effects, sample-level variations and therefore better comparing intensities across studied biological samples \cite{Wang2021}. In this work, quantile normalisation (as described by \cite{Bolstad2003}) was performed using the \texttt{normalize.quantiles} function from the \texttt{preprocessCore} \texttt{R} package \citep{Bolstad2021}.

\subsection{Multiple imputation methods} \label{section:MI}
Several methods for imputing missing values in mass spectrometry-based proteomics datasets were developed in the last decade. However, the recent benchmarks of imputation algorithms do not reach a consensus (as shown in Supplementary data, Table S1.1). This is mainly due to the complex nature of the underlying missing values mechanism. In this work, we chose to focus on some of the most commonly used methods \ref{Tab:imp}.

\begin{center}
\begin{table}[h]
\begin{tabular}{|c|c|c|}
\hline
\textbf{Method} & \textbf{Implementation} & \textbf{References} \\ \hline
k Nearest Neighbours & \begin{tabular}[c]{@{}c@{}}impute.knn\\ (impute R package)\end{tabular} & \begin{tabular}[c]{@{}c@{}}Hastie et al. (2021)\\ Hastie et al. (1999)\\ Troyanskaya et al. (2001)\end{tabular} \\ \hline
\begin{tabular}[c]{@{}c@{}}Maximum Likelihood\\ Estimation\end{tabular} & \begin{tabular}[c]{@{}c@{}}impute.mle\\ (imp4p R package)\end{tabular} & \begin{tabular}[c]{@{}c@{}}Giai-Gianetto (2020)\\ Schafer (1997)\\ Van Buuren (2011)\end{tabular} \\ \hline
\begin{tabular}[c]{@{}c@{}}Bayesian Linear \\ Regression\end{tabular} & \begin{tabular}[c]{@{}c@{}}mice\\ (mice R package)\end{tabular} & \begin{tabular}[c]{@{}c@{}}Van Buuren (2021)\\Rubin (1987)\\ Schafer (1997)\end{tabular} \\ \hline
\begin{tabular}[c]{@{}c@{}}Principal Component\\ Analysis\end{tabular} & \begin{tabular}[c]{@{}c@{}}impute.pca\\ (imp4p R package)\end{tabular} & \begin{tabular}[c]{@{}c@{}}Giai-Gianetto (2020)\\ Josse \& Husson (2013)\end{tabular} \\ \hline
Random Forests & \begin{tabular}[c]{@{}c@{}}impute.RF\\ (imp4p R package)\end{tabular} & \begin{tabular}[c]{@{}c@{}}Giai-Gianetto (2020)\\ Stekhoven \& Buehlmann (2012)\end{tabular} \\ \hline
\end{tabular}
\caption{Overview of the imputation methods used in the evaluation of the \texttt{mi4p} workflow.}
\label{Tab:imp}
\end{table}
\end{center}

The $k$-Nearest Neighbours method imputes missing values by averaging the $k$-nearest observations of the given missing value in terms of Euclidean distance. This method was described by \cite{Hastie1999} and \cite{Troyanskaya} and implemented in \cite{Hastie2021}. 
The Maximum Likelihood Estimation method imputed missing values using the EM algorithm proposed by \cite{Schafer} and implemented by \cite{imp4p}.
The Bayesian linear regression method imputes missing values using the normal model and following the method described by \cite{Rubin1987} and implemented by \cite{mice}. 
The Principal Component Analysis imputes missing values using the algorithm proposed by \cite{Josse2013} and implemented by \cite{imp4p}. 
The Random Forests method imputes missing values using the algorithm proposed by \cite{MissForest} and implemented by \cite{imp4p}.

\subsection{Estimation}
The objective of multiple imputation is to estimate from $\textcolor{colD}{D}$ drawn datasets the vector of parameters of interest and its variance-covariance matrix. Notably, accounting for multiple-imputation-based variability is possible thanks to Rubin's rules, which provide an accurate estimation of these parameters.\par
Let us consider a $\textcolor{colD}{D}$-time imputed dataset with $\textcolor{colP}{P}$ individuals (corresponding to $\textcolor{colP}{P}$ peptides or proteins) and $\textcolor{colN}{N}$ observations (corresponding to $\textcolor{colN}{N}$ biological samples), divided between $\textcolor{colI}{I}$ groups (corresponding to $\textcolor{colI}{I}$ conditions to be compared). Let $\beta_{\textcolor{colP}{P}}$ be the vector of parameters of interest, such as :
\begin{equation}
    \beta_{\textcolor{colP}{P}} = \left(\beta_{\textcolor{colP}{P},1}, \dots, \beta_{\textcolor{colP}{P},I}\right)
\end{equation}
The first Rubin's rule provides the combined estimator of $\beta_{\textcolor{colP}{P}}$:
\begin{equation}
    \hat{\beta}_{\textcolor{colP}{p}} = \frac{1}{\textcolor{colD}{D}}\sum_{\textcolor{colD}{d}}^{\textcolor{colD}{D}}\hat{\beta}_{\textcolor{colP}{p},\textcolor{colD}{d}}
\end{equation}
where $\hat{\beta}_{\textcolor{colP}{p},\textcolor{colD}{d}}$ is the estimator of $\beta_{\textcolor{colP}{P}}$ in the $\textcolor{colD}{d}$-imputed dataset.\par
The second Rubin's rule gives the combined estimator of the variance-covariance matrix for each estimated vector of parameters of interest for peptide $\textcolor{colP}{p}$ through the $\textcolor{colD}{D}$ imputed datasets such as:
\begin{equation}
    \hat{\Sigma}_{\textcolor{colP}{p}} = \frac{1}{\textcolor{colD}{D}} \sum_{\textcolor{colD}{d}=1}^{\textcolor{colD}{D}} W_{\textcolor{colD}{d}} + \frac{\textcolor{colD}{D}+1}{\textcolor{colD}{D}(\textcolor{colD}{D}-1)} \sum_{\textcolor{colD}{d}=1}^{\textcolor{colD}{D}} (\hat{\beta}_{\textcolor{colP}{p},\textcolor{colD}{d}} - \hat{\beta}_{\textcolor{colP}{p}})^T(\hat{\beta}_{\textcolor{colP}{p},\textcolor{colD}{d}} - \hat{\beta}_{\textcolor{colP}{p}})
\end{equation}
where $W_{\textcolor{colD}{d}}$ denotes the variance-covariance matrix of $\hat{\beta}_{\textcolor{colP}{p},\textcolor{colD}{d}}$, {\it i.e.} the variability of the vector of parameters of interest as estimated in the $\textcolor{colD}{d}$-th imputed dataset.

\subsection{Projection}
State-of-the-art tests, including Student's $t$-test, Welch's $t$-test and moderated $t$-test, rely on the variance estimation. Here, the variability induced by multiple imputation is described by a variance-covariance matrix. Therefore, a projection step is required to get a unidimensional variance parameter. In our work, we chose to perform projection using the following formula :
\begin{equation}
    \hat{s}_{\textcolor{colP}{p}} = \max_k\left({\hat{\Sigma}_{\textcolor{colP}{p},(k,k)}\mathbf{X}^t\mathbf{X}}\right)
\end{equation}
where $\hat{\Sigma}_{\textcolor{colP}{p},(k,k)}$ is the $k$-th diagonal element of the matrix $\hat{\Sigma}_{\textcolor{colP}{p}}$ and $\mathbf{X}$ is the design matrix.

\subsection{Testing}
In our work, we focus our methodology on the moderated $t$-test introduced by \cite{Smyth}. This testing technique relies on the empirical Bayes procedure, commonly used in microarray data analysis, and to a more recent extent for differential analysis in quantitative proteomics \cite{DAPAR}. The moderated $t$-test procedure relies on the following Bayesian hierarchical model:
\begin{align}
    \hat{s}^2_{\textcolor{colP}{p}} \mid \sigma^2_{\textcolor{colP}{p}} \sim \frac{\sigma^2_{\textcolor{colP}{p}}}{d_{\textcolor{colP}{p}}} \times \chi_{d_{\textcolor{colP}{p}}}^2 \quad\mbox{ and }\quad
    \frac{1}{\sigma^2_{\textcolor{colP}{p}}} \sim \frac{1}{d_0 \times s_{0}^2} \times \chi_{d_0}^2
\end{align}
where $\sigma^2_{\textcolor{colP}{p}}$ is the peptide-wise variance, $d_0$ and $s_0$ are hyperparameters to be estimated \citep{Phipson}. From there, a so-called moderated variance estimator $\hat{s}^2_{{\textcolor{colP}{p}}[mod]}$ of the variance $\sigma^2_{\textcolor{colP}{p}}$ is derived:
\begin{equation}
    \hat{s}^2_{{\textcolor{colP}{p}}[mod]} = \frac{d_{\textcolor{colP}{p}} \times \hat{s}_{\textcolor{colP}{p}}^2 + d_0 \times s^2_0}{d_{\textcolor{colP}{p}} + d_0}
\end{equation}
This estimator is then computed in the test statistic associated to the null hypothesis $\mathcal{H}_0: \beta_{\textcolor{colP}{p}\textcolor{colI}{i}} = 0$ (see Equation \ref{TestStatistic}). Therefore, the results of this testing procedure account both for the specific structure of the data and the uncertainty caused by the multiple imputation step. 

\begin{equation} \label{TestStatistic}
   T_{\textcolor{colP}{p}\textcolor{colI}{i}[mod]}= \frac{\hat{\beta}_{\textcolor{colP}{p}\textcolor{colI}{i}}}{\hat{s}^2_{{\textcolor{colP}{p}}[mod]}\sqrt{(X^TX)^{-1}_{k,k}}}
\end{equation}
with $(X^T\Omega_{\textcolor{colP}{p}}X)^{-1}_{k,k}$ the $k$-th diagonal element in the matrix $(X^T\Omega_{\textcolor{colP}{p}}X)^{-1}$.
Under the $\mathcal{H}_0$ hypothesis, $T_{\textcolor{colP}{p}\textcolor{colI}{i}[mod]}$ follows a Student distribution with $d_{\textcolor{colP}{p}}+d_0$ degrees of freedom.

As many tests as the number of peptides considered are performed. Hence, the proportion of falsely rejected hypotheses has to be controlled. Here, the False Discovery Rate control procedure from \cite{FDR:BH} was performed using the \texttt{cp4p} \texttt{R} package by \cite{cp4p}.

\subsection{Aggregation}
The methodology implemented in the \texttt{mi4p} \texttt{R} package can be applied to peptide-level quantification data as well as protein-level quantification data. However, we were interested in evaluating our method on a peptide-level dataset and inferring results at a protein level, as it is common practice in proteomics. Therefore, for intensity aggregation, we chose to sum all unique peptides' intensities for each protein. The detailed pipeline for intensity aggregation is described in Supplementary data in Section~S2.

\subsection{Measures of performance}
We compared our methodology to the \texttt{limma} testing pipeline implemented in the state-of-the-art \texttt{ProStaR} software, through the \texttt{DAPAR} \texttt{R} package. To assess the performances of both methods, we used the following measures: sensitivity (also known as true positive rate or recall), specificity (also known as true negative rate), precision (also known as positive predictive value), $F$-score and Matthews correlation coefficient. In our work, we define a true positive (respectively negative) as a peptide/protein that is correctly considered as (not) differentially expressed by the testing procedure. Similarly, we define a false positive (respectively negative) as a peptide/protein that is falsely considered as (not) differentially expressed by the testing procedure. The expressions of the previously mentioned performance indicators are given in Supplementary data in Section~S3.

\section{Results and Discussion}
We highlight here results obtained using the maximum likelihood estimation imputation method. Results from other imputation methods on simulated data can be found in Supplementary data (Tables~S4.3 to~S4.6, S5.8 to~S5.11 and~S6.13 to S6.16). For each experiment, simulated or real, the performances of each method are based on adjusted $p$-values, with a 5\% significance level and using a 1\% Benjamini-Hochberg False Discovery Rate.

\subsection{Simulated datasets}

\begin{figure}[ht!]
    \centering
    \includegraphics[height=0.4\textheight]{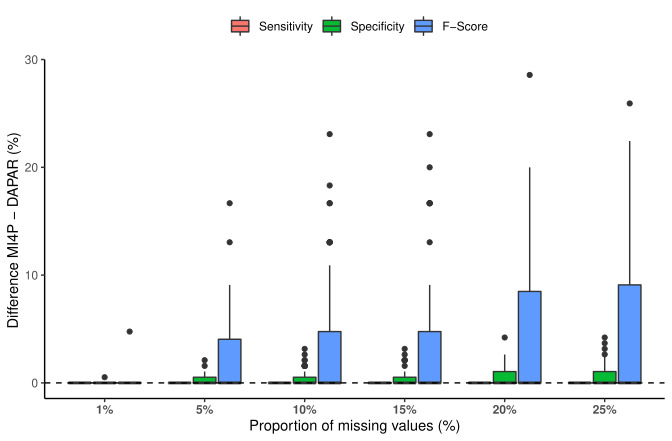}
    \caption{Distributions of differences in sensitivity, specificity and F-score for the first set of simulations.}
    \label{fig:Bplot:Sim1}
\end{figure}

Figure \ref{fig:Bplot:Sim1} describes the evolution of the distribution of differences in sensitivity and specificity between \texttt{mi4p} and \texttt{DAPAR} depending on the proportion of missing values in the first set of simulations. For a small proportion of missing values (1\%), where the imputation process induces little variability, performances in terms of sensitivity, specificity and $F$-Score are equivalent between both methods. No improvement nor deterioration was observed for sensitivity, as it remains at 100\% regardless of the missing value proportion. Specificity and $F$-Score are improved with the \texttt{mi4p} workflow above 5\% missing values. The same observations can be drawn for precision and Matthews coefficient correlation (see Figure~S4.1 in Supplementary data). Detailed results can be found in Table~S4.2.\par

\begin{figure}[hb!]
    \centering
    \includegraphics[height=0.4\textheight]{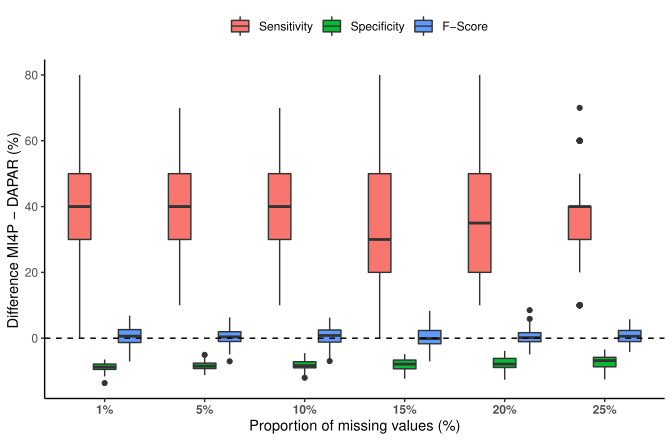}
    \caption{Distributions of differences in sensitivity, specificity and F-score for the second set of simulations.}
    \label{fig:Bplot:SimL16}
\end{figure}

Figure \ref{fig:Bplot:SimL16} describes the evolution of the distribution of differences in sensitivity and specificity between \texttt{mi4p} and \texttt{DAPAR} depending on the proportion of missing values in the second set of simulations. For all proportions of missing values, we observe a trade-off between sensitivity and specificity. A slight loss in specificity (remaining above 99\%) provides a greater gain in terms of sensitivity. The mean $F$-score across the 100 datasets is also increased with the \texttt{mi4p} workflow than with the \texttt{DAPAR} one. The Matthews correlation coefficient highlights the gain in performances (as illustrated in Supplementary data, Figure~S5.2).

Extending the simulation model from fixed effects to random effects using the last set of simulations provides similar results, as shown in Supplementary data (Figure~S6.3 and Tables~S6.12 to~S6.16).

\subsection{Real datasets}
The trade-off suggested by the simulation study is confirmed by the results obtained on the real datasets. In the \textit{Saccharomyces cerevisiae} + UPS1 experiment, a decrease of 70\% in the number of false positives is observed, improving the specificity and precision (Table~S8.23 in Supplementary data). However, this costs in the number of true positives (see Table \ref{Table:Y+UPS:1of3:impMLE:adjp:comp}), decreasing of sensitivity. The same trend is observed in the \textit{Arabidopsis thaliana} + UPS1 experiment; the number of false positives is decreased by 50\% (see Table~\ref{Table:A+UPS:1of3:impMLE:adjp:comp} and Table S8.17), thus improving specificity and precision at the cost of sensitivity. The loss in sensitivity is bigger in the highest points of the range in both experiments. The structure of the calibrated datasets used here can explain these observations. Indeed, the quantitative dataset considered takes into account all samples from all conditions, while the testing procedure focuses on one-vs-one comparisons. Two issues can be raised:
\begin{itemize}
    \item The data preprocessing step can lead to more data filtering than necessary. For instance, we chose to use the filtering criterion such that rows with at least one quantified value in each condition were kept. The more conditions are considered, the more stringent the rule is, possibly leading to a poorer dataset (with fewer observations) for the conditions of interest. 
    \item The imputation process is done on the whole dataset, as well as the estimation step. Then, while projecting the variance-covariance matrix, the estimated variance (later used in the test statistic) is the same for all comparisons. Thus, if one is interested in comparing conditions with fewer missing values, the variance estimator will be penalised by the presence of conditions with more missing values in the initial dataset.
\end{itemize} 

\begin{table}[ht]
\centering
\begin{tabular}{|c|c|c|c|c|c|}
\hline
\textbf{\begin{tabular}[c]{@{}c@{}}Condition\\vs. 25fmol\end{tabular}} &
\begin{tabular}[c]{@{}c@{}}\textbf{True}\\\textbf{positives}\end{tabular} & \begin{tabular}[c]{@{}c@{}}\textbf{False}\\\textbf{positives}\end{tabular} &
\textbf{Sensitivity} & \textbf{Specificity} & \textbf{F-Score} \\ \hline
\textbf{0.5fmol} & -2.7\% & -67.2\%     & -2.7\% & +1.6\%   & +53.6\% \\ \hline
\textbf{1fmol}   & -1.6\% & -71.1\%     & -0.5\% & +0.9\%   & +37.8\% \\ \hline
\textbf{2.5fmol} & -3.2\% & -75.8\%     & -3.3\% & +0.7\%   & +26.9\% \\ \hline
\textbf{5fmol}   & -14.3\% & -78.7\%    & -14.3\% & +0.5\%  & +11.4\%  \\ \hline
\textbf{10fmol}  & -41.9\% & -75.2\%    & -41.9\% & +0.5\%  & -14.4\% \\ \hline
\end{tabular}
\caption{Performance of the \texttt{mi4p} methodology expressed in percentage with respect to \texttt{DAPAR} workflow, on \textit{Saccharomyces cerevisiae} + UPS1 experiment, with Match Between Runs and at least 1 out of 3 quantified values in each condition. Missing values (6\%) were imputed using the maximum likelihood estimation method.}
\label{Table:Y+UPS:1of3:impMLE:adjp:comp}
\end{table}

\begin{table}[ht]
\centering
\begin{tabular}{|c|c|c|c|c|c|}
\hline
\textbf{\begin{tabular}[c]{@{}c@{}}Condition\\vs. 10fmol\end{tabular}} & \begin{tabular}[c]{@{}c@{}}\textbf{True}\\\textbf{positives}\end{tabular} & \begin{tabular}[c]{@{}c@{}}\textbf{False}\\\textbf{positives}\end{tabular} & 
\textbf{Sensitivity} & \textbf{Specificity} &
\textbf{F-Score} \\ \hline

\textbf{0.05fmol} & -2.3\%  & -43\%     & -2.3\%  & +15\%       & +62.7\%  \\ \hline
\textbf{0.25fmol} & -1.5\%  & -43\%     & -1.4\%  & +13.9\%     & +65.3\%  \\ \hline
\textbf{0.5fmol}  & -1.5\%  & -50.6\%   & -1.4\%  & +10.8\%     & +81.4\%  \\ \hline
\textbf{1.25fmol} & -2.3\%  & -62.6\%   & -2.3\%  & +10.9\%     & +119.8\% \\ \hline
\textbf{2.5fmol}  & -25.6\% & -69.3\%   & -25.5\% & +2.4\%      & +45.9\%  \\ \hline
\textbf{5fmol}    & -30.3\% & -65.2\%   & -30.4\% & +5.5\%      & +56.1\%  \\ \hline

\end{tabular}
\caption{Performance of the \texttt{mi4p} methodology expressed in percentage with respect to \texttt{DAPAR} workflow, on \textit{Arabidopsis thaliana} + UPS1 experiment, with at least 1 out of 3 quantified values in each condition. Missing values (6\%) were imputed using the maximum likelihood estimation method.}
\label{Table:A+UPS:1of3:impMLE:adjp:comp}
\end{table}

This phenomenon is illustrated in Table~S8.18, where solely the two highest points of the range have been compared, only using the quantitative data from those two conditions. More peptides have been taken into account for the statistical analysis. This strategy leads to better scores for precision, $F$-score and Matthews correlation coefficient compared to the previous framework.

As far as data extracted without the Match Between Runs algorithm are concerned, the results were equivalent in both methods considered in the \textit{Arabidopsis thaliana} + UPS1 experiment (as illustrated in Tables~S8.20 and S8.21). Furthermore, the same observations can be drawn from datasets filtered with the criterion of a minimum of 2 out of 3 observed values in each group for the \textit{Arabidopsis thaliana} + UPS1 experiment (Tables~S8.19 and S8.21) as well as for the \textit{Saccharomyces cerevisiae} + UPS1 experiment (Table~S8.24). These observations translate a loss of global information in the dataset, as filtering criteria lead to fewer peptides considered with fewer missing values per peptide.

The \texttt{mi4p} methodology also provides better results at the protein-level (after aggregation) in terms of specificity, precision, $F$-score and Matthews correlation coefficient, with a minor loss in sensitivity (Table~S8.25). In particular, a decrease of 63.2\% to 80\% in the number of false positives is observed with a lower loss on the number of true positives and on sensitivity (up to 2.6\%) for the \textit{Saccharomyces cerevisiae} + UPS1 experiment, as illustrated in Table~\ref{Table:Y+UPS:1of3:Aggreg:impMLE:adjp:comp}. As far as the \textit{Arabidopsis thaliana} + UPS1 experiment is concerned, the same trend is observed (Table~S8.22). Indeed, the number of false positives is decreased by 31\% to 66.8\%, with a maximum loss in the number of true positives of 9.8\%, as illustrated in Table~\ref{Table:A+UPS:1of3:Aggreg:impMLE:adjp:comp}.

\begin{table}[]
\centering
\begin{tabular}{|c|c|c|c|c|c|}
\hline
\textbf{\begin{tabular}[c]{@{}c@{}}Condition\\vs. 10fmol\end{tabular}} & \begin{tabular}[c]{@{}c@{}}\textbf{True}\\\textbf{positives}\end{tabular} & \begin{tabular}[c]{@{}c@{}}\textbf{False}\\\textbf{positives}\end{tabular} & 
\textbf{Sensitivity} & \textbf{Specificity} &
\textbf{F-Score} \\ \hline
\textbf{0.05fmol} & 0\% & -27.6\% & 0\% & +18.3\% & +34.2\% \\ \hline
\textbf{0.25fmol} & 0\% & -25.7\% & 0\% & +18.1\% & +31\% \\ \hline
\textbf{0.5fmol} & 0\% & -31\% & 0\% & +15.2\% & +39.5\% \\ \hline
\textbf{1.25fmol} & 0\% & -65.3\% & 0\% & +12.1 & +119.2\% \\ \hline
\textbf{2.5fmol} & -2.4\% & -66.8\% & -2.4\% & +5.8\% & +88.3\% \\ \hline
\textbf{5fmol} & -9.8\% & -57.3\% & -9.8\% & +12.9\% & +78.9\% \\ \hline
\end{tabular}
\caption{Performance of the \texttt{mi4p} methodology (with the aggregation step) expressed in percentage with respect to \texttt{DAPAR} workflow, on Arabidopsis + UPS1 experiment, with at least 1 out of 3 quantified values in each condition. Missing values were imputed using the Maximum Likelihood Estimation method.}
\label{Table:A+UPS:1of3:Aggreg:impMLE:adjp:comp}
\end{table}

\begin{table}[]
\centering
\begin{tabular}{|c|c|c|c|c|c|}
\hline
\textbf{\begin{tabular}[c]{@{}c@{}}Condition\\ vs. 25fmol\end{tabular}} &
\begin{tabular}[c]{@{}c@{}}\textbf{True}\\\textbf{positives}\end{tabular} & \begin{tabular}[c]{@{}c@{}}\textbf{False}\\\textbf{positives}\end{tabular} &
\textbf{Sensitivity} & \textbf{Specificity} & \textbf{F-Score} \\ \hline
\textbf{0.5fmol} & 0\% & -73.3\% & 0\% & +2.9\% & +61.1\% \\ \hline
\textbf{1fmol} & -2.4\% & -80\% & -2.4\% & +2.3\% & +51.4\% \\ \hline
\textbf{2.5fmol} & 0\% & -70.4\% & 0\% & +0.8\% & +20.9\% \\ \hline
\textbf{5fmol} & -2.4\% & -63.2\% & -2.4\% & +0.5\% & +11.6\% \\ \hline
\textbf{10fmol} & -2.6\% & -69.6\% & -2.6\% & +0.7\% & +16.5\% \\ \hline
\end{tabular}
\caption{Performance of the \texttt{mi4p} methodology (with the aggregation step) expressed in percentage with respect to \texttt{DAPAR} workflow, on Yeast + UPS1 experiment, with at least 1 out of 3 quantified values in each condition. Missing values were imputed using the Maximum Likelihood Estimation method.}
\label{Table:Y+UPS:1of3:Aggreg:impMLE:adjp:comp}
\end{table}

\newpage
\section{Conclusion}
In this work, we presented as a key step of a workflow a rigorous multiple imputation method by estimating both the parameters of interest and their variability. We considered this variability downstream of the statistical analysis by including it in the moderated $t$-test statistic. The methodology was implemented in the \texttt{R} statistical language through a package called \texttt{mi4p}. Its performance was compared on both simulated and real datasets to the state-of-the-art methodologies, such as the package \texttt{DAPAR}, using confusion matrix-based indicators. The results showed a trade-off between those indicators. In real datasets, the methodology reduces the number of false positives in exchange for a minor reduction of the number of true positives. The results are similar among all imputation methods considered, especially when the proportion of missing values is small. Our methodology with an additional aggregation step provides better results with a minor loss in sensitivity and can be of interest for proteomicists who will benefit from results at the protein level while using peptide-level quantification data.  

\section*{Acknowledgements}
The authors wish to thank Leslie Muller and Nicolas Pythoud for providing the real proteomics datasets used in this work, as well as Thomas Burger and Quentin Giai-Gianetto for their help on the \texttt{DAPAR} and \texttt{imp4p} \texttt{R} packages. The real datasets were deposited with the ProteomeXchange Consortium via the PRIDE partner repository with the dataset identifiers PXD003841 and PXD027800 \citep{Deutsch2017}.

\section*{Funding}
This work was funded through a PhD grant (2018-2021) awarded to MC and received by FB and CC from the Agence Nationale de la Recherche (ANR) through the Labex IRMIA [ANR-11-LABX-0055\_IRMIA].

\end{document}


\maketitle

\tableofcontents

\listoffigures

\listoftables

\newpage

\section{State of the art on imputation in quantitative proteomics}
Table \ref{tab:biblio} gives an overview of the recent literature on imputation methods in quantitative proteomics. Imputation methods are abbreviated as follows.

\begin{itemize}
    \item \textbf{BPCA:} Bayesian principal component analysis
    \item \textbf{CAM:} Convex analysis of mixtures
    \item \textbf{FCS:} Fully conditional specification
    \item \textbf{FRMF:} Fused regularisation matrix factorisation
    \item \textbf{kNN:} k-nearest neighbours
    \item \textbf{LLS:} Local least-squares
    \item \textbf{LOD1:} Half of the global minimum
    \item \textbf{LOD2:} Half of the peptide minimum
    \item \textbf{LSA:} Least-squares adaptive
    \item \textbf{MBI:} Model-based imputation
    \item \textbf{MCMC:} Monte-Carlo Markov chains
    \item \textbf{MI:} Multiple imputation
    \item \textbf{mice:} Multiple imputation using chained equations
    \item \textbf{MinDet:} Deterministic minimum
    \item \textbf{MinProb:} Probabilistic minimum
    \item \textbf{MLE:} Maximum likelihood estimation
    \item \textbf{NIPALS:} Non-linear estimation by iterative partial least squares
    \item \textbf{PCA:} Principal component analysis
    \item \textbf{PPCA:} Probabilistic principal component analysis
    \item \textbf{pwKNN:} Protein-wise k-nearest neighbours
    \item \textbf{QRLIC:} Quantile regression imputation of left-censored
    \item \textbf{SLSA:} Structured least squares algorithm
    \item \textbf{SVD:} Singular value decomposition
    \item \textbf{SVT:} Singular value thresholding
    \item \textbf{swKNN:} Sample-wise k-nearest neighbours
    \item \textbf{REM:} Regularised expectation maximisation
    \item \textbf{RF:} Random forests
    \item \textbf{RTI:} Random tail imputation
\end{itemize}

\begin{landscape}
\begin{table}[t]
\centering
\makebox[\textheight][c]{\resizebox{1.6\textwidth}{!}{\begin{tabular}{|l|l|l|}
\hline
\multicolumn{1}{|c|}{\textsc{Authors}} &
  \multicolumn{1}{c|}{\textsc{Methods}} &
  \multicolumn{1}{c|}{\textsc{Datasets}} \\ \hline
\cite{karpievitchNormalizationMissingValue2012} &
  \textbf{Single imputation:} MLE &
  \begin{tabular}[c]{@{}l@{}}\textbf{Simulated dataset:}\\ 10 samples, 2 groups, 1400 proteins\end{tabular} \\ \hline
\cite{choiMSstatsPackageStatistical2014} &
  \textbf{Single imputation:} Accelerated Failure Time model &
   \\ \hline
\cite{webb-robertsonReviewEvaluationDiscussion2015} &
  \begin{tabular}[c]{@{}l@{}}\textbf{Single imputation:}\\ Single-Value Approaches (LOD1, LOD2, RTI)\\ Local Similarity Approaches (KNN, LLS, LSA, REM, MBI)\\ Global-Structure Approaches (PPCA and BPCA)\end{tabular} &
  \begin{tabular}[c]{@{}l@{}}\textbf{Real datasets:} \\ Mouse plasma + Shewanella oneidensis, 60 samples, 1518 peptides\\ Human Plasma, 71 samples, 48 vs 23 T2D, 6729 peptides\\ Mouse Lung, 32 samples, 6295 peptides\end{tabular} \\ \hline
\cite{tyanovaPerseusComputationalPlatform2016} &
  \textbf{Single imputation:} Gaussian distribution, constant &
   \\ \hline
\cite{lazarAccountingMultipleNatures2016} &
  \textbf{Single imputation:} kNN, SVD, MLE, MinDet, MinProb &
  \begin{tabular}[c]{@{}l@{}}\textbf{Simulated dataset:} \cite{karpievitchNormalizationMissingValue2012}\\ 1000 peptides, 20 replicates\\ \textbf{Real dataset:} \cite{zhangProteomicProfilesHuman2014}\end{tabular} \\ \hline
\cite{yinMultipleImputationAnalysis2016} &
  \textbf{Multiple imputation:} MCMC + FCS &
  \begin{tabular}[c]{@{}l@{}}\textbf{Real dataset:} \\ Framingham Heart Study Offspring cohort\\ 861 plasma proteins, 135 samples\\ MCAR amputation on the 261 entirely observed proteins\\ Application to 544 partially unobserved proteins (40\% missing values)\end{tabular} \\ \hline
\cite{wieczorekDAPARProStaRSoftware2017} &
  \textbf{Single imputation:} kNN, MLE, BPCA, Quantile regression &
   \\ \hline
\cite{changPANDAviewEasytouseTool2018} &
  \begin{tabular}[c]{@{}l@{}}\textbf{Single imputation:} kNN\\ \textbf{Multiple imputation:} mice\end{tabular} &
   \\ \hline
\cite{liGMSimputeGeneralizedTwostep2020} &
  \begin{tabular}[c]{@{}l@{}}\textbf{Single imputation:} \\ Two-step lasso method, kNN, TR-kNN, RF, DanteR, Min\end{tabular} &
   \\ \hline
\end{tabular}}}
\end{table}
\end{landscape}

\begin{landscape}
\begin{table}[t]
\centering
\makebox[\textheight][c]{\resizebox{1.6\textwidth}{!}{\begin{tabular}{|l|l|l|}
\hline
\multicolumn{1}{|c|}{\textsc{Authors}} &
  \multicolumn{1}{c|}{\textsc{Methods}} &
  \multicolumn{1}{c|}{\textsc{Datasets}} \\ \hline
\cite{goeminneMSqRobTakesMissing2020} &
  Hurdle model. &
  \textbf{Real dataset:} Paulovich et al. 2010 \\ \hline
\cite{gianettoPeptidelevelMultipleImputation2020} &
  \begin{tabular}[c]{@{}l@{}}\textbf{Multiple imputation:} \\ MI, PCA, MLE, kNN, IGCDA, RF, SLSA\end{tabular} &
  \textbf{Simulated dataset:} Ramus et al. 2016 \\ \hline
\cite{liuProperImputationMissing2020} &
  \begin{tabular}[c]{@{}l@{}}\textbf{Single imputation:} \\ BPCA, kNN, MinProb, MLE, QRLIC, SVD, DetMin\end{tabular} &
  \begin{tabular}[c]{@{}l@{}}\textbf{Real datasets:}  1-4 groups, 9-56 samples, 1847-6932 proteins\\ Available on PRIDE repositories\\ \\ \textbf{Simulated datasets:} Based on the real datasets\\ 3 groups, 27-60 samples, 2800-3500 proteins\end{tabular} \\ \hline
\cite{jinComparativeStudyEvaluating2021} &
  \begin{tabular}[c]{@{}l@{}}\textbf{Single imputation:} \\ left-censored methods, kNN, LLS, RF, SVD, BPCA\end{tabular} &
  \begin{tabular}[c]{@{}l@{}}\textbf{Real datasets:}\\ (E.coli + Yeast) + UPS, 7 groups, 56 samples\\ Immune cell dataset, 3 vs 4 samples\\ Amputation of complete cases\end{tabular} \\ \hline
\cite{shenComparativeAssessmentOutlook2021} &
  \begin{tabular}[c]{@{}l@{}}\textbf{Single imputation:} \\ swKNN, pwKNN, Min/2, Mean, PPCA, NIPALS, SVD, \\ SVT, FRMF, CAM\end{tabular} &
  \begin{tabular}[c]{@{}l@{}}\textbf{Real dataset:}\\ Herrington et al. 2018\\ Amputation of complete cases from real datasets\end{tabular} \\ \hline
\cite{songMissingValueImputation2021} &
  \textbf{Single imputation:} Xgboost, mean, kNN, BPCA, LLS, RF &
  \begin{tabular}[c]{@{}l@{}}\textbf{Real datasets:}\\ Kinases expression of human colon \\ and rectal cancer cell line : 65 samples, 235 kinases\\ Proteome about the interstitial lung disease : 11 samples, \\ random draw of 500 completely observed proteins\\ Ovarian cancer proteome dataset : 25 samples, \\ random draw of 400 completely observed proteins\end{tabular} \\ \hline
\end{tabular}}}
\caption{State of the art on imputation methods used in quantitative proteomics and type of data used.}
\label{tab:biblio}
\end{table}
\end{landscape}

\section{Aggregation of peptides' intensities}
The methodology implemented in the \texttt{mi4p} \texttt{R} package can be applied to peptide-level quantification data as well as protein-level quantification data. However, we were interested in evaluating our method on a peptide-level dataset and inferring results at a protein level, as it is common practice in proteomics. Therefore, for intensity aggregation, we chose to sum all unique peptides' intensities for each protein. We then adjusted our pipeline as follows:
\begin{enumerate}
    \item Out-filtration of non-unique peptides from the peptide-level quantification dataset.
    \item Normalisation of the $\textrm{log2}$-transformed peptide intensities.
    \item Multiple imputation of $\textrm{log2}$-transformed peptide intensities.
    \item Aggregation by summing all peptides intensities (non-$\textrm{log2}$-transformed) from a given protein in each imputed dataset.
    \item $\textrm{log2}$-transformation of protein intensities.
    \item Estimation of variance-covariance matrix.
    \item Projection of the estimated variance-covariance matrix.
    \item Moderated $t$-testing on the combined protein-level dataset
\end{enumerate}

\section{Indicators of performance} \label{sec:Perf}
Let $TP$, $TN$, $FP$ and $FN$ respectively denote the numbers of true positives, true negatives, false positives, and false negatives.

\begin{equation}
    \mathrm{Sensitivity} = \frac{TP}{TP + FN}
\end{equation}
\begin{equation}
    \mathrm{Specificity} = \frac{TN}{TN + FP}
\end{equation}
\begin{equation}
    \mathrm{Precision} = \frac{TP}{TP + FP}
\end{equation}
\begin{equation}
    F\mathrm{-Score} = \frac{TP}{TP + \frac{1}{2} \times (FP + FN)}
\end{equation}
\begin{equation}
    \mathrm{MCC} = \frac{TP \times TN - FP \times FN}{\sqrt{\left(TP+FP\right)\left(TP+FN\right)\left(TN+FP\right)\left(TN+FP\right)}}
\end{equation}

\section{Results on the first set of simulations}

\subsection{Simulation design}
We considered an experimental design where the distributions of the two groups to be compared scarcely overlap. This design led to a fixed effect one-way ANOVA model, which can be written as:
\begin{equation}
    y_{ij} = \mu + \delta_{ij} + \epsilon_{ij}
\end{equation}
with $\mu = 100$, $\delta_{ij} = 100$ if  $1 \leq i \leq 10$ and $j=2$ and $\delta_{ij} = 0$ otherwise and $\epsilon_{ijk} \sim \mathcal{N}(0,1)$. Here, $y_{ij}$ represents the log-transformed abundance of peptide $i$ in the $j$-th sample.
Thus, we generated $100$ datasets by considering $200$ individuals and $10$ variables, divided into $2$ groups of $5$ variables, using the following steps:
\begin{enumerate}
    \item For the first 10 rows of the data frame, set as differentially expressed, draw the first 5 observations (first group) from a Gaussian distribution with a mean of 100 and a standard deviation of 1. Then draw the remaining 5 observations (second group) from a Gaussian distribution with a mean of 200 and a standard deviation of 1.
    \item For the remaining 190 rows, set as non-differentially expressed, draw the first 5 observations as well as the last 5 observations from a Gaussian distribution with a mean of 100 and a standard deviation of 1.
\end{enumerate}

\newpage
\subsection{Performance evaluation}
This subsection provides the evaluation of the \texttt{mi4p} workflow compared to the \texttt{DAPAR} workflow on the first set of simulations. The performance is described using the indicators detailed in Section \ref{sec:Perf}.

\begin{figure}[ht]
    \centering
    \makebox[\textwidth][c]{\includegraphics[width=1.2\textwidth]{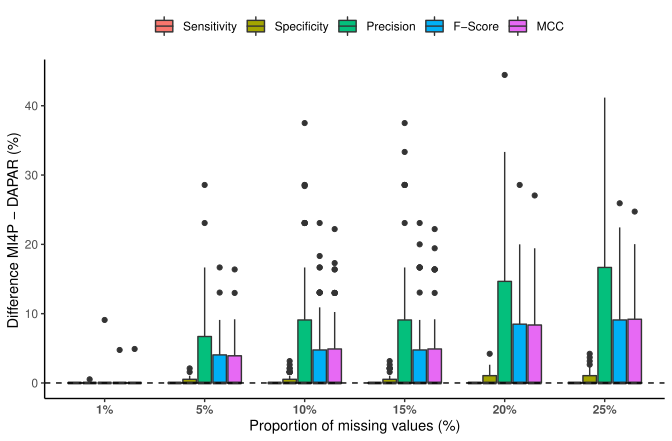}}
    \caption{Distribution of the difference of performance between \texttt{mi4p} and \texttt{DAPAR} workflows on the first set of simulations imputed using maximum likelihood estimation.}
    \label{fig:Sim1:Bplot}
\end{figure}

The following tables provide results expressed as the mean of the given indicator over the 100 simulated datasets $\pm$ the mean of the standard deviations of the given indicator over the 100 simulated datasets. Results are based on adjusted p-values using the Benjamini-Hochberg procedure \citep{benjaminiControllingFalseDiscovery1995} and a false discovery rate of 1\%.

\begin{landscape}
\begin{table}[ht]
\centering
\small
\begin{tabular}{|c|c|c|c|c|c|c|c|c|c|c|}
\hline
\multicolumn{1}{|c|}{\textbf{\%MV}} & \textbf{Method}   & 
\begin{tabular}[c]{@{}c@{}}\textbf{True}\\\textbf{positives}\end{tabular} & \begin{tabular}[c]{@{}c@{}}\textbf{False}\\\textbf{positives}\end{tabular} & \begin{tabular}[c]{@{}c@{}}\textbf{True}\\\textbf{negatives}\end{tabular} & \begin{tabular}[c]{@{}c@{}}\textbf{False}\\\textbf{negatives}\end{tabular} & 
\begin{tabular}[c]{@{}c@{}}\textbf{Sensitivity}\\\textbf{(\%)}\end{tabular} &
\begin{tabular}[c]{@{}c@{}}\textbf{Specificity}\\\textbf{(\%)}\end{tabular} &
\begin{tabular}[c]{@{}c@{}}\textbf{Precision}\\\textbf{(\%)}\end{tabular} & 
\begin{tabular}[c]{@{}c@{}}\textbf{F-score}\\\textbf{(\%)}\end{tabular} & 
\begin{tabular}[c]{@{}c@{}}\textbf{MCC}\\\textbf{(\%)}\end{tabular}\\ \hline
\multirow{2}{*}{\textbf{1\%}}  & \textbf{DAPAR}    & 10 $\pm$ 0           & 0.5 $\pm$ 0.7         & 189.5 $\pm$ 0.7      & 0 $\pm$ 0             & 100 $\pm$ 0       & 99.8 $\pm$ 0.4    & 95.9 $\pm$ 5.7  & 97.8 $\pm$ 3.1 & 97.8 $\pm$ 3.1 \\ \cline{2-11} 
                               & \textbf{MI4P} & 10 $\pm$ 0           & 0.5 $\pm$ 0.7         & 189.5 $\pm$ 0.7      & 0 $\pm$ 0             & 100 $\pm$ 0       & 99.8 $\pm$ 0.4    & 96 $\pm$ 5.7    & 97.9 $\pm$ 3.1 & 97.8 $\pm$ 3.1 \\ \hline
\multirow{2}{*}{\textbf{5\%}}  & \textbf{DAPAR}    & 10 $\pm$ 0           & 0.8 $\pm$ 1           & 189.2 $\pm$ 1        & 0 $\pm$ 0             & 100 $\pm$ 0       & 99.6 $\pm$ 0.5    & 92.9 $\pm$ 7.6  & 96.2 $\pm$ 4.2 & 96.1 $\pm$ 4.2 \\ \cline{2-11} 
                               & \textbf{MI4P} & 10 $\pm$ 0           & 0.5 $\pm$ 0.7         & 189.5 $\pm$ 0.7      & 0 $\pm$ 0             & 100 $\pm$ 0       & 99.8 $\pm$ 0.4    & 95.9 $\pm$ 6.1  & 97.8 $\pm$ 3.3 & 97.8 $\pm$ 3.4 \\ \hline
\multirow{2}{*}{\textbf{10\%}} & \textbf{DAPAR}    & 10 $\pm$ 0           & 1.2 $\pm$ 1.3         & 188.8 $\pm$ 1.3      & 0 $\pm$ 0             & 100 $\pm$ 0       & 99.4 $\pm$ 0.7    & 90.3 $\pm$ 9.3  & 94.6 $\pm$ 5.4 & 94.6 $\pm$ 5.3 \\ \cline{2-11} 
                               & \textbf{MI4P} & 10 $\pm$ 0           & 0.6 $\pm$ 0.8         & 189.4 $\pm$ 0.8      & 0 $\pm$ 0             & 100 $\pm$ 0       & 99.7 $\pm$ 0.4    & 95.3 $\pm$ 6.8  & 97.5 $\pm$ 3.7 & 97.4 $\pm$ 3.8 \\ \hline
\multirow{2}{*}{\textbf{15\%}} & \textbf{DAPAR}    & 10 $\pm$ 0           & 1.3 $\pm$ 1.3         & 188.7 $\pm$ 1.3      & 0 $\pm$ 0             & 100 $\pm$ 0       & 99.3 $\pm$ 0.7    & 89.6 $\pm$ 9.4  & 94.2 $\pm$ 5.4 & 94.2 $\pm$ 5.4 \\ \cline{2-11} 
                               & \textbf{MI4P} & 10 $\pm$ 0           & 0.6 $\pm$ 1           & 189.4 $\pm$ 1        & 0 $\pm$ 0             & 100 $\pm$ 0       & 99.7 $\pm$ 0.5    & 95.3 $\pm$ 7.4  & 97.4 $\pm$ 4.2 & 97.4 $\pm$ 4.2 \\ \hline
\multirow{2}{*}{\textbf{20\%}} & \textbf{DAPAR}    & 10 $\pm$ 0           & 2.2 $\pm$ 1.7         & 187.7 $\pm$ 1.7      & 0 $\pm$ 0             & 100 $\pm$ 0       & 98.8 $\pm$ 0.9    & 83.1 $\pm$ 10.9 & 90.4 $\pm$ 6.6 & 90.5 $\pm$ 6.4 \\ \cline{2-11} 
                               & \textbf{MI4P} & 10 $\pm$ 0           & 1.3 $\pm$ 1.7         & 188.6 $\pm$ 1.8      & 0 $\pm$ 0             & 100 $\pm$ 0       & 99.3 $\pm$ 0.9    & 89.8 $\pm$ 11.4 & 94.2 $\pm$ 6.7 & 94.3 $\pm$ 6.6 \\ \hline
\multirow{2}{*}{\textbf{25\%}} & \textbf{DAPAR}    & 10 $\pm$ 0.2         & 2.9 $\pm$ 2.1         & 186.8 $\pm$ 2.2      & 0 $\pm$ 0             & 100 $\pm$ 0       & 98.5 $\pm$ 1.1    & 79.7 $\pm$ 12.5 & 88.2 $\pm$ 7.9 & 88.3 $\pm$ 7.5 \\ \cline{2-11} 
                               & \textbf{MI4P} & 10 $\pm$ 0.2         & 1.6 $\pm$ 1.8         & 188 $\pm$ 2.1        & 0 $\pm$ 0             & 100 $\pm$ 0       & 99.2 $\pm$ 1      & 88.3 $\pm$ 12   & 93.3 $\pm$ 7.2 & 93.4 $\pm$ 7   \\ \hline

\end{tabular}
\caption{Performance evaluation on the first set of simulations imputed using maximum likelihood estimation.}
\label{Table:Sim1:impMLE:adjp}
\end{table}
\end{landscape}

\begin{landscape}
\begin{table}
\centering
\small
\begin{tabular}{|c|c|c|c|c|c|c|c|c|c|c|}
\hline
\multicolumn{1}{|c|}{\textbf{\%MV}} & \textbf{Method}   & 
\begin{tabular}[c]{@{}c@{}}\textbf{True}\\\textbf{positives}\end{tabular} & \begin{tabular}[c]{@{}c@{}}\textbf{False}\\\textbf{positives}\end{tabular} & \begin{tabular}[c]{@{}c@{}}\textbf{True}\\\textbf{negatives}\end{tabular} & \begin{tabular}[c]{@{}c@{}}\textbf{False}\\\textbf{negatives}\end{tabular} & 
\begin{tabular}[c]{@{}c@{}}\textbf{Sensitivity}\\\textbf{(\%)}\end{tabular} &
\begin{tabular}[c]{@{}c@{}}\textbf{Specificity}\\\textbf{(\%)}\end{tabular} &
\begin{tabular}[c]{@{}c@{}}\textbf{Precision}\\\textbf{(\%)}\end{tabular} & 
\begin{tabular}[c]{@{}c@{}}\textbf{F-score}\\\textbf{(\%)}\end{tabular} & 
\begin{tabular}[c]{@{}c@{}}\textbf{MCC}\\\textbf{(\%)}\end{tabular}\\ \hline
\multirow{2}{*}{\textbf{1\%}}  & \textbf{DAPAR}    & 10 $\pm$ 0           & 0.4 $\pm$ 0.6         & 189.6 $\pm$ 0.6      & 0 $\pm$ 0             & 100 $\pm$ 0       & 99.8 $\pm$ 0.3    & 96.3 $\pm$ 5.4  & 98 $\pm$ 2.9   & 98 $\pm$ 2.9   \\ \cline{2-11} 
                               & \textbf{MI4P} & 10 $\pm$ 0           & 0.4 $\pm$ 0.6         & 189.6 $\pm$ 0.6      & 0 $\pm$ 0             & 100 $\pm$ 0       & 99.8 $\pm$ 0.3    & 96.3 $\pm$ 5.4  & 98 $\pm$ 2.9   & 98 $\pm$ 2.9   \\ \hline
\multirow{2}{*}{\textbf{5\%}}  & \textbf{DAPAR}    & 10 $\pm$ 0           & 0.3 $\pm$ 0.5         & 189.7 $\pm$ 0.5      & 0 $\pm$ 0             & 100 $\pm$ 0       & 99.9 $\pm$ 0.3    & 97.7 $\pm$ 4.5  & 98.8 $\pm$ 2.4 & 98.7 $\pm$ 2.5 \\ \cline{2-11} 
                               & \textbf{MI4P} & 10 $\pm$ 0           & 0.3 $\pm$ 0.5         & 189.7 $\pm$ 0.5      & 0 $\pm$ 0             & 100 $\pm$ 0       & 99.9 $\pm$ 0.3    & 97.7 $\pm$ 4.5  & 98.8 $\pm$ 2.4 & 98.7 $\pm$ 2.5 \\ \hline
\multirow{2}{*}{\textbf{10\%}} & \textbf{DAPAR}    & 10 $\pm$ 0           & 0.3 $\pm$ 0.6         & 189.7 $\pm$ 0.6      & 0 $\pm$ 0             & 100 $\pm$ 0       & 99.8 $\pm$ 0.3    & 97.2 $\pm$ 4.9  & 98.5 $\pm$ 2.6 & 98.5 $\pm$ 2.7 \\ \cline{2-11} 
                               & \textbf{MI4P} & 10 $\pm$ 0           & 0.3 $\pm$ 0.6         & 189.7 $\pm$ 0.6      & 0 $\pm$ 0             & 100 $\pm$ 0       & 99.8 $\pm$ 0.3    & 97.2 $\pm$ 4.9  & 98.5 $\pm$ 2.6 & 98.5 $\pm$ 2.7 \\ \hline
\multirow{2}{*}{\textbf{15\%}} & \textbf{DAPAR}    & 10 $\pm$ 0.1         & 0.2 $\pm$ 0.6         & 189.8 $\pm$ 0.6      & 0 $\pm$ 0.1           & 99.9 $\pm$ 1      & 99.9 $\pm$ 0.3    & 97.9 $\pm$ 4.7  & 98.8 $\pm$ 2.6 & 98.8 $\pm$ 2.6 \\ \cline{2-11} 
                               & \textbf{MI4P} & 10 $\pm$ 0.1         & 0.2 $\pm$ 0.6         & 189.8 $\pm$ 0.6      & 0 $\pm$ 0.1           & 99.9 $\pm$ 1      & 99.9 $\pm$ 0.3    & 97.9 $\pm$ 4.7  & 98.8 $\pm$ 2.6 & 98.8 $\pm$ 2.6 \\ \hline
\multirow{2}{*}{\textbf{20\%}} & \textbf{DAPAR}    & 9.9 $\pm$ 0.2        & 0.4 $\pm$ 0.7         & 189.6 $\pm$ 0.7      & 0.1 $\pm$ 0.2         & 99.4 $\pm$ 2.4    & 99.8 $\pm$ 0.4    & 96.2 $\pm$ 5.8  & 97.6 $\pm$ 3.3 & 97.6 $\pm$ 3.3 \\ \cline{2-11} 
                               & \textbf{MI4P} & 9.9 $\pm$ 0.2        & 0.4 $\pm$ 0.7         & 189.6 $\pm$ 0.7      & 0.1 $\pm$ 0.2         & 99.4 $\pm$ 2.4    & 99.8 $\pm$ 0.4    & 96.2 $\pm$ 5.8  & 97.6 $\pm$ 3.3 & 97.6 $\pm$ 3.3 \\ \hline
\multirow{2}{*}{\textbf{25\%}} & \textbf{DAPAR}    & 9.8 $\pm$ 0.5        & 0.9 $\pm$ 1           & 189.1 $\pm$ 1        & 0.2 $\pm$ 0.5         & 97.7 $\pm$ 4.7    & 99.5 $\pm$ 0.5    & 92.7 $\pm$ 7.8  & 94.9 $\pm$ 4.7 & 94.8 $\pm$ 4.8 \\ \cline{2-11} 
                               & \textbf{MI4P} & 9.8 $\pm$ 0.5        & 0.9 $\pm$ 1           & 189.1 $\pm$ 1        & 0.2 $\pm$ 0.5         & 97.7 $\pm$ 4.7    & 99.5 $\pm$ 0.5    & 92.7 $\pm$ 7.8  & 94.9 $\pm$ 4.7 & 94.8 $\pm$ 4.8 \\ \hline

\end{tabular}
\caption{Performance evaluation on the first set of simulations imputed using $k$-nearest neighbours.}
\label{Table:Sim1:impKNN:adjp}
\end{table}
\end{landscape}

\begin{landscape}
\begin{table}[ht]
\centering
\small
\begin{tabular}{|c|c|c|c|c|c|c|c|c|c|c|}
\hline
\multicolumn{1}{|c|}{\textbf{\%MV}} & \textbf{Method}   & 
\begin{tabular}[c]{@{}c@{}}\textbf{True}\\\textbf{positives}\end{tabular} & \begin{tabular}[c]{@{}c@{}}\textbf{False}\\\textbf{positives}\end{tabular} & \begin{tabular}[c]{@{}c@{}}\textbf{True}\\\textbf{negatives}\end{tabular} & \begin{tabular}[c]{@{}c@{}}\textbf{False}\\\textbf{negatives}\end{tabular} & 
\begin{tabular}[c]{@{}c@{}}\textbf{Sensitivity}\\\textbf{(\%)}\end{tabular} &
\begin{tabular}[c]{@{}c@{}}\textbf{Specificity}\\\textbf{(\%)}\end{tabular} &
\begin{tabular}[c]{@{}c@{}}\textbf{Precision}\\\textbf{(\%)}\end{tabular} & 
\begin{tabular}[c]{@{}c@{}}\textbf{F-score}\\\textbf{(\%)}\end{tabular} & 
\begin{tabular}[c]{@{}c@{}}\textbf{MCC}\\\textbf{(\%)}\end{tabular}\\ \hline
\multirow{2}{*}{\textbf{1\%}}  & \textbf{DAPAR}    & 10 $\pm$ 0.2         & 0.2 $\pm$ 0.4         & 189.8 $\pm$ 0.4      & 0 $\pm$ 0.2           & 99.8 $\pm$ 2      & 99.9 $\pm$ 0.2    & 98.4 $\pm$ 3.7  & 99 $\pm$ 2.2   & 99 $\pm$ 2.2   \\ \cline{2-11} 
                               & \textbf{MI4P} & 9.9 $\pm$ 0.3        & 0.2 $\pm$ 0.4         & 189.8 $\pm$ 0.4      & 0.1 $\pm$ 0.3         & 99.3 $\pm$ 2.9    & 99.9 $\pm$ 0.2    & 98.3 $\pm$ 4    & 98.7 $\pm$ 2.8 & 98.7 $\pm$ 2.8 \\ \hline
\multirow{2}{*}{\textbf{5\%}}  & \textbf{DAPAR}    & 10 $\pm$ 0.2         & 0.2 $\pm$ 0.4         & 189.8 $\pm$ 0.4      & 0 $\pm$ 0.2           & 99.6 $\pm$ 2      & 99.9 $\pm$ 0.2    & 98.6 $\pm$ 3.7  & 99 $\pm$ 2.1   & 99 $\pm$ 2.2   \\ \cline{2-11} 
                               & \textbf{MI4P} & 9.7 $\pm$ 0.5        & 0.2 $\pm$ 0.4         & 189.8 $\pm$ 0.4      & 0.3 $\pm$ 0.5         & 96.9 $\pm$ 5.4    & 99.9 $\pm$ 0.2    & 97.9 $\pm$ 4.1  & 97.3 $\pm$ 3.4 & 97.2 $\pm$ 3.5 \\ \hline
\multirow{2}{*}{\textbf{10\%}} & \textbf{DAPAR}    & 10 $\pm$ 0           & 0.2 $\pm$ 0.5         & 189.8 $\pm$ 0.5      & 0 $\pm$ 0             & 100 $\pm$ 0       & 99.9 $\pm$ 0.2    & 97.8 $\pm$ 4.1  & 98.9 $\pm$ 2.1 & 98.8 $\pm$ 2.2 \\ \cline{2-11} 
                               & \textbf{MI4P} & 9.6 $\pm$ 0.7        & 0.1 $\pm$ 0.3         & 189.9 $\pm$ 0.3      & 0.4 $\pm$ 0.7         & 95.5 $\pm$ 6.9    & 100 $\pm$ 0.1     & 99.2 $\pm$ 2.6  & 97.2 $\pm$ 4   & 97.1 $\pm$ 4   \\ \hline
\multirow{2}{*}{\textbf{15\%}} & \textbf{DAPAR}    & 10 $\pm$ 0           & 0.3 $\pm$ 0.6         & 189.7 $\pm$ 0.6      & 0 $\pm$ 0             & 100 $\pm$ 0       & 99.8 $\pm$ 0.3    & 97.2 $\pm$ 4.9  & 98.5 $\pm$ 2.6 & 98.5 $\pm$ 2.7 \\ \cline{2-11} 
                               & \textbf{MI4P} & 9.2 $\pm$ 0.9        & 0 $\pm$ 0.2           & 190 $\pm$ 0.2        & 0.8 $\pm$ 0.9         & 91.7 $\pm$ 8.8    & 100 $\pm$ 0.1     & 99.6 $\pm$ 1.8  & 95.3 $\pm$ 4.9 & 95.3 $\pm$ 4.8 \\ \hline
\multirow{2}{*}{\textbf{20\%}} & \textbf{DAPAR}    & 10 $\pm$ 0           & 0.6 $\pm$ 0.8         & 189.4 $\pm$ 0.8      & 0 $\pm$ 0             & 100 $\pm$ 0       & 99.7 $\pm$ 0.4    & 94.6 $\pm$ 6.4  & 97.1 $\pm$ 3.5 & 97.1 $\pm$ 3.6 \\ \cline{2-11} 
                               & \textbf{MI4P} & 8.9 $\pm$ 1          & 0 $\pm$ 0.1           & 190 $\pm$ 0.1        & 1.1 $\pm$ 1           & 89.1 $\pm$ 10.3   & 100 $\pm$ 0.1     & 99.9 $\pm$ 1    & 93.9 $\pm$ 6.1 & 93.9 $\pm$ 5.9 \\ \hline
\multirow{2}{*}{\textbf{25\%}} & \textbf{DAPAR}    & 10 $\pm$ 0.1         & 1.2 $\pm$ 1.1         & 188.8 $\pm$ 1.1      & 0 $\pm$ 0.1           & 99.9 $\pm$ 1      & 99.4 $\pm$ 0.6    & 90.3 $\pm$ 8    & 94.7 $\pm$ 4.6 & 94.6 $\pm$ 4.6 \\ \cline{2-11} 
                               & \textbf{MI4P} & 8.9 $\pm$ 1.1        & 0 $\pm$ 0             & 190 $\pm$ 0          & 1.1 $\pm$ 1.1         & 89.3 $\pm$ 11.1   & 100 $\pm$ 0       & 100 $\pm$ 0     & 94 $\pm$ 6.7   & 94.1 $\pm$ 6.4 \\ \hline
\end{tabular}
\caption{Performance evaluation on the first set of simulations imputed using Bayesian linear regression.}
\label{Table:Sim1:impNORM:adjp}
\end{table}
\end{landscape}

\begin{landscape}
\begin{table}[ht]
\centering
\small
\begin{tabular}{|c|c|c|c|c|c|c|c|c|c|c|}
\hline
\multicolumn{1}{|c|}{\textbf{\%MV}} & \textbf{Method}   & 
\begin{tabular}[c]{@{}c@{}}\textbf{True}\\\textbf{positives}\end{tabular} & \begin{tabular}[c]{@{}c@{}}\textbf{False}\\\textbf{positives}\end{tabular} & \begin{tabular}[c]{@{}c@{}}\textbf{True}\\\textbf{negatives}\end{tabular} & \begin{tabular}[c]{@{}c@{}}\textbf{False}\\\textbf{negatives}\end{tabular} & 
\begin{tabular}[c]{@{}c@{}}\textbf{Sensitivity}\\\textbf{(\%)}\end{tabular} &
\begin{tabular}[c]{@{}c@{}}\textbf{Specificity}\\\textbf{(\%)}\end{tabular} &
\begin{tabular}[c]{@{}c@{}}\textbf{Precision}\\\textbf{(\%)}\end{tabular} & 
\begin{tabular}[c]{@{}c@{}}\textbf{F-score}\\\textbf{(\%)}\end{tabular} & 
\begin{tabular}[c]{@{}c@{}}\textbf{MCC}\\\textbf{(\%)}\end{tabular}\\ \hline
\multirow{2}{*}{\textbf{1\%}}  & \textbf{DAPAR}    & 10 $\pm$ 0           & 0.5 $\pm$ 0.7         & 189.5 $\pm$ 0.7      & 0 $\pm$ 0             & 100 $\pm$ 0       & 99.7 $\pm$ 0.4    & 95.8 $\pm$ 6    & 97.8 $\pm$ 3.3 & 97.7 $\pm$ 3.3 \\ \cline{2-11} 
                               & \textbf{MI4P} & 10 $\pm$ 0           & 0.5 $\pm$ 0.7         & 189.5 $\pm$ 0.7      & 0 $\pm$ 0             & 100 $\pm$ 0       & 99.7 $\pm$ 0.4    & 95.8 $\pm$ 6    & 97.8 $\pm$ 3.3 & 97.7 $\pm$ 3.3 \\ \hline
\multirow{2}{*}{\textbf{5\%}}  & \textbf{DAPAR}    & 10 $\pm$ 0           & 0.6 $\pm$ 0.8         & 189.4 $\pm$ 0.8      & 0 $\pm$ 0             & 100 $\pm$ 0       & 99.7 $\pm$ 0.4    & 94.6 $\pm$ 6.8  & 97.1 $\pm$ 3.7 & 97 $\pm$ 3.7   \\ \cline{2-11} 
                               & \textbf{MI4P} & 10 $\pm$ 0           & 0.6 $\pm$ 0.8         & 189.4 $\pm$ 0.8      & 0 $\pm$ 0             & 100 $\pm$ 0       & 99.7 $\pm$ 0.4    & 94.6 $\pm$ 6.8  & 97.1 $\pm$ 3.7 & 97 $\pm$ 3.7   \\ \hline
\multirow{2}{*}{\textbf{10\%}} & \textbf{DAPAR}    & 10 $\pm$ 0           & 1 $\pm$ 1.1           & 189 $\pm$ 1.1        & 0 $\pm$ 0             & 100 $\pm$ 0       & 99.5 $\pm$ 0.6    & 91.8 $\pm$ 8.3  & 95.5 $\pm$ 4.7 & 95.5 $\pm$ 4.7 \\ \cline{2-11} 
                               & \textbf{MI4P} & 10 $\pm$ 0           & 1 $\pm$ 1.1           & 189 $\pm$ 1.1        & 0 $\pm$ 0             & 100 $\pm$ 0       & 99.5 $\pm$ 0.6    & 91.8 $\pm$ 8.3  & 95.5 $\pm$ 4.7 & 95.5 $\pm$ 4.7 \\ \hline
\multirow{2}{*}{\textbf{15\%}} & \textbf{DAPAR}    & 10 $\pm$ 0           & 1.2 $\pm$ 1.2         & 188.8 $\pm$ 1.2      & 0 $\pm$ 0             & 100 $\pm$ 0       & 99.4 $\pm$ 0.6    & 90.1 $\pm$ 8.9  & 94.5 $\pm$ 5.1 & 94.5 $\pm$ 5.1 \\ \cline{2-11} 
                               & \textbf{MI4P} & 10 $\pm$ 0           & 1.2 $\pm$ 1.2         & 188.8 $\pm$ 1.2      & 0 $\pm$ 0             & 100 $\pm$ 0       & 99.4 $\pm$ 0.6    & 90.1 $\pm$ 8.9  & 94.5 $\pm$ 5.1 & 94.5 $\pm$ 5.1 \\ \hline
\multirow{2}{*}{\textbf{20\%}} & \textbf{DAPAR}    & 10 $\pm$ 0           & 1.9 $\pm$ 1.5         & 188 $\pm$ 1.5        & 0 $\pm$ 0             & 100 $\pm$ 0       & 99 $\pm$ 0.8      & 85.1 $\pm$ 9.8  & 91.6 $\pm$ 5.9 & 91.6 $\pm$ 5.7 \\ \cline{2-11} 
                               & \textbf{MI4P} & 10 $\pm$ 0           & 1.9 $\pm$ 1.5         & 188 $\pm$ 1.5        & 0 $\pm$ 0             & 100 $\pm$ 0       & 99 $\pm$ 0.8      & 85.4 $\pm$ 9.8  & 91.8 $\pm$ 5.9 & 91.8 $\pm$ 5.7 \\ \hline
\multirow{2}{*}{\textbf{25\%}} & \textbf{DAPAR}    & 10 $\pm$ 0.2         & 2.5 $\pm$ 1.6         & 187.2 $\pm$ 1.7      & 0 $\pm$ 0             & 100 $\pm$ 0       & 98.7 $\pm$ 0.9    & 81 $\pm$ 10.5   & 89.1 $\pm$ 6.4 & 89.2 $\pm$ 6.1 \\ \cline{2-11} 
                               & \textbf{MI4P} & 10 $\pm$ 0.2         & 2.6 $\pm$ 1.6         & 186.8 $\pm$ 2        & 0 $\pm$ 0             & 100 $\pm$ 0       & 98.6 $\pm$ 0.9    & 80.5 $\pm$ 10.5 & 88.8 $\pm$ 6.4 & 88.9 $\pm$ 6.2 \\ \hline
\end{tabular}
\caption{Performance evaluation on the first set of simulations imputed using principal component analysis.}
\label{Table:Sim1:impPCA:adjp}
\end{table}
\end{landscape}

\begin{landscape}
\begin{table}[ht]
\centering
\small
\begin{tabular}{|c|c|c|c|c|c|c|c|c|c|c|}
\hline
\multicolumn{1}{|c|}{\textbf{\%MV}} & \textbf{Method}   & 
\begin{tabular}[c]{@{}c@{}}\textbf{True}\\\textbf{positives}\end{tabular} & \begin{tabular}[c]{@{}c@{}}\textbf{False}\\\textbf{positives}\end{tabular} & \begin{tabular}[c]{@{}c@{}}\textbf{True}\\\textbf{negatives}\end{tabular} & \begin{tabular}[c]{@{}c@{}}\textbf{False}\\\textbf{negatives}\end{tabular} & 
\begin{tabular}[c]{@{}c@{}}\textbf{Sensitivity}\\\textbf{(\%)}\end{tabular} &
\begin{tabular}[c]{@{}c@{}}\textbf{Specificity}\\\textbf{(\%)}\end{tabular} &
\begin{tabular}[c]{@{}c@{}}\textbf{Precision}\\\textbf{(\%)}\end{tabular} & 
\begin{tabular}[c]{@{}c@{}}\textbf{F-score}\\\textbf{(\%)}\end{tabular} & 
\begin{tabular}[c]{@{}c@{}}\textbf{MCC}\\\textbf{(\%)}\end{tabular}\\ \hline
\multirow{2}{*}{\textbf{1\%}} & \textbf{DAPAR} & 10 $\pm$ 0 & 0.5 $\pm$ 0.7 & 189.5 $\pm$ 0.7 & 0 $\pm$ 0 & 100 $\pm$ 0 & 99.8 $\pm$ 0.4 & 96 $\pm$ 6 & 97.9 $\pm$ 3.3 & 97.8 $\pm$ 3.3 \\ \cline{2-11} 
 & \textbf{MI4P} & 10 $\pm$ 0 & 0.5 $\pm$ 0.7 & 189.5 $\pm$ 0.7 & 0 $\pm$ 0 & 100 $\pm$ 0 & 99.8 $\pm$ 0.4 & 96 $\pm$ 6 & 97.9 $\pm$ 3.3 & 97.8 $\pm$ 3.3 \\ \hline
\multirow{2}{*}{\textbf{5\%}} & \textbf{DAPAR} & 10 $\pm$ 0 & 0.4 $\pm$ 0.6 & 189.6 $\pm$ 0.6 & 0 $\pm$ 0 & 100 $\pm$ 0 & 99.8 $\pm$ 0.3 & 96 $\pm$ 5.3 & 97.9 $\pm$ 2.8 & 97.8 $\pm$ 2.9 \\ \cline{2-11} 
 & \textbf{MI4P} & 10 $\pm$ 0 & 0.4 $\pm$ 0.6 & 189.6 $\pm$ 0.6 & 0 $\pm$ 0 & 100 $\pm$ 0 & 99.8 $\pm$ 0.3 & 96 $\pm$ 5.3 & 97.9 $\pm$ 2.8 & 97.8 $\pm$ 2.9 \\ \hline
\multirow{2}{*}{\textbf{10\%}} & \textbf{DAPAR} & 10 $\pm$ 0 & 0.5 $\pm$ 0.8 & 189.5 $\pm$ 0.8 & 0 $\pm$ 0 & 100 $\pm$ 0 & 99.7 $\pm$ 0.4 & 95.8 $\pm$ 6.7 & 97.7 $\pm$ 3.7 & 97.7 $\pm$ 3.7 \\ \cline{2-11} 
 & \textbf{MI4P} & 10 $\pm$ 0.1 & 0.5 $\pm$ 0.8 & 189.5 $\pm$ 0.8 & 0 $\pm$ 0.1 & 99.8 $\pm$ 1.4 & 99.7 $\pm$ 0.4 & 95.9 $\pm$ 6.4 & 97.7 $\pm$ 3.6 & 97.6 $\pm$ 3.6 \\ \hline
\multirow{2}{*}{\textbf{15\%}} & \textbf{DAPAR} & 10 $\pm$ 0 & 0.3 $\pm$ 0.6 & 189.7 $\pm$ 0.6 & 0 $\pm$ 0 & 100 $\pm$ 0 & 99.8 $\pm$ 0.3 & 97.2 $\pm$ 5.3 & 98.5 $\pm$ 2.9 & 98.5 $\pm$ 2.9 \\ \cline{2-11} 
 & \textbf{MI4P} & 10 $\pm$ 0.1 & 0.4 $\pm$ 0.7 & 189.6 $\pm$ 0.7 & 0 $\pm$ 0.1 & 99.8 $\pm$ 1.4 & 99.8 $\pm$ 0.3 & 96.8 $\pm$ 5.5 & 98.2 $\pm$ 3 & 98.1 $\pm$ 3 \\ \hline
\multirow{2}{*}{\textbf{20\%}} & \textbf{DAPAR} & 10 $\pm$ 0.1 & 0.4 $\pm$ 0.6 & 189.5 $\pm$ 0.7 & 0 $\pm$ 0.1 & 99.8 $\pm$ 1.4 & 99.8 $\pm$ 0.3 & 96.3 $\pm$ 5.4 & 97.9 $\pm$ 3 & 97.9 $\pm$ 3.1 \\ \cline{2-11} 
 & \textbf{MI4P} & 10 $\pm$ 0.1 & 0.4 $\pm$ 0.6 & 189.4 $\pm$ 0.8 & 0 $\pm$ 0.1 & 99.8 $\pm$ 1.4 & 99.8 $\pm$ 0.3 & 96 $\pm$ 5.4 & 97.8 $\pm$ 3 & 97.7 $\pm$ 3.1 \\ \hline
\multirow{2}{*}{\textbf{25\%}} & \textbf{DAPAR} & 10 $\pm$ 0.2 & 0.3 $\pm$ 0.6 & 189.4 $\pm$ 0.9 & 0 $\pm$ 0.1 & 99.9 $\pm$ 1 & 99.8 $\pm$ 0.3 & 97.5 $\pm$ 5 & 98.6 $\pm$ 2.7 & 98.6 $\pm$ 2.8 \\ \cline{2-11} 
 & \textbf{MI4P} & 9.9 $\pm$ 0.2 & 0.3 $\pm$ 0.6 & 189.1 $\pm$ 1.3 & 0 $\pm$ 0.2 & 99.7 $\pm$ 1.7 & 99.9 $\pm$ 0.3 & 97.5 $\pm$ 4.9 & 98.5 $\pm$ 2.7 & 98.5 $\pm$ 2.8 \\ \hline
\end{tabular}
\caption{Performance evaluation on the first set of simulations imputed using random forests.}
\label{Table:Sim1:impRF:adjp}
\end{table}
\end{landscape}

\section{Results on the second set of simulations}

\subsection{Simulation design}
Secondly, we considered an experimental design, where the distributions of the two groups to be compared might highly overlap. Hence, we based it on the random hierarchical ANOVA model by \cite{lazarAccountingMultipleNatures2016}, derived from \cite{karpievitchNormalizationMissingValue2012}. The simulation design follows the following model:
\begin{equation}
    y_{ij} = P_{i} + G_{ik} + \epsilon_{ijk}
\end{equation}
where $y_{ij}$ is the log-transformed abundance of peptide $i$ in the $j$-th sample, $P_{i}$ is the mean value of peptide $i$, $G_{ik}$ is the mean difference between the condition groups, and $\epsilon_{ij}$ is the random error term, which stands for the peptide-wise variance.
We generated 100 datasets by considering 1000 individuals and 20 variables, divided into 2 groups of 10 variables, using the following steps: 
\begin{enumerate}
    \item Generate the peptide-wise effect $P_{i}$ by drawing 1000 observations from a Gaussian distribution with a mean of 1.5 and a standard deviation of 0.5.
    \item Generate the group effect $G_{ik}$ by drawing 200 observations (for the 200 individuals set as differentially expressed) from a Gaussian distribution with a mean of 1.5 and a standard deviation of 0.5 and 800 observations fixed to 0.
    \item Build the first group dataset by replicating 10 times the sum of $P_{i}$ and the random error term, drawn from a Gaussian distribution of mean 0 and standard deviation 0.5.
    \item Build the second group dataset by replicating 10 times the sum of $P_{i}$, $G_{ik}$ and the random error term drawn from a Gaussian distribution of mean 0 and standard deviation 0.5.
    \item Bind both datasets to get the complete dataset.
\end{enumerate}

\newpage
\subsection{Performance evaluation}
This subsection provides the evaluation of the \texttt{mi4p} workflow compared to the \texttt{DAPAR} workflow on the second set of simulations. The performance is described using the indicators detailed in Section \ref{sec:Perf}.

\begin{figure}[ht]
    \centering
    \makebox[\textwidth][c]{\includegraphics[width=1.2\textwidth]{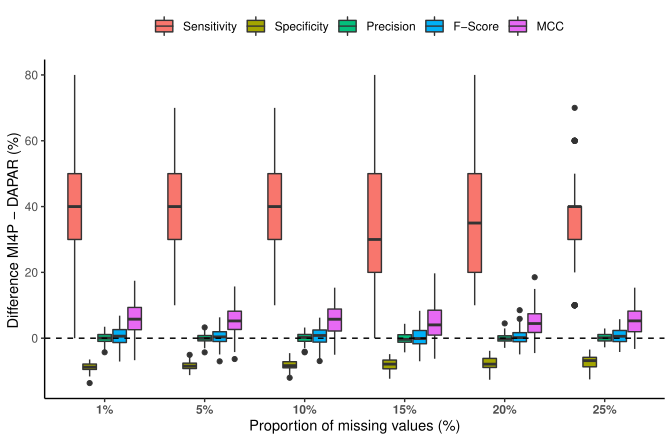}}
    \caption{Distribution of the difference of performance between \texttt{mi4p} and \texttt{DAPAR} workflows on the second set of simulations imputed using maximum likelihood estimation.}
    \label{fig:L16bis:Bplot}
\end{figure}

The following tables provide results expressed as the mean of the given indicator over the 100 simulated datasets $\pm$ the mean of the standard deviations of the given indicator over the 100 simulated datasets. Results are based on adjusted p-values using the Benjamini-Hochberg procedure \citep{benjaminiControllingFalseDiscovery1995} and a false discovery rate of 1\%.

\begin{landscape}
\begin{table}[ht]
\centering
\begin{tabular}{|c|c|c|c|c|c|c|c|c|c|c|}
\hline
\multicolumn{1}{|c|}{\textbf{\%MV}} & \textbf{Method}   & 
\begin{tabular}[c]{@{}c@{}}\textbf{True}\\\textbf{positives}\end{tabular} & \begin{tabular}[c]{@{}c@{}}\textbf{False}\\\textbf{positives}\end{tabular} & \begin{tabular}[c]{@{}c@{}}\textbf{True}\\\textbf{negatives}\end{tabular} & \begin{tabular}[c]{@{}c@{}}\textbf{False}\\\textbf{negatives}\end{tabular} & 
\begin{tabular}[c]{@{}c@{}}\textbf{Sensitivity}\\\textbf{(\%)}\end{tabular} &
\begin{tabular}[c]{@{}c@{}}\textbf{Specificity}\\\textbf{(\%)}\end{tabular} &
\begin{tabular}[c]{@{}c@{}}\textbf{Precision}\\\textbf{(\%)}\end{tabular} & 
\begin{tabular}[c]{@{}c@{}}\textbf{F-score}\\\textbf{(\%)}\end{tabular} & 
\begin{tabular}[c]{@{}c@{}}\textbf{MCC}\\\textbf{(\%)}\end{tabular}\\ \hline
\multirow{2}{*}{\textbf{1\%}} & \textbf{DAPAR} & 80.8 $\pm$ 11.4 & 1.9 $\pm$ 1.5 & 798.1 $\pm$ 1.5 & 119.2 $\pm$ 11.4 & 40.4 $\pm$ 5.7 & 99.8 $\pm$ 0.2 & 97.8 $\pm$ 1.6 & 56.9 $\pm$ 5.9 & 58.2 $\pm$ 4.5 \\ \cline{2-11} 
 & \textbf{MI4P} & 166.9 $\pm$ 5 & 6.3 $\pm$ 2.7 & 793.7 $\pm$ 2.7 & 33.1 $\pm$ 5 & 83.4 $\pm$ 2.5 & 99.2 $\pm$ 0.3 & 96.4 $\pm$ 1.4 & 89.4 $\pm$ 1.5 & 87.4 $\pm$ 1.6 \\ \hline
\multirow{2}{*}{\textbf{5\%}} & \textbf{DAPAR} & 80.8 $\pm$ 12.1 & 2.4 $\pm$ 1.8 & 797.6 $\pm$ 1.8 & 119.2 $\pm$ 12.1 & 40.4 $\pm$ 6.1 & 99.7 $\pm$ 0.2 & 97.3 $\pm$ 1.9 & 56.8 $\pm$ 6.1 & 58 $\pm$ 4.6 \\ \cline{2-11} 
 & \textbf{MI4P} & 164.2 $\pm$ 6.1 & 6.1 $\pm$ 3.5 & 793.9 $\pm$ 3.5 & 35.8 $\pm$ 6.1 & 82.1 $\pm$ 3 & 99.2 $\pm$ 0.4 & 96.5 $\pm$ 1.9 & 88.7 $\pm$ 1.5 & 86.6 $\pm$ 1.6 \\ \hline
\multirow{2}{*}{\textbf{10\%}} & \textbf{DAPAR} & 78.8 $\pm$ 11.9 & 2.4 $\pm$ 1.6 & 797.6 $\pm$ 1.6 & 121.2 $\pm$ 11.9 & 39.4 $\pm$ 5.9 & 99.7 $\pm$ 0.2 & 97.1 $\pm$ 1.8 & 55.8 $\pm$ 6.1 & 57.1 $\pm$ 4.7 \\ \cline{2-11} 
 & \textbf{MI4P} & 160.7 $\pm$ 7.8 & 5.6 $\pm$ 3.8 & 794.4 $\pm$ 3.8 & 39.3 $\pm$ 7.8 & 80.4 $\pm$ 3.9 & 99.3 $\pm$ 0.5 & 96.7 $\pm$ 2.1 & 87.7 $\pm$ 1.9 & 85.6 $\pm$ 2 \\ \hline
\multirow{2}{*}{\textbf{15\%}} & \textbf{DAPAR} & 80.3 $\pm$ 11.4 & 3.3 $\pm$ 1.9 & 796.7 $\pm$ 1.9 & 119.7 $\pm$ 11.4 & 40.1 $\pm$ 5.7 & 99.6 $\pm$ 0.2 & 96.1 $\pm$ 2.1 & 56.4 $\pm$ 5.8 & 57.3 $\pm$ 4.6 \\ \cline{2-11} 
 & \textbf{MI4P} & 159 $\pm$ 8.8 & 6.7 $\pm$ 5.1 & 793.3 $\pm$ 5.1 & 41 $\pm$ 8.8 & 79.5 $\pm$ 4.4 & 99.2 $\pm$ 0.6 & 96.2 $\pm$ 2.7 & 86.9 $\pm$ 2.1 & 84.7 $\pm$ 2.2 \\ \hline
\multirow{2}{*}{\textbf{20\%}} & \textbf{DAPAR} & 81.3 $\pm$ 11.6 & 4 $\pm$ 2.1 & 796 $\pm$ 2.1 & 118.7 $\pm$ 11.6 & 40.7 $\pm$ 5.8 & 99.5 $\pm$ 0.3 & 95.4 $\pm$ 2.4 & 56.8 $\pm$ 5.9 & 57.4 $\pm$ 4.7 \\ \cline{2-11} 
 & \textbf{MI4P} & 158 $\pm$ 9.8 & 7.2 $\pm$ 5.4 & 792.8 $\pm$ 5.4 & 42 $\pm$ 9.8 & 79 $\pm$ 4.9 & 99.1 $\pm$ 0.7 & 95.8 $\pm$ 2.9 & 86.5 $\pm$ 2.3 & 84.2 $\pm$ 2.3 \\ \hline
\multirow{2}{*}{\textbf{25\%}} & \textbf{DAPAR} & 82.5 $\pm$ 12.3 & 4.7 $\pm$ 2.7 & 795.3 $\pm$ 2.7 & 117.5 $\pm$ 12.3 & 41.2 $\pm$ 6.2 & 99.4 $\pm$ 0.3 & 94.7 $\pm$ 2.8 & 57.2 $\pm$ 6 & 57.5 $\pm$ 4.8 \\ \cline{2-11} 
 & \textbf{MI4P} & 154.5 $\pm$ 10.4 & 6.9 $\pm$ 6.2 & 793.1 $\pm$ 6.2 & 45.5 $\pm$ 10.4 & 77.3 $\pm$ 5.2 & 99.1 $\pm$ 0.8 & 96 $\pm$ 3.3 & 85.4 $\pm$ 2.5 & 83.1 $\pm$ 2.4 \\ \hline
\end{tabular}
\caption{Performance evaluation on the second set of simulations imputed using maximum likelihood estimation.}
\label{Table:L16bis:impMLE:adjp}
\end{table}
\end{landscape}

\begin{landscape}
\begin{table}[ht]
\centering
\begin{tabular}{|c|c|c|c|c|c|c|c|c|c|c|}
\hline
\multicolumn{1}{|c|}{\textbf{\%MV}} & \textbf{Method}   & 
\begin{tabular}[c]{@{}c@{}}\textbf{True}\\\textbf{positives}\end{tabular} & \begin{tabular}[c]{@{}c@{}}\textbf{False}\\\textbf{positives}\end{tabular} & \begin{tabular}[c]{@{}c@{}}\textbf{True}\\\textbf{negatives}\end{tabular} & \begin{tabular}[c]{@{}c@{}}\textbf{False}\\\textbf{negatives}\end{tabular} & 
\begin{tabular}[c]{@{}c@{}}\textbf{Sensitivity}\\\textbf{(\%)}\end{tabular} &
\begin{tabular}[c]{@{}c@{}}\textbf{Specificity}\\\textbf{(\%)}\end{tabular} &
\begin{tabular}[c]{@{}c@{}}\textbf{Precision}\\\textbf{(\%)}\end{tabular} & 
\begin{tabular}[c]{@{}c@{}}\textbf{F-score}\\\textbf{(\%)}\end{tabular} & 
\begin{tabular}[c]{@{}c@{}}\textbf{MCC}\\\textbf{(\%)}\end{tabular}\\ \hline
\multirow{2}{*}{\textbf{1\%}} & \textbf{DAPAR} & 80.5 $\pm$ 12.1 & 1.8 $\pm$ 1.4 & 798.2 $\pm$ 1.4 & 119.5 $\pm$ 12.1 & 40.2 $\pm$ 6 & 99.8 $\pm$ 0.2 & 97.9 $\pm$ 1.6 & 56.8 $\pm$ 6.3 & 58.1 $\pm$ 4.9 \\ \cline{2-11} 
 & \textbf{MI4P} & 167.9 $\pm$ 4.8 & 6.6 $\pm$ 2.5 & 793.4 $\pm$ 2.5 & 32 $\pm$ 4.8 & 84 $\pm$ 2.4 & 99.2 $\pm$ 0.3 & 96.2 $\pm$ 1.4 & 89.7 $\pm$ 1.4 & 87.6 $\pm$ 1.7 \\ \hline
\multirow{2}{*}{\textbf{5\%}} & \textbf{DAPAR} & 79.6 $\pm$ 12.4 & 1.9 $\pm$ 1.7 & 798.1 $\pm$ 1.7 & 120.4 $\pm$ 12.4 & 39.8 $\pm$ 6.2 & 99.8 $\pm$ 0.2 & 97.8 $\pm$ 1.9 & 56.2 $\pm$ 6.5 & 57.7 $\pm$ 5 \\ \cline{2-11} 
 & \textbf{MI4P} & 169.6 $\pm$ 4.3 & 6.7 $\pm$ 2.8 & 793.3 $\pm$ 2.8 & 30.4 $\pm$ 4.3 & 84.8 $\pm$ 2.2 & 99.2 $\pm$ 0.4 & 96.2 $\pm$ 1.5 & 90.1 $\pm$ 1.4 & 88.1 $\pm$ 1.6 \\ \hline
\multirow{2}{*}{\textbf{10\%}} & \textbf{DAPAR} & 78.2 $\pm$ 13.5 & 2 $\pm$ 1.7 & 798 $\pm$ 1.7 & 121.8 $\pm$ 13.5 & 39.1 $\pm$ 6.8 & 99.8 $\pm$ 0.2 & 97.7 $\pm$ 1.8 & 55.5 $\pm$ 7.1 & 57.1 $\pm$ 5.4 \\ \cline{2-11} 
 & \textbf{MI4P} & 170.8 $\pm$ 4.3 & 6.3 $\pm$ 2.8 & 793.7 $\pm$ 2.8 & 29.2 $\pm$ 4.3 & 85.4 $\pm$ 2.2 & 99.2 $\pm$ 0.4 & 96.5 $\pm$ 1.5 & 90.6 $\pm$ 1.4 & 88.7 $\pm$ 1.6 \\ \hline
\multirow{2}{*}{\textbf{15\%}} & \textbf{DAPAR} & 79 $\pm$ 14.1 & 2 $\pm$ 1.7 & 798 $\pm$ 1.7 & 121 $\pm$ 14.1 & 39.5 $\pm$ 7 & 99.8 $\pm$ 0.2 & 97.6 $\pm$ 1.8 & 55.9 $\pm$ 7.3 & 57.4 $\pm$ 5.6 \\ \cline{2-11} 
 & \textbf{MI4P} & 171.6 $\pm$ 4.5 & 6.2 $\pm$ 3.1 & 793.8 $\pm$ 3.1 & 28.4 $\pm$ 4.5 & 85.8 $\pm$ 2.2 & 99.2 $\pm$ 0.4 & 96.5 $\pm$ 1.7 & 90.8 $\pm$ 1.4 & 89 $\pm$ 1.7 \\ \hline
\multirow{2}{*}{\textbf{20\%}} & \textbf{DAPAR} & 77.2 $\pm$ 16.8 & 1.9 $\pm$ 1.6 & 798.1 $\pm$ 1.6 & 122.8 $\pm$ 16.8 & 38.6 $\pm$ 8.4 & 99.8 $\pm$ 0.2 & 97.7 $\pm$ 1.9 & 54.7 $\pm$ 9.8 & 56.4 $\pm$ 7.9 \\ \cline{2-11} 
 & \textbf{MI4P} & 171.1 $\pm$ 4.7 & 5.7 $\pm$ 2.7 & 794.3 $\pm$ 2.7 & 28.9 $\pm$ 4.7 & 85.5 $\pm$ 2.3 & 99.3 $\pm$ 0.3 & 96.8 $\pm$ 1.5 & 90.8 $\pm$ 1.4 & 89 $\pm$ 1.7 \\ \hline
\multirow{2}{*}{\textbf{25\%}} & \textbf{DAPAR} & 74.4 $\pm$ 16.8 & 1.8 $\pm$ 1.7 & 798.2 $\pm$ 1.7 & 125.6 $\pm$ 16.8 & 37.2 $\pm$ 8.4 & 99.8 $\pm$ 0.2 & 97.7 $\pm$ 1.9 & 53.3 $\pm$ 9.8 & 55.3 $\pm$ 7.8 \\ \cline{2-11} 
 & \textbf{MI4P} & 170.3 $\pm$ 4.9 & 5.9 $\pm$ 2.9 & 794.1 $\pm$ 2.9 & 29.7 $\pm$ 4.9 & 85.1 $\pm$ 2.5 & 99.3 $\pm$ 0.4 & 96.7 $\pm$ 1.6 & 90.5 $\pm$ 1.5 & 88.6 $\pm$ 1.8 \\ \hline
\end{tabular}
\caption{Performance evaluation on the second set of simulations imputed using $k$-nearest neighbours method.}
\label{Table:L16bis:impKNN:adjp}
\end{table}
\end{landscape}

\begin{landscape}
\begin{table}[ht]
\centering
\begin{tabular}{|c|c|c|c|c|c|c|c|c|c|c|}
\hline
\multicolumn{1}{|c|}{\textbf{\%MV}} & \textbf{Method}   & 
\begin{tabular}[c]{@{}c@{}}\textbf{True}\\\textbf{positives}\end{tabular} & \begin{tabular}[c]{@{}c@{}}\textbf{False}\\\textbf{positives}\end{tabular} & \begin{tabular}[c]{@{}c@{}}\textbf{True}\\\textbf{negatives}\end{tabular} & \begin{tabular}[c]{@{}c@{}}\textbf{False}\\\textbf{negatives}\end{tabular} & 
\begin{tabular}[c]{@{}c@{}}\textbf{Sensitivity}\\\textbf{(\%)}\end{tabular} &
\begin{tabular}[c]{@{}c@{}}\textbf{Specificity}\\\textbf{(\%)}\end{tabular} &
\begin{tabular}[c]{@{}c@{}}\textbf{Precision}\\\textbf{(\%)}\end{tabular} & 
\begin{tabular}[c]{@{}c@{}}\textbf{F-score}\\\textbf{(\%)}\end{tabular} & 
\begin{tabular}[c]{@{}c@{}}\textbf{MCC}\\\textbf{(\%)}\end{tabular}\\ \hline
\multirow{2}{*}{\textbf{1\%}} & \textbf{DAPAR} & 80.7 $\pm$ 11.9 & 1.9 $\pm$ 1.6 & 798.1 $\pm$ 1.6 & 119.3 $\pm$ 11.9 & 40.4 $\pm$ 6 & 99.8 $\pm$ 0.2 & 97.8 $\pm$ 1.8 & 56.8 $\pm$ 6.1 & 58.2 $\pm$ 4.7 \\ \cline{2-11} 
 & \textbf{MI4P} & 165.7 $\pm$ 5 & 5.4 $\pm$ 2.4 & 794.6 $\pm$ 2.4 & 34.3 $\pm$ 5 & 82.8 $\pm$ 2.5 & 99.3 $\pm$ 0.3 & 96.9 $\pm$ 1.3 & 89.3 $\pm$ 1.5 & 87.3 $\pm$ 1.7 \\ \hline
\multirow{2}{*}{\textbf{5\%}} & \textbf{DAPAR} & 80.5 $\pm$ 12.5 & 2.3 $\pm$ 1.7 & 797.7 $\pm$ 1.7 & 119.5 $\pm$ 12.5 & 40.3 $\pm$ 6.2 & 99.7 $\pm$ 0.2 & 97.3 $\pm$ 1.8 & 56.6 $\pm$ 6.4 & 57.9 $\pm$ 4.9 \\ \cline{2-11} 
 & \textbf{MI4P} & 157.3 $\pm$ 5.5 & 2.5 $\pm$ 1.7 & 797.5 $\pm$ 1.7 & 42.6 $\pm$ 5.5 & 78.7 $\pm$ 2.8 & 99.7 $\pm$ 0.2 & 98.5 $\pm$ 1 & 87.4 $\pm$ 1.7 & 85.5 $\pm$ 1.7 \\ \hline
\multirow{2}{*}{\textbf{10\%}} & \textbf{DAPAR} & 79.6 $\pm$ 12.8 & 2.7 $\pm$ 2 & 797.3 $\pm$ 2 & 120.4 $\pm$ 12.8 & 39.8 $\pm$ 6.4 & 99.7 $\pm$ 0.2 & 96.9 $\pm$ 2.1 & 56.1 $\pm$ 6.5 & 57.3 $\pm$ 5 \\ \cline{2-11} 
 & \textbf{MI4P} & 156.2 $\pm$ 5.7 & 2.4 $\pm$ 1.6 & 797.6 $\pm$ 1.6 & 43.8 $\pm$ 5.7 & 78.1 $\pm$ 2.8 & 99.7 $\pm$ 0.2 & 98.5 $\pm$ 1 & 87.1 $\pm$ 1.8 & 85.2 $\pm$ 1.9 \\ \hline
\multirow{2}{*}{\textbf{15\%}} & \textbf{DAPAR} & 80.6 $\pm$ 15 & 3.2 $\pm$ 2.4 & 796.8 $\pm$ 2.4 & 119.4 $\pm$ 15 & 40.3 $\pm$ 7.5 & 99.6 $\pm$ 0.3 & 96.3 $\pm$ 2.5 & 56.3 $\pm$ 8.3 & 57.3 $\pm$ 6.6 \\ \cline{2-11} 
 & \textbf{MI4P} & 150.7 $\pm$ 6.7 & 1.6 $\pm$ 1.2 & 798.4 $\pm$ 1.2 & 49.3 $\pm$ 6.7 & 75.3 $\pm$ 3.4 & 99.8 $\pm$ 0.1 & 98.9 $\pm$ 0.8 & 85.5 $\pm$ 2.2 & 83.6 $\pm$ 2.2 \\ \hline
\multirow{2}{*}{\textbf{20\%}} & \textbf{DAPAR} & 80.5 $\pm$ 15.3 & 3.9 $\pm$ 2.6 & 796.1 $\pm$ 2.6 & 119.5 $\pm$ 15.3 & 40.3 $\pm$ 7.6 & 99.5 $\pm$ 0.3 & 95.5 $\pm$ 2.7 & 56.2 $\pm$ 8.1 & 57 $\pm$ 6.3 \\ \cline{2-11} 
 & \textbf{MI4P} & 144 $\pm$ 6.9 & 0.9 $\pm$ 1 & 799.1 $\pm$ 1 & 56 $\pm$ 6.9 & 72 $\pm$ 3.4 & 99.9 $\pm$ 0.1 & 99.4 $\pm$ 0.7 & 83.4 $\pm$ 2.3 & 81.7 $\pm$ 2.3 \\ \hline
\multirow{2}{*}{\textbf{25\%}} & \textbf{DAPAR} & 79.7 $\pm$ 17.6 & 4.6 $\pm$ 3.2 & 795.4 $\pm$ 3.2 & 120.3 $\pm$ 17.6 & 39.9 $\pm$ 8.8 & 99.4 $\pm$ 0.4 & 94.8 $\pm$ 2.8 & 55.5 $\pm$ 9.5 & 56.3 $\pm$ 7.3 \\ \cline{2-11} 
 & \textbf{MI4P} & 137.2 $\pm$ 6.7 & 0.6 $\pm$ 0.8 & 799.4 $\pm$ 0.8 & 62.8 $\pm$ 6.7 & 68.6 $\pm$ 3.3 & 99.9 $\pm$ 0.1 & 99.6 $\pm$ 0.6 & 81.2 $\pm$ 2.4 & 79.5 $\pm$ 2.3 \\ \hline
\end{tabular}
\caption{Performance evaluation on the second set of simulations imputed using Bayesian linear regression.}
\label{Table:L16bis:impNORM:adjp}
\end{table}
\end{landscape}

\begin{landscape}
\begin{table}[ht]
\centering
\begin{tabular}{|c|c|c|c|c|c|c|c|c|c|c|}
\hline
\multicolumn{1}{|c|}{\textbf{\%MV}} & \textbf{Method}   & 
\begin{tabular}[c]{@{}c@{}}\textbf{True}\\\textbf{positives}\end{tabular} & \begin{tabular}[c]{@{}c@{}}\textbf{False}\\\textbf{positives}\end{tabular} & \begin{tabular}[c]{@{}c@{}}\textbf{True}\\\textbf{negatives}\end{tabular} & \begin{tabular}[c]{@{}c@{}}\textbf{False}\\\textbf{negatives}\end{tabular} & 
\begin{tabular}[c]{@{}c@{}}\textbf{Sensitivity}\\\textbf{(\%)}\end{tabular} &
\begin{tabular}[c]{@{}c@{}}\textbf{Specificity}\\\textbf{(\%)}\end{tabular} &
\begin{tabular}[c]{@{}c@{}}\textbf{Precision}\\\textbf{(\%)}\end{tabular} & 
\begin{tabular}[c]{@{}c@{}}\textbf{F-score}\\\textbf{(\%)}\end{tabular} & 
\begin{tabular}[c]{@{}c@{}}\textbf{MCC}\\\textbf{(\%)}\end{tabular}\\ \hline
\multirow{2}{*}{\textbf{1\%}} & \textbf{DAPAR} & 80.6 $\pm$ 11.8 & 1.9 $\pm$ 1.5 & 798.1 $\pm$ 1.5 & 119.4 $\pm$ 11.8 & 40.3 $\pm$ 5.9 & 99.8 $\pm$ 0.2 & 97.8 $\pm$ 1.7 & 56.8 $\pm$ 6.1 & 58.1 $\pm$ 4.8 \\ \cline{2-11} 
 & \textbf{MI4P} & 168.1 $\pm$ 4.8 & 6.8 $\pm$ 2.7 & 793.2 $\pm$ 2.7 & 31.9 $\pm$ 4.8 & 84 $\pm$ 2.4 & 99.2 $\pm$ 0.3 & 96.1 $\pm$ 1.5 & 89.7 $\pm$ 1.5 & 87.6 $\pm$ 1.7 \\ \hline
\multirow{2}{*}{\textbf{5\%}} & \textbf{DAPAR} & 80.9 $\pm$ 12.6 & 2.4 $\pm$ 1.8 & 797.6 $\pm$ 1.8 & 119.1 $\pm$ 12.6 & 40.4 $\pm$ 6.3 & 99.7 $\pm$ 0.2 & 97.2 $\pm$ 2 & 56.8 $\pm$ 6.5 & 58 $\pm$ 5 \\ \cline{2-11} 
 & \textbf{MI4P} & 170 $\pm$ 4.6 & 7.6 $\pm$ 2.9 & 792.5 $\pm$ 2.9 & 30 $\pm$ 4.6 & 85 $\pm$ 2.3 & 99.1 $\pm$ 0.4 & 95.8 $\pm$ 1.6 & 90 $\pm$ 1.4 & 88 $\pm$ 1.6 \\ \hline
\multirow{2}{*}{\textbf{10\%}} & \textbf{DAPAR} & 79.9 $\pm$ 13 & 2.8 $\pm$ 1.9 & 797.2 $\pm$ 1.9 & 120.1 $\pm$ 13 & 40 $\pm$ 6.5 & 99.7 $\pm$ 0.2 & 96.8 $\pm$ 2 & 56.2 $\pm$ 6.6 & 57.4 $\pm$ 5.1 \\ \cline{2-11} 
 & \textbf{MI4P} & 172.1 $\pm$ 4.6 & 8.2 $\pm$ 3 & 791.8 $\pm$ 3 & 27.9 $\pm$ 4.6 & 86.1 $\pm$ 2.3 & 99 $\pm$ 0.4 & 95.5 $\pm$ 1.5 & 90.5 $\pm$ 1.4 & 88.5 $\pm$ 1.6 \\ \hline
\multirow{2}{*}{\textbf{15\%}} & \textbf{DAPAR} & 81.8 $\pm$ 12.9 & 3.6 $\pm$ 2.5 & 796.4 $\pm$ 2.5 & 118.2 $\pm$ 12.9 & 40.9 $\pm$ 6.4 & 99.6 $\pm$ 0.3 & 95.9 $\pm$ 2.5 & 57 $\pm$ 6.5 & 57.8 $\pm$ 5.1 \\ \cline{2-11} 
 & \textbf{MI4P} & 174.2 $\pm$ 4 & 9.4 $\pm$ 3.6 & 790.6 $\pm$ 3.6 & 25.8 $\pm$ 4 & 87.1 $\pm$ 2 & 98.8 $\pm$ 0.5 & 94.9 $\pm$ 1.9 & 90.8 $\pm$ 1.3 & 88.8 $\pm$ 1.6 \\ \hline
\multirow{2}{*}{\textbf{20\%}} & \textbf{DAPAR} & 82.1 $\pm$ 15.4 & 4.4 $\pm$ 2.6 & 795.6 $\pm$ 2.6 & 117.9 $\pm$ 15.4 & 41 $\pm$ 7.7 & 99.5 $\pm$ 0.3 & 95.1 $\pm$ 2.7 & 56.8 $\pm$ 8 & 57.4 $\pm$ 6.2 \\ \cline{2-11} 
 & \textbf{MI4P} & 175.6 $\pm$ 4.1 & 11.3 $\pm$ 4.1 & 788.7 $\pm$ 4.1 & 24.4 $\pm$ 4.1 & 87.8 $\pm$ 2.1 & 98.6 $\pm$ 0.5 & 94 $\pm$ 2 & 90.8 $\pm$ 1.5 & 88.7 $\pm$ 1.8 \\ \hline
\multirow{2}{*}{\textbf{25\%}} & \textbf{DAPAR} & 83.3 $\pm$ 14.6 & 5.3 $\pm$ 2.9 & 794.7 $\pm$ 2.9 & 116.7 $\pm$ 14.6 & 41.6 $\pm$ 7.3 & 99.3 $\pm$ 0.4 & 94.1 $\pm$ 2.8 & 57.3 $\pm$ 7.3 & 57.5 $\pm$ 5.8 \\ \cline{2-11} 
 & \textbf{MI4P} & 176.3 $\pm$ 4.5 & 13 $\pm$ 3.8 & 787 $\pm$ 3.8 & 23.7 $\pm$ 4.5 & 88.1 $\pm$ 2.3 & 98.4 $\pm$ 0.5 & 93.2 $\pm$ 1.9 & 90.6 $\pm$ 1.5 & 88.4 $\pm$ 1.8 \\ \hline
\end{tabular}
\caption{Performance evaluation on the second set of simulations imputed using principal component analysis.}
\label{Table:L16bis:impPCA:adjp}
\end{table}
\end{landscape}

\begin{landscape}
\begin{table}[ht]
\centering
\begin{tabular}{|c|c|c|c|c|c|c|c|c|c|c|}
\hline
\multicolumn{1}{|c|}{\textbf{\%MV}} & \textbf{Method}   & 
\begin{tabular}[c]{@{}c@{}}\textbf{True}\\\textbf{positives}\end{tabular} & \begin{tabular}[c]{@{}c@{}}\textbf{False}\\\textbf{positives}\end{tabular} & \begin{tabular}[c]{@{}c@{}}\textbf{True}\\\textbf{negatives}\end{tabular} & \begin{tabular}[c]{@{}c@{}}\textbf{False}\\\textbf{negatives}\end{tabular} & 
\begin{tabular}[c]{@{}c@{}}\textbf{Sensitivity}\\\textbf{(\%)}\end{tabular} &
\begin{tabular}[c]{@{}c@{}}\textbf{Specificity}\\\textbf{(\%)}\end{tabular} &
\begin{tabular}[c]{@{}c@{}}\textbf{Precision}\\\textbf{(\%)}\end{tabular} & 
\begin{tabular}[c]{@{}c@{}}\textbf{F-score}\\\textbf{(\%)}\end{tabular} & 
\begin{tabular}[c]{@{}c@{}}\textbf{MCC}\\\textbf{(\%)}\end{tabular}\\ \hline
\multirow{2}{*}{\textbf{1\%}} & \textbf{DAPAR} & 80.8 $\pm$ 11.7 & 1.9 $\pm$ 1.5 & 798.1 $\pm$ 1.5 & 119.2 $\pm$ 11.7 & 40.4 $\pm$ 5.8 & 99.8 $\pm$ 0.2 & 97.8 $\pm$ 1.7 & 56.9 $\pm$ 6 & 58.2 $\pm$ 4.7 \\ \cline{2-11} 
 & \textbf{MI4P} & 168 $\pm$ 4.7 & 6.8 $\pm$ 2.7 & 793.2 $\pm$ 2.7 & 32 $\pm$ 4.7 & 84 $\pm$ 2.4 & 99.2 $\pm$ 0.3 & 96.1 $\pm$ 1.4 & 89.6 $\pm$ 1.4 & 87.6 $\pm$ 1.7 \\ \hline
\multirow{2}{*}{\textbf{5\%}} & \textbf{DAPAR} & 80.7 $\pm$ 12.7 & 2.4 $\pm$ 1.9 & 797.6 $\pm$ 1.9 & 119.3 $\pm$ 12.7 & 40.3 $\pm$ 6.3 & 99.7 $\pm$ 0.2 & 97.2 $\pm$ 2 & 56.7 $\pm$ 6.5 & 57.9 $\pm$ 5 \\ \cline{2-11} 
 & \textbf{MI4P} & 169.9 $\pm$ 4.4 & 7.5 $\pm$ 3 & 792.5 $\pm$ 3 & 30.1 $\pm$ 4.4 & 85 $\pm$ 2.2 & 99.1 $\pm$ 0.4 & 95.8 $\pm$ 1.6 & 90 $\pm$ 1.4 & 88 $\pm$ 1.6 \\ \hline
\multirow{2}{*}{\textbf{10\%}} & \textbf{DAPAR} & 79.9 $\pm$ 12.5 & 2.7 $\pm$ 1.8 & 797.3 $\pm$ 1.8 & 120.1 $\pm$ 12.5 & 40 $\pm$ 6.3 & 99.7 $\pm$ 0.2 & 96.8 $\pm$ 2 & 56.3 $\pm$ 6.4 & 57.5 $\pm$ 5 \\ \cline{2-11} 
 & \textbf{MI4P} & 171.6 $\pm$ 4.6 & 8.1 $\pm$ 3.1 & 792 $\pm$ 3.1 & 28.4 $\pm$ 4.6 & 85.8 $\pm$ 2.3 & 99 $\pm$ 0.4 & 95.5 $\pm$ 1.6 & 90.4 $\pm$ 1.5 & 88.4 $\pm$ 1.7 \\ \hline
\multirow{2}{*}{\textbf{15\%}} & \textbf{DAPAR} & 81.4 $\pm$ 13.8 & 3.5 $\pm$ 2.4 & 796.5 $\pm$ 2.4 & 118.6 $\pm$ 13.8 & 40.7 $\pm$ 6.9 & 99.6 $\pm$ 0.3 & 96 $\pm$ 2.4 & 56.8 $\pm$ 7.1 & 57.6 $\pm$ 5.5 \\ \cline{2-11} 
 & \textbf{MI4P} & 173.5 $\pm$ 4 & 9.3 $\pm$ 3.8 & 790.7 $\pm$ 3.8 & 26.5 $\pm$ 4 & 86.8 $\pm$ 2 & 98.8 $\pm$ 0.5 & 94.9 $\pm$ 1.9 & 90.6 $\pm$ 1.4 & 88.6 $\pm$ 1.7 \\ \hline
\multirow{2}{*}{\textbf{20\%}} & \textbf{DAPAR} & 82.1 $\pm$ 13.5 & 4.4 $\pm$ 2.6 & 795.6 $\pm$ 2.6 & 117.9 $\pm$ 13.5 & 41.1 $\pm$ 6.8 & 99.4 $\pm$ 0.3 & 95 $\pm$ 2.6 & 57 $\pm$ 6.9 & 57.5 $\pm$ 5.4 \\ \cline{2-11} 
 & \textbf{MI4P} & 174.4 $\pm$ 4.1 & 10.9 $\pm$ 3.9 & 789.1 $\pm$ 3.9 & 25.6 $\pm$ 4.1 & 87.2 $\pm$ 2 & 98.6 $\pm$ 0.5 & 94.1 $\pm$ 2 & 90.5 $\pm$ 1.4 & 88.4 $\pm$ 1.7 \\ \hline
\multirow{2}{*}{\textbf{25\%}} & \textbf{DAPAR} & 82.2 $\pm$ 16 & 5 $\pm$ 2.9 & 795 $\pm$ 2.9 & 117.8 $\pm$ 16 & 41.1 $\pm$ 8 & 99.4 $\pm$ 0.4 & 94.4 $\pm$ 2.8 & 56.8 $\pm$ 8.5 & 57.2 $\pm$ 6.7 \\ \cline{2-11} 
 & \textbf{MI4P} & 174.7 $\pm$ 4.5 & 12.4 $\pm$ 4 & 787.6 $\pm$ 4 & 25.3 $\pm$ 4.5 & 87.3 $\pm$ 2.2 & 98.5 $\pm$ 0.5 & 93.4 $\pm$ 1.9 & 90.3 $\pm$ 1.5 & 88 $\pm$ 1.8 \\ \hline
\end{tabular}
\caption{Performance evaluation on the second set of simulations imputed using random forests.}
\label{Table:L16bis:impRF:adjp}
\end{table}
\end{landscape}

\section{Results on the third set of simulations}
\subsection{Simulation design}
Finally, we considered an experimental design similar to the second one, but with random effects $P_{i}$ and $G_{ik}$. The 100 datasets were generated as follows. 
\begin{enumerate}
    \item For the first group, replicate 10 times (for the 10 variables in this group) a draw from a mixture of 2 Gaussian distributions. The first one has the following parameters: a mean of 1.5 and a standard deviation of 0.5 (corresponds to $P_{i}$). The second one has the following parameters: a mean of 0 and a standard deviation of 0.5 (corresponds to $\epsilon_{ij}$).
    \item For the second group replicate 10 times (for the 10 variables in this group) a draw from a mixture of the following 3 distributions.
\begin{enumerate}
    \item The first one is a Gaussian distribution with the following parameters: a mean of 1.5 and a standard deviation of 0.5 (corresponds to $P_{i}$).
    \item The second one is the mixture of a Gaussian distribution with a mean of 1.5 and a standard deviation of 0.5 for the 200 first rows (set as differentially expressed) and a zero vector for the remaining 800 rows (set as not differentially expressed). This mixture illustrates the $G_{ik}$ term in the previous model.
    \item The third distribution has the following parameters: a mean of 0 and a standard deviation of 0.5 (corresponds to $\epsilon_{ij}$).
\end{enumerate}
\end{enumerate}

\newpage
\subsection{Performance evaluation}
This subsection provides the evaluation of the \texttt{mi4p} workflow compared to the \texttt{DAPAR} workflow on the first set of simulations. The performance is described using the indicators detailed in Section \ref{sec:Perf}.

\begin{figure}[ht]
    \centering
    \makebox[\textwidth][c]{\includegraphics[width=1.2\textwidth]{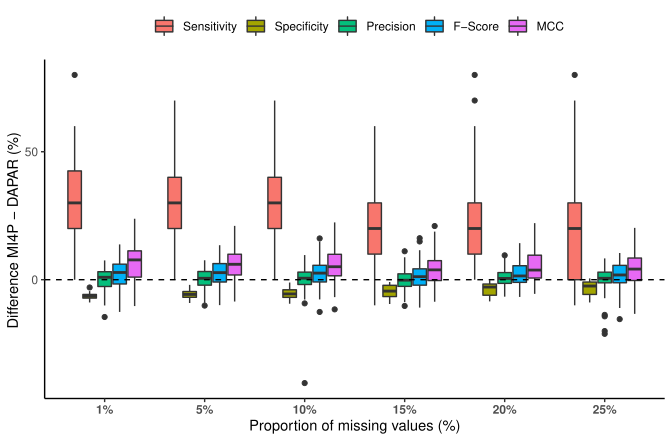}}
    \caption{Distribution of the difference of performance between \texttt{mi4p} and \texttt{DAPAR} workflows on the third set of simulations imputed using maximum likelihood estimation.}
    \label{fig:L16:Bplot}
\end{figure}

The following tables provide results expressed as the mean of the given indicator over the 100 simulated datasets $\pm$ the mean of the standard deviations of the given indicator over the 100 simulated datasets. Results are based on adjusted p-values using the Benjamini-Hochberg procedure \citep{benjaminiControllingFalseDiscovery1995} and a false discovery rate of 1\%.

\begin{landscape}
\begin{table}
\centering
\small
\begin{tabular}{|c|c|c|c|c|c|c|c|c|c|c|}
\hline
\multicolumn{1}{|c|}{\textbf{\%MV}}& \textbf{Method}   & \begin{tabular}[c]{@{}c@{}}\textbf{True}\\\textbf{positives}\end{tabular} & \begin{tabular}[c]{@{}c@{}}\textbf{False}\\\textbf{positives}\end{tabular} & \begin{tabular}[c]{@{}c@{}}\textbf{True}\\\textbf{negatives}\end{tabular} & \begin{tabular}[c]{@{}c@{}}\textbf{False}\\\textbf{negatives}\end{tabular} & 
\begin{tabular}[c]{@{}c@{}}\textbf{Sensitivity}\\\textbf{(\%)}\end{tabular} &
\begin{tabular}[c]{@{}c@{}}\textbf{Specificity}\\\textbf{(\%)}\end{tabular} &
\begin{tabular}[c]{@{}c@{}}\textbf{Precision}\\\textbf{(\%)}\end{tabular} & 
\begin{tabular}[c]{@{}c@{}}\textbf{F-score}\\\textbf{(\%)}\end{tabular} & 
\begin{tabular}[c]{@{}c@{}}\textbf{MCC}\\\textbf{(\%)}\end{tabular}\\ \hline
\multirow{2}{*}{\textbf{1\%}} & \textbf{DAPAR} & 25.6 $\pm$ 10.7 & 0.5 $\pm$ 0.8 & 799.5 $\pm$ 0.8 & 174.4 $\pm$ 10.7 & 12.8 $\pm$ 5.4 & 99.9 $\pm$ 0.1 & 98.3 $\pm$ 2.4 & 22.2 $\pm$ 8.4 & 31.2 $\pm$ 7.4 \\ \cline{2-11} 
 & \textbf{MI4P} & 91 $\pm$ 10.6 & 2.7 $\pm$ 1.8 & 797.3 $\pm$ 1.8 & 109 $\pm$ 10.6 & 45.5 $\pm$ 5.3 & 99.7 $\pm$ 0.2 & 97.2 $\pm$ 1.8 & 61.8 $\pm$ 4.9 & 61.9 $\pm$ 4 \\ \hline
\multirow{2}{*}{\textbf{5\%}} & \textbf{DAPAR} & 25.6 $\pm$ 10.2 & 0.4 $\pm$ 0.7 & 799.6 $\pm$ 0.7 & 174.4 $\pm$ 10.2 & 12.8 $\pm$ 5.1 & 99.9 $\pm$ 0.1 & 98.5 $\pm$ 2.4 & 22.3 $\pm$ 7.9 & 31.4 $\pm$ 6.8 \\ \cline{2-11} 
 & \textbf{MI4P} & 83 $\pm$ 13.6 & 2.1 $\pm$ 1.8 & 797.9 $\pm$ 1.8 & 117 $\pm$ 13.6 & 41.5 $\pm$ 6.8 & 99.7 $\pm$ 0.2 & 97.6 $\pm$ 1.9 & 57.9 $\pm$ 6.7 & 59 $\pm$ 5.1 \\ \hline
\multirow{2}{*}{\textbf{10\%}} & \textbf{DAPAR} & 25.9 $\pm$ 10.8 & 0.6 $\pm$ 0.7 & 799.4 $\pm$ 0.7 & 174.1 $\pm$ 10.8 & 13 $\pm$ 5.4 & 99.9 $\pm$ 0.1 & 96.1 $\pm$ 14 & 22.5 $\pm$ 8.6 & 31.1 $\pm$ 8.3 \\ \cline{2-11} 
 & \textbf{MI4P} & 80.2 $\pm$ 18.2 & 2.3 $\pm$ 2.1 & 797.7 $\pm$ 2.1 & 119.8 $\pm$ 18.2 & 40.1 $\pm$ 9.1 & 99.7 $\pm$ 0.3 & 97.5 $\pm$ 2 & 56.2 $\pm$ 9.2 & 57.6 $\pm$ 6.9 \\ \hline
\multirow{2}{*}{\textbf{15\%}} & \textbf{DAPAR} & 26.6 $\pm$ 11.5 & 0.8 $\pm$ 1 & 799.2 $\pm$ 1 & 173.4 $\pm$ 11.5 & 13.3 $\pm$ 5.7 & 99.9 $\pm$ 0.1 & 96.5 $\pm$ 10.3 & 23 $\pm$ 9 & 31.5 $\pm$ 8.2 \\ \cline{2-11} 
 & \textbf{MI4P} & 71.9 $\pm$ 22.7 & 2.1 $\pm$ 2.3 & 797.9 $\pm$ 2.3 & 128.1 $\pm$ 22.7 & 35.9 $\pm$ 11.3 & 99.7 $\pm$ 0.3 & 97.7 $\pm$ 2.3 & 51.4 $\pm$ 12.3 & 54 $\pm$ 9.1 \\ \hline
\multirow{2}{*}{\textbf{20\%}} & \textbf{DAPAR} & 28.5 $\pm$ 12.1 & 1.1 $\pm$ 1.3 & 798.9 $\pm$ 1.3 & 171.5 $\pm$ 12.1 & 14.2 $\pm$ 6.1 & 99.9 $\pm$ 0.2 & 95.4 $\pm$ 10.4 & 24.3 $\pm$ 9.3 & 32.3 $\pm$ 8.5 \\ \cline{2-11} 
 & \textbf{MI4P} & 67.1 $\pm$ 22.4 & 1.9 $\pm$ 2.3 & 798.1 $\pm$ 2.3 & 132.9 $\pm$ 22.4 & 33.6 $\pm$ 11.2 & 99.8 $\pm$ 0.3 & 97.8 $\pm$ 2.3 & 48.8 $\pm$ 12.4 & 52 $\pm$ 9.2 \\ \hline
\multirow{2}{*}{\textbf{25\%}} & \textbf{DAPAR} & 26.9 $\pm$ 12.4 & 1.3 $\pm$ 1.4 & 798.7 $\pm$ 1.4 & 173.1 $\pm$ 12.4 & 13.4 $\pm$ 6.2 & 99.8 $\pm$ 0.2 & 96.2 $\pm$ 4 & 23 $\pm$ 9.7 & 31.1 $\pm$ 8.6 \\ \cline{2-11} 
 & \textbf{MI4P} & 61.2 $\pm$ 24 & 2 $\pm$ 2.8 & 798 $\pm$ 2.8 & 138.8 $\pm$ 24 & 30.6 $\pm$ 12 & 99.7 $\pm$ 0.4 & 97.7 $\pm$ 2.8 & 45.2 $\pm$ 13.6 & 49.2 $\pm$ 10 \\ \hline
\end{tabular}
\caption{Performance evaluation on the third set of simulation imputed using maximum likelihood estimation}
\label{Table:L16Sim:impMLE:adjp}
\end{table}
\end{landscape}

\begin{landscape}
\begin{table}
\centering
\small
\begin{tabular}{|c|c|c|c|c|c|c|c|c|c|c|}
\hline
\multicolumn{1}{|c|}{\textbf{\%MV}}& \textbf{Method}   & \begin{tabular}[c]{@{}c@{}}\textbf{True}\\\textbf{positives}\end{tabular} & \begin{tabular}[c]{@{}c@{}}\textbf{False}\\\textbf{positives}\end{tabular} & \begin{tabular}[c]{@{}c@{}}\textbf{True}\\\textbf{negatives}\end{tabular} & \begin{tabular}[c]{@{}c@{}}\textbf{False}\\\textbf{negatives}\end{tabular} & 
\begin{tabular}[c]{@{}c@{}}\textbf{Sensitivity}\\\textbf{(\%)}\end{tabular} &
\begin{tabular}[c]{@{}c@{}}\textbf{Specificity}\\\textbf{(\%)}\end{tabular} &
\begin{tabular}[c]{@{}c@{}}\textbf{Precision}\\\textbf{(\%)}\end{tabular} & 
\begin{tabular}[c]{@{}c@{}}\textbf{F-score}\\\textbf{(\%)}\end{tabular} & 
\begin{tabular}[c]{@{}c@{}}\textbf{MCC}\\\textbf{(\%)}\end{tabular}\\ \hline
\multirow{2}{*}{\textbf{1\%}} & \textbf{DAPAR} & 26 $\pm$ 10.4 & 0.5 $\pm$ 0.8 & 799.5 $\pm$ 0.8 & 174 $\pm$ 10.4 & 13 $\pm$ 5.2 & 99.9 $\pm$ 0.1 & 98.5 $\pm$ 2.3 & 22.5 $\pm$ 8.1 & 31.5 $\pm$ 7 \\ \cline{2-11} 
 & \textbf{MI4P} & 95.8 $\pm$ 9.8 & 3.1 $\pm$ 1.9 & 796.9 $\pm$ 1.9 & 104.2 $\pm$ 9.8 & 47.9 $\pm$ 4.9 & 99.6 $\pm$ 0.2 & 96.9 $\pm$ 1.8 & 64 $\pm$ 4.4 & 63.6 $\pm$ 3.7 \\ \hline
\multirow{2}{*}{\textbf{5\%}} & \textbf{DAPAR} & 25.4 $\pm$ 11.1 & 0.4 $\pm$ 0.7 & 799.6 $\pm$ 0.7 & 174.6 $\pm$ 11.1 & 12.7 $\pm$ 5.5 & 99.9 $\pm$ 0.1 & 98.5 $\pm$ 2.5 & 22.1 $\pm$ 8.7 & 31.1 $\pm$ 7.5 \\ \cline{2-11} 
 & \textbf{MI4P} & 98 $\pm$ 9.9 & 2.9 $\pm$ 1.8 & 797.1 $\pm$ 1.8 & 102 $\pm$ 9.9 & 49 $\pm$ 4.9 & 99.6 $\pm$ 0.2 & 97.1 $\pm$ 1.7 & 65 $\pm$ 4.4 & 64.6 $\pm$ 3.7 \\ \hline
\multirow{2}{*}{\textbf{10\%}} & \textbf{DAPAR} & 24.5 $\pm$ 10.6 & 0.6 $\pm$ 0.9 & 799.4 $\pm$ 0.9 & 175.5 $\pm$ 10.6 & 12.3 $\pm$ 5.3 & 99.9 $\pm$ 0.1 & 95.8 $\pm$ 14.1 & 21.4 $\pm$ 8.4 & 30.2 $\pm$ 7.9 \\ \cline{2-11} 
 & \textbf{MI4P} & 101.1 $\pm$ 9.5 & 3.2 $\pm$ 1.8 & 796.8 $\pm$ 1.8 & 98.9 $\pm$ 9.5 & 50.6 $\pm$ 4.8 & 99.6 $\pm$ 0.2 & 97 $\pm$ 1.6 & 66.3 $\pm$ 4.1 & 65.6 $\pm$ 3.5 \\ \hline
\multirow{2}{*}{\textbf{15\%}} & \textbf{DAPAR} & 25.1 $\pm$ 12.2 & 0.4 $\pm$ 0.7 & 799.6 $\pm$ 0.7 & 174.9 $\pm$ 12.2 & 12.5 $\pm$ 6.1 & 99.9 $\pm$ 0.1 & 96.4 $\pm$ 14.1 & 21.7 $\pm$ 9.7 & 30.4 $\pm$ 9.2 \\ \cline{2-11} 
 & \textbf{MI4P} & 103.8 $\pm$ 10.9 & 2.6 $\pm$ 1.4 & 797.4 $\pm$ 1.4 & 96.2 $\pm$ 10.9 & 51.9 $\pm$ 5.4 & 99.7 $\pm$ 0.2 & 97.6 $\pm$ 1.3 & 67.6 $\pm$ 4.7 & 66.8 $\pm$ 4 \\ \hline
\multirow{2}{*}{\textbf{20\%}} & \textbf{DAPAR} & 24.7 $\pm$ 13.2 & 0.4 $\pm$ 0.7 & 799.6 $\pm$ 0.7 & 175.3 $\pm$ 13.2 & 12.3 $\pm$ 6.6 & 99.9 $\pm$ 0.1 & 95.6 $\pm$ 17.1 & 21.3 $\pm$ 10.4 & 29.9 $\pm$ 10.1 \\ \cline{2-11} 
 & \textbf{MI4P} & 106.2 $\pm$ 11.9 & 2.7 $\pm$ 1.7 & 797.3 $\pm$ 1.7 & 93.8 $\pm$ 11.9 & 53.1 $\pm$ 5.9 & 99.7 $\pm$ 0.2 & 97.6 $\pm$ 1.4 & 68.6 $\pm$ 5 & 67.7 $\pm$ 4.3 \\ \hline
\multirow{2}{*}{\textbf{25\%}} & \textbf{DAPAR} & 24.7 $\pm$ 12.3 & 0.6 $\pm$ 0.9 & 799.4 $\pm$ 0.9 & 175.3 $\pm$ 12.3 & 12.3 $\pm$ 6.2 & 99.9 $\pm$ 0.1 & 96.8 $\pm$ 10.3 & 21.4 $\pm$ 9.7 & 30.1 $\pm$ 8.9 \\ \cline{2-11} 
 & \textbf{MI4P} & 105.4 $\pm$ 11.1 & 2.9 $\pm$ 1.9 & 797.1 $\pm$ 1.9 & 94.6 $\pm$ 11.1 & 52.7 $\pm$ 5.5 & 99.6 $\pm$ 0.2 & 97.4 $\pm$ 1.6 & 68.2 $\pm$ 4.7 & 67.3 $\pm$ 4 \\ \hline

\end{tabular}
\caption{Performance evaluation on the third set of simulations imputed using $k$-nearest neighbours method.}
\label{Table:L16Sim:impKNN:adjp}
\end{table}
\end{landscape}

\begin{landscape}
\begin{table}
\centering
\small
\begin{tabular}{|c|c|c|c|c|c|c|c|c|c|c|}
\hline
\multicolumn{1}{|c|}{\textbf{\%MV}}& \textbf{Method}   & \begin{tabular}[c]{@{}c@{}}\textbf{True}\\\textbf{positives}\end{tabular} & \begin{tabular}[c]{@{}c@{}}\textbf{False}\\\textbf{positives}\end{tabular} & \begin{tabular}[c]{@{}c@{}}\textbf{True}\\\textbf{negatives}\end{tabular} & \begin{tabular}[c]{@{}c@{}}\textbf{False}\\\textbf{negatives}\end{tabular} & 
\begin{tabular}[c]{@{}c@{}}\textbf{Sensitivity}\\\textbf{(\%)}\end{tabular} &
\begin{tabular}[c]{@{}c@{}}\textbf{Specificity}\\\textbf{(\%)}\end{tabular} &
\begin{tabular}[c]{@{}c@{}}\textbf{Precision}\\\textbf{(\%)}\end{tabular} & 
\begin{tabular}[c]{@{}c@{}}\textbf{F-score}\\\textbf{(\%)}\end{tabular} & 
\begin{tabular}[c]{@{}c@{}}\textbf{MCC}\\\textbf{(\%)}\end{tabular}\\ \hline
\multirow{2}{*}{\textbf{1\%}} & \textbf{DAPAR} & 25.8 $\pm$ 10.6 & 0.5 $\pm$ 0.8 & 799.5 $\pm$ 0.8 & 174.2 $\pm$ 10.6 & 12.9 $\pm$ 5.3 & 99.9 $\pm$ 0.1 & 98.3 $\pm$ 2.5 & 22.4 $\pm$ 8.4 & 31.3 $\pm$ 7.4 \\ \cline{2-11} 
 & \textbf{MI4P} & 87.9 $\pm$ 9.5 & 2.2 $\pm$ 1.6 & 797.8 $\pm$ 1.6 & 112.1 $\pm$ 9.5 & 43.9 $\pm$ 4.8 & 99.7 $\pm$ 0.2 & 97.6 $\pm$ 1.7 & 60.4 $\pm$ 4.5 & 60.9 $\pm$ 3.7 \\ \hline
\multirow{2}{*}{\textbf{5\%}} & \textbf{DAPAR} & 25.6 $\pm$ 10.7 & 0.5 $\pm$ 0.7 & 799.5 $\pm$ 0.7 & 174.4 $\pm$ 10.7 & 12.8 $\pm$ 5.4 & 99.9 $\pm$ 0.1 & 98.4 $\pm$ 2.4 & 22.3 $\pm$ 8.4 & 31.3 $\pm$ 7.3 \\ \cline{2-11} 
 & \textbf{MI4P} & 63.1 $\pm$ 10.4 & 0.5 $\pm$ 0.7 & 799.5 $\pm$ 0.7 & 136.9 $\pm$ 10.4 & 31.5 $\pm$ 5.2 & 99.9 $\pm$ 0.1 & 99.2 $\pm$ 1.1 & 47.6 $\pm$ 6.1 & 51.4 $\pm$ 4.6 \\ \hline
\multirow{2}{*}{\textbf{10\%}} & \textbf{DAPAR} & 24.4 $\pm$ 11.5 & 0.6 $\pm$ 0.8 & 799.4 $\pm$ 0.8 & 175.6 $\pm$ 11.5 & 12.2 $\pm$ 5.7 & 99.9 $\pm$ 0.1 & 96 $\pm$ 14.1 & 21.2 $\pm$ 9.2 & 29.9 $\pm$ 8.8 \\ \cline{2-11} 
 & \textbf{MI4P} & 37.2 $\pm$ 11.3 & 0.1 $\pm$ 0.3 & 799.9 $\pm$ 0.3 & 162.8 $\pm$ 11.3 & 18.6 $\pm$ 5.6 & 100 $\pm$ 0 & 99.7 $\pm$ 0.9 & 31 $\pm$ 8.1 & 38.8 $\pm$ 6.4 \\ \hline
\multirow{2}{*}{\textbf{15\%}} & \textbf{DAPAR} & 24.9 $\pm$ 12.4 & 0.7 $\pm$ 0.9 & 799.3 $\pm$ 0.9 & 175.1 $\pm$ 12.4 & 12.5 $\pm$ 6.2 & 99.9 $\pm$ 0.1 & 95.7 $\pm$ 14 & 21.6 $\pm$ 9.7 & 30.1 $\pm$ 9.2 \\ \cline{2-11} 
 & \textbf{MI4P} & 17.6 $\pm$ 11.7 & 0 $\pm$ 0.2 & 800 $\pm$ 0.2 & 182.4 $\pm$ 11.7 & 8.8 $\pm$ 5.8 & 100 $\pm$ 0 & 92.9 $\pm$ 25.6 & 15.6 $\pm$ 9.8 & 24.5 $\pm$ 11.1 \\ \hline
\multirow{2}{*}{\textbf{20\%}} & \textbf{DAPAR} & 23.3 $\pm$ 12.4 & 0.7 $\pm$ 1 & 799.3 $\pm$ 1 & 176.7 $\pm$ 12.4 & 11.6 $\pm$ 6.2 & 99.9 $\pm$ 0.1 & 96.3 $\pm$ 10.5 & 20.2 $\pm$ 9.8 & 28.9 $\pm$ 9.2 \\ \cline{2-11} 
 & \textbf{MI4P} & 6.4 $\pm$ 6.9 & 0 $\pm$ 0 & 800 $\pm$ 0 & 193.6 $\pm$ 6.9 & 3.2 $\pm$ 3.5 & 100 $\pm$ 0 & 74 $\pm$ 44.1 & 6 $\pm$ 6.3 & 12.8 $\pm$ 9.8 \\ \hline
\multirow{2}{*}{\textbf{25\%}} & \textbf{DAPAR} & 24.1 $\pm$ 11.8 & 0.8 $\pm$ 1.2 & 799.2 $\pm$ 1.2 & 175.8 $\pm$ 11.8 & 12.1 $\pm$ 5.9 & 99.9 $\pm$ 0.1 & 97.4 $\pm$ 3.5 & 21 $\pm$ 9.3 & 29.7 $\pm$ 8.2 \\ \cline{2-11} 
 & \textbf{MI4P} & 1.7 $\pm$ 3.2 & 0 $\pm$ 0 & 800 $\pm$ 0 & 198.3 $\pm$ 3.2 & 0.9 $\pm$ 1.6 & 100 $\pm$ 0 & 43 $\pm$ 49.8 & 1.7 $\pm$ 3 & 5 $\pm$ 6.8 \\ \hline

\end{tabular}
\caption{Performance evaluation on the third set of simulation imputed using Bayesian linear regression.}
\label{Table:L16Sim:impNORM:adjp}
\end{table}
\end{landscape}

\begin{landscape}
\begin{table}
\centering
\small
\begin{tabular}{|c|c|c|c|c|c|c|c|c|c|c|}
\hline
\multicolumn{1}{|c|}{\textbf{\%MV}}& \textbf{Method}   & \begin{tabular}[c]{@{}c@{}}\textbf{True}\\\textbf{positives}\end{tabular} & \begin{tabular}[c]{@{}c@{}}\textbf{False}\\\textbf{positives}\end{tabular} & \begin{tabular}[c]{@{}c@{}}\textbf{True}\\\textbf{negatives}\end{tabular} & \begin{tabular}[c]{@{}c@{}}\textbf{False}\\\textbf{negatives}\end{tabular} & 
\begin{tabular}[c]{@{}c@{}}\textbf{Sensitivity}\\\textbf{(\%)}\end{tabular} &
\begin{tabular}[c]{@{}c@{}}\textbf{Specificity}\\\textbf{(\%)}\end{tabular} &
\begin{tabular}[c]{@{}c@{}}\textbf{Precision}\\\textbf{(\%)}\end{tabular} & 
\begin{tabular}[c]{@{}c@{}}\textbf{F-score}\\\textbf{(\%)}\end{tabular} & 
\begin{tabular}[c]{@{}c@{}}\textbf{MCC}\\\textbf{(\%)}\end{tabular}\\ \hline
\multirow{2}{*}{\textbf{1\%}} & \textbf{DAPAR} & 25.8 $\pm$ 10.2 & 0.5 $\pm$ 0.8 & 799.5 $\pm$ 0.8 & 174.2 $\pm$ 10.2 & 12.9 $\pm$ 5.1 & 99.9 $\pm$ 0.1 & 98.3 $\pm$ 2.4 & 22.4 $\pm$ 8 & 31.4 $\pm$ 7 \\ \cline{2-11} 
 & \textbf{MI4P} & 95.7 $\pm$ 9.9 & 3.2 $\pm$ 1.8 & 796.8 $\pm$ 1.8 & 104.3 $\pm$ 9.9 & 47.9 $\pm$ 4.9 & 99.6 $\pm$ 0.2 & 96.8 $\pm$ 1.7 & 63.9 $\pm$ 4.4 & 63.5 $\pm$ 3.7 \\ \hline
\multirow{2}{*}{\textbf{5\%}} & \textbf{DAPAR} & 24.9 $\pm$ 10.4 & 0.5 $\pm$ 0.7 & 799.5 $\pm$ 0.7 & 175.2 $\pm$ 10.4 & 12.4 $\pm$ 5.2 & 99.9 $\pm$ 0.1 & 98.2 $\pm$ 2.5 & 21.7 $\pm$ 8.3 & 30.6 $\pm$ 7.5 \\ \cline{2-11} 
 & \textbf{MI4P} & 97.7 $\pm$ 9.5 & 3 $\pm$ 1.8 & 797 $\pm$ 1.8 & 102.3 $\pm$ 9.5 & 48.8 $\pm$ 4.7 & 99.6 $\pm$ 0.2 & 97 $\pm$ 1.7 & 64.8 $\pm$ 4.2 & 64.4 $\pm$ 3.6 \\ \hline
\multirow{2}{*}{\textbf{10\%}} & \textbf{DAPAR} & 24.5 $\pm$ 10.6 & 0.6 $\pm$ 0.9 & 799.4 $\pm$ 0.9 & 175.5 $\pm$ 10.6 & 12.3 $\pm$ 5.3 & 99.9 $\pm$ 0.1 & 95.8 $\pm$ 14.1 & 21.4 $\pm$ 8.4 & 30.2 $\pm$ 7.9 \\ \cline{2-11} 
 & \textbf{MI4P} & 101.1 $\pm$ 9.5 & 3.2 $\pm$ 1.8 & 796.8 $\pm$ 1.8 & 98.9 $\pm$ 9.5 & 50.6 $\pm$ 4.8 & 99.6 $\pm$ 0.2 & 97 $\pm$ 1.6 & 66.3 $\pm$ 4.1 & 65.6 $\pm$ 3.5 \\ \hline
\multirow{2}{*}{\textbf{15\%}} & \textbf{DAPAR} & 24.2 $\pm$ 12.4 & 0.7 $\pm$ 0.9 & 799.3 $\pm$ 0.9 & 175.8 $\pm$ 12.4 & 12.1 $\pm$ 6.2 & 99.9 $\pm$ 0.1 & 95.7 $\pm$ 14 & 21 $\pm$ 9.7 & 29.6 $\pm$ 9.1 \\ \cline{2-11} 
 & \textbf{MI4P} & 104.6 $\pm$ 10.1 & 3.4 $\pm$ 2.1 & 796.6 $\pm$ 2.1 & 95.4 $\pm$ 10.1 & 52.3 $\pm$ 5.1 & 99.6 $\pm$ 0.3 & 96.9 $\pm$ 1.8 & 67.8 $\pm$ 4.3 & 66.8 $\pm$ 3.7 \\ \hline
\multirow{2}{*}{\textbf{20\%}} & \textbf{DAPAR} & 23.6 $\pm$ 12.2 & 0.7 $\pm$ 0.9 & 799.3 $\pm$ 0.9 & 176.4 $\pm$ 12.2 & 11.8 $\pm$ 6.1 & 99.9 $\pm$ 0.1 & 94.7 $\pm$ 17.1 & 20.5 $\pm$ 9.7 & 29 $\pm$ 9.7 \\ \cline{2-11} 
 & \textbf{MI4P} & 110 $\pm$ 10.1 & 3.7 $\pm$ 2.1 & 796.3 $\pm$ 2.1 & 90 $\pm$ 10.1 & 55 $\pm$ 5.1 & 99.5 $\pm$ 0.3 & 96.8 $\pm$ 1.7 & 70 $\pm$ 4.2 & 68.7 $\pm$ 3.6 \\ \hline
\multirow{2}{*}{\textbf{25\%}} & \textbf{DAPAR} & 24.7 $\pm$ 11.3 & 0.8 $\pm$ 1.2 & 799.2 $\pm$ 1.2 & 175.3 $\pm$ 11.3 & 12.3 $\pm$ 5.7 & 99.9 $\pm$ 0.1 & 97.2 $\pm$ 3.6 & 21.4 $\pm$ 8.9 & 30.2 $\pm$ 7.7 \\ \cline{2-11} 
 & \textbf{MI4P} & 113.6 $\pm$ 9.3 & 4.4 $\pm$ 2.3 & 795.6 $\pm$ 2.3 & 86.4 $\pm$ 9.3 & 56.8 $\pm$ 4.6 & 99.4 $\pm$ 0.3 & 96.3 $\pm$ 1.7 & 71.3 $\pm$ 3.6 & 69.7 $\pm$ 3.2 \\ \hline
\end{tabular}
\caption{Performance evaluation on the third set of simulation imputed using principal component analysis.}
\label{Table:L16Sim:impPCA:adjp}
\end{table}
\end{landscape}

\begin{landscape}
\begin{table}
\centering
\small
\begin{tabular}{|c|c|c|c|c|c|c|c|c|c|c|}
\hline
\multicolumn{1}{|c|}{\textbf{\%MV}}& \textbf{Method}   & \begin{tabular}[c]{@{}c@{}}\textbf{True}\\\textbf{positives}\end{tabular} & \begin{tabular}[c]{@{}c@{}}\textbf{False}\\\textbf{positives}\end{tabular} & \begin{tabular}[c]{@{}c@{}}\textbf{True}\\\textbf{negatives}\end{tabular} & \begin{tabular}[c]{@{}c@{}}\textbf{False}\\\textbf{negatives}\end{tabular} & 
\begin{tabular}[c]{@{}c@{}}\textbf{Sensitivity}\\\textbf{(\%)}\end{tabular} &
\begin{tabular}[c]{@{}c@{}}\textbf{Specificity}\\\textbf{(\%)}\end{tabular} &
\begin{tabular}[c]{@{}c@{}}\textbf{Precision}\\\textbf{(\%)}\end{tabular} & 
\begin{tabular}[c]{@{}c@{}}\textbf{F-score}\\\textbf{(\%)}\end{tabular} & 
\begin{tabular}[c]{@{}c@{}}\textbf{MCC}\\\textbf{(\%)}\end{tabular}\\ \hline
\multirow{2}{*}{\textbf{1\%}} & \textbf{DAPAR} & 25.7 $\pm$ 10.2 & 0.5 $\pm$ 0.7 & 799.5 $\pm$ 0.7 & 174.3 $\pm$ 10.2 & 12.8 $\pm$ 5.1 & 99.9 $\pm$ 0.1 & 98.5 $\pm$ 2.3 & 22.3 $\pm$ 8 & 31.3 $\pm$ 7 \\ \cline{2-11} 
 & \textbf{MI4P} & 95.8 $\pm$ 9.8 & 3.1 $\pm$ 1.9 & 796.9 $\pm$ 1.9 & 104.2 $\pm$ 9.8 & 47.9 $\pm$ 4.9 & 99.6 $\pm$ 0.2 & 96.9 $\pm$ 1.8 & 63.9 $\pm$ 4.4 & 63.6 $\pm$ 3.7 \\ \hline
\multirow{2}{*}{\textbf{5\%}} & \textbf{DAPAR} & 25.2 $\pm$ 10.5 & 0.5 $\pm$ 0.7 & 799.5 $\pm$ 0.7 & 174.8 $\pm$ 10.5 & 12.6 $\pm$ 5.2 & 99.9 $\pm$ 0.1 & 98.4 $\pm$ 2.5 & 21.9 $\pm$ 8.2 & 31 $\pm$ 7.1 \\ \cline{2-11} 
 & \textbf{MI4P} & 97.7 $\pm$ 9.8 & 3 $\pm$ 1.8 & 797 $\pm$ 1.8 & 102.3 $\pm$ 9.8 & 48.8 $\pm$ 4.9 & 99.6 $\pm$ 0.2 & 97.1 $\pm$ 1.7 & 64.8 $\pm$ 4.3 & 64.4 $\pm$ 3.6 \\ \hline
\multirow{2}{*}{\textbf{10\%}} & \textbf{DAPAR} & 24.4 $\pm$ 11.4 & 0.5 $\pm$ 0.8 & 799.5 $\pm$ 0.8 & 175.6 $\pm$ 11.4 & 12.2 $\pm$ 5.7 & 99.9 $\pm$ 0.1 & 95.2 $\pm$ 17.1 & 21.2 $\pm$ 9.1 & 29.9 $\pm$ 9.1 \\ \cline{2-11} 
 & \textbf{MI4P} & 102.2 $\pm$ 9.9 & 2.9 $\pm$ 1.7 & 797.1 $\pm$ 1.7 & 97.8 $\pm$ 9.9 & 51.1 $\pm$ 4.9 & 99.6 $\pm$ 0.2 & 97.3 $\pm$ 1.6 & 66.9 $\pm$ 4.3 & 66.1 $\pm$ 3.7 \\ \hline
\multirow{2}{*}{\textbf{15\%}} & \textbf{DAPAR} & 25.4 $\pm$ 12.7 & 0.5 $\pm$ 0.8 & 799.5 $\pm$ 0.8 & 174.6 $\pm$ 12.7 & 12.7 $\pm$ 6.3 & 99.9 $\pm$ 0.1 & 96.4 $\pm$ 14.1 & 21.9 $\pm$ 10 & 30.5 $\pm$ 9.5 \\ \cline{2-11} 
 & \textbf{MI4P} & 105.7 $\pm$ 10.1 & 2.7 $\pm$ 1.6 & 797.3 $\pm$ 1.6 & 94.3 $\pm$ 10.1 & 52.8 $\pm$ 5.1 & 99.7 $\pm$ 0.2 & 97.5 $\pm$ 1.4 & 68.4 $\pm$ 4.3 & 67.5 $\pm$ 3.7 \\ \hline
\multirow{2}{*}{\textbf{20\%}} & \textbf{DAPAR} & 25.1 $\pm$ 12.5 & 0.4 $\pm$ 0.7 & 799.5 $\pm$ 0.7 & 174.9 $\pm$ 12.5 & 12.5 $\pm$ 6.3 & 99.9 $\pm$ 0.1 & 95.6 $\pm$ 17.1 & 21.7 $\pm$ 9.8 & 30.4 $\pm$ 9.5 \\ \cline{2-11} 
 & \textbf{MI4P} & 110.8 $\pm$ 10.2 & 3 $\pm$ 1.9 & 797 $\pm$ 1.9 & 89.2 $\pm$ 10.2 & 55.4 $\pm$ 5.1 & 99.6 $\pm$ 0.2 & 97.4 $\pm$ 1.5 & 70.5 $\pm$ 4.1 & 69.3 $\pm$ 3.5 \\ \hline
\multirow{2}{*}{\textbf{25\%}} & \textbf{DAPAR} & 26.7 $\pm$ 12.1 & 0.7 $\pm$ 1 & 799.3 $\pm$ 1 & 173.3 $\pm$ 12.1 & 13.3 $\pm$ 6 & 99.9 $\pm$ 0.1 & 97.8 $\pm$ 3.2 & 23 $\pm$ 9.5 & 31.6 $\pm$ 8.4 \\ \cline{2-11} 
 & \textbf{MI4P} & 113.9 $\pm$ 9.8 & 3.4 $\pm$ 2 & 796.6 $\pm$ 2 & 86.1 $\pm$ 9.8 & 57 $\pm$ 4.9 & 99.6 $\pm$ 0.3 & 97.1 $\pm$ 1.6 & 71.7 $\pm$ 3.9 & 70.2 $\pm$ 3.4 \\ \hline

\end{tabular}
\caption{Performance evaluation on the third set of simulation imputed using random forests.}
\label{Table:L16Sim:impRF:adjp}
\end{table}
\end{landscape}

\section{Real datasets generation}

\subsection{Complex total cell lysates (\textit{Saccharomyces cerevisiae} and \textit{Arabidopsis thaliana}) spiked UPS1 standard protein mixtures}

We consider a first real dataset from \cite{mullerBenchmarkingSamplePreparation2016}. The experiment involved six peptide mixtures, composed of a constant yeast (\textit{Saccharomyces cerevisiae}) background, into which  increasing amounts of UPS1 standard proteins mixtures (Sigma) were spiked at 0.5, 1, 2.5, 5, 10 and 25 fmol, respectively. 
In a second well-calibrated dataset, yeast was replaced by a more complex total lysate of \textit{Arabidopsis thaliana} in which UPS1 was spiked in 7 different amounts, namely 0.05, 0.25, 0.5, 1.25, 2.5, 5 and 10 fmol. For each mixture, technical triplicates were constituted.
The \textit{Saccharomyces cerevisiae} dataset was acquired on a nanoLC-MS/MS coupling composed of nanoAcquity UPLC device (Waters) coupled to a Q-Exactive Plus mass spectrometer (Thermo Scientific, Bremen, Germany) as extensively described in \cite{mullerBenchmarkingSamplePreparation2016}. The \textit{Arabidopsis thaliana} dataset was acquired on a nanoLC-MS/MS coupling composed of nanoAcquity UPLC device (Waters) coupled to a Q-Exactive HF-X mass spectrometer (Thermo Scientific, Bremen, Germany) as described hereafter.

\subsection{Data preprocessing}
For the \textit{Saccharomyces cerevisiae} and \textit{Arabidopsis thaliana} datasets, Maxquant software was used to identify peptides and derive extracted ion chromatograms. Peaks were assigned with the Andromeda search engine with full trypsin specificity. The database used for the searches was concatenated in house with the \textit{Saccharomyces cerevisiae} entries extracted from the UniProtKB-SwissProt  database (16 April 2015, 7806 entries) or the \textit{Arabidopsis thaliana} entries (09 April 2019, 15 818 entries) and those of the  UPS1 proteins (48 entries). The minimum peptide length required was seven amino acids and a maximum of one missed  cleavage was allowed. Default mass tolerances parameters were used. The maximum false discovery rate was 1\% at peptide and protein levels with the use of a decoy strategy.
For the \textit{Arabidopsis thaliana} + UPS1 experiment, data were extracted both with and without Match Between Runs and 2 pre-filtering criteria were applied prior to statistical analysis: only peptides with at least 1 out of 3 quantified values in each condition on one hand and 2 out of 3 on the other hand were kept. Thus, 4 datasets derived from the \textit{Arabidopsis thaliana} + UPS1 were considered. 
For the \textit{Saccharomyces cerevisiae} + UPS1 experiment, the same filtering criteria were applied, but only on data extracted with Match Between Runs, leading to 2 datasets considered.

\subsection{Supplemental methods for \textit{Arabidopsis thaliana} dataset}
Peptide separation was performed on an ACQUITY UPLC BEH130 C18 column (250 mm × 75 µm with 1.7 µm diameter particles) and a Symmetry C18 precolumn (20 mm ×180 µm with 5 µm diameter particles; Waters). The solvent system consisted of 0.1\% FA in water (solvent A) and 0.1\% FA in ACN (solvent B). The samples were loaded into the enrichment column over 3 min at 5 µL/min with 99\% of solvent A and 1\% of solvent B. The peptides were eluted at 400 nL/min with the following gradient of solvent B: from 3 to 20\% over 63 min, 20 to 40\% over 19 min, and 40 to 90\% over 1 min.
The MS capillary voltage was set to 2 kV at 250 °C. The system was operated in a data-dependent acquisition mode with automatic switching between MS (mass range 375–1500 m/z with R = 120 000, automatic gain control fixed at 3 × 106 ions, and a maximum injection time set at 60 ms) and MS/MS (mass range 200–2000 m/z with R = 15 000, automatic gain control fixed at 1× 105, and the maximal injection time set to 60 ms) modes. The twenty most abundant peptides were selected on each MS spectrum for further isolation and higher energy collision dissociation fragmentation, excluding unassigned and monocharged ions. The dynamic exclusion time was set to 40s.

\section{Results on real datasets}
This section provides the evaluation of the \texttt{mi4p} workflow compared to the \texttt{DAPAR} workflow on the real datasets considered. The performance is described using the indicators detailed in Section \ref{sec:Perf}. Results are based on adjusted p-values using the Benjamini-Hochberg procedure \citep{benjaminiControllingFalseDiscovery1995} and a false discovery rate of 1\%. Missing values were imputed using maximum likelihood estimation.

\subsection{\textit{Arabidopsis thaliana} + UPS1 experiment}

\begin{landscape}
\begin{table}
\begin{tabular}{|c|c|c|c|c|c|c|c|c|c|c|}
\hline
\textbf{\begin{tabular}[c]{@{}c@{}}Condition\\ (vs 10fmol)\end{tabular}} & \textbf{Method} & \begin{tabular}[c]{@{}c@{}}\textbf{True}\\\textbf{positives}\end{tabular} & \begin{tabular}[c]{@{}c@{}}\textbf{False}\\\textbf{positives}\end{tabular} & \begin{tabular}[c]{@{}c@{}}\textbf{True}\\\textbf{negatives}\end{tabular} & \begin{tabular}[c]{@{}c@{}}\textbf{False}\\\textbf{negatives}\end{tabular} & 
\begin{tabular}[c]{@{}c@{}}\textbf{Sensitivity}\\\textbf{(\%)}\end{tabular} &
\begin{tabular}[c]{@{}c@{}}\textbf{Specificity}\\\textbf{(\%)}\end{tabular} &
\begin{tabular}[c]{@{}c@{}}\textbf{Precision}\\\textbf{(\%)}\end{tabular} & 
\begin{tabular}[c]{@{}c@{}}\textbf{F-score}\\\textbf{(\%)}\end{tabular} & 
\begin{tabular}[c]{@{}c@{}}\textbf{MCC}\\\textbf{(\%)}\end{tabular}\\ \hline
\multirow{2}{*}{\textbf{0.05fmol}} & \textbf{DAPAR}    & 132 & 3677 & 10507 & 5  & 96.4 & 74.1 & 3.5  & 6.7  & 15.5 \\ \cline{2-11} 
                                   & \textbf{MI4P} & 129 & 2095 & 12089 & 8  & 94.2 & 85.2 & 5.8  & 10.9 & 21.3 \\ \hline
\multirow{2}{*}{\textbf{0.25fmol}} & \textbf{DAPAR}    & 135 & 3466 & 10718 & 2  & 98.5 & 75.6 & 3.7  & 7.2  & 16.6 \\ \cline{2-11} 
                                   & \textbf{MI4P} & 133 & 1974 & 12210 & 4  & 97.1 & 86.1 & 6.3  & 11.9 & 22.9 \\ \hline
\multirow{2}{*}{\textbf{0.5fmol}}  & \textbf{DAPAR}    & 134 & 2495 & 11689 & 3  & 97.8 & 82.4 & 5.1  & 9.7  & 20.2 \\ \cline{2-11} 
                                   & \textbf{MI4P} & 132 & 1233 & 12951 & 5  & 96.4 & 91.3 & 9.7  & 17.6 & 29.1 \\ \hline
\multirow{2}{*}{\textbf{1.25fmol}} & \textbf{DAPAR}    & 132 & 2118 & 12066 & 5  & 96.4 & 85.1 & 5.9  & 11.1 & 21.8 \\ \cline{2-11} 
                                   & \textbf{MI4P} & 129 & 792  & 13392 & 8  & 94.2 & 94.4 & 14   & 24.4 & 35.1 \\ \hline
\multirow{2}{*}{\textbf{2.5fmol}}  & \textbf{DAPAR}    & 125 & 473  & 13711 & 12 & 91.2 & 96.7 & 20.9 & 34   & 42.8 \\ \cline{2-11} 
                                   & \textbf{MI4P} & 93  & 145  & 14039 & 44 & 67.9 & 99   & 39.1 & 49.6 & 50.9 \\ \hline
\multirow{2}{*}{\textbf{5fmol}}    & \textbf{DAPAR}    & 122 & 1100 & 13084 & 15 & 89.1 & 92.2 & 10   & 18   & 28.3 \\ \cline{2-11} 
                                   & \textbf{MI4P} & 85  & 383  & 13801 & 52 & 62   & 97.3 & 18.2 & 28.1 & 32.5 \\ \hline
\end{tabular}
\caption{Performance evaluation on the \textit{Arabidopsis thaliana} + UPS1 dataset, filtered with at least 1 quantified value in each condition.}
\label{Table:A+UPS:1of3:impMLE:adjp}
\end{table}

\begin{table}
\centering
\begin{tabular}{|c|c|c|c|c|c|c|c|c|c|c|}
\hline
\textbf{\begin{tabular}[c]{@{}c@{}}Condition\\ (vs 10fmol)\end{tabular}} & \textbf{Method}   & \begin{tabular}[c]{@{}c@{}}\textbf{True}\\\textbf{positives}\end{tabular} & \begin{tabular}[c]{@{}c@{}}\textbf{False}\\\textbf{positives}\end{tabular} & \begin{tabular}[c]{@{}c@{}}\textbf{True}\\\textbf{negatives}\end{tabular} & \begin{tabular}[c]{@{}c@{}}\textbf{False}\\\textbf{negatives}\end{tabular} & 
\begin{tabular}[c]{@{}c@{}}\textbf{Sensitivity}\\\textbf{(\%)}\end{tabular} &
\begin{tabular}[c]{@{}c@{}}\textbf{Specificity}\\\textbf{(\%)}\end{tabular} &
\begin{tabular}[c]{@{}c@{}}\textbf{Precision}\\\textbf{(\%)}\end{tabular} & 
\begin{tabular}[c]{@{}c@{}}\textbf{F-score}\\\textbf{(\%)}\end{tabular} & 
\begin{tabular}[c]{@{}c@{}}\textbf{MCC}\\\textbf{(\%)}\end{tabular}\\ \hline
\multirow{2}{*}{\textbf{5fmol}}  & \textbf{DAPAR} & 372 & 226 & 15522 & 196 & 65.5 & 98.6 & 62.2 & 63.8 & 62.5 \\\cline{2-11} 
& \textbf{MI4P} & 348 & 179 & 15569 & 220 & 61.3 & 98.9 & 66 & 63.6 & 62.3 \\ \hline
\end{tabular}
\caption{Performance evaluation on the \textit{Arabidopsis thaliana} + UPS1 dataset, filtered with at least 1 quantified value in each condition and focusing only on the comparison 5fmol vs. 10fmol.}
\label{Table:A+UPS:6vs7:1of3:impMLE:adjp}
\end{table}
\end{landscape}

\begin{landscape}
\begin{table}[ht]
\begin{tabular}{|c|c|c|c|c|c|c|c|c|c|c|}
\hline
\textbf{\begin{tabular}[c]{@{}c@{}}Condition\\ (vs 10fmol)\end{tabular}} & \textbf{Method}   & \begin{tabular}[c]{@{}c@{}}\textbf{True}\\\textbf{positives}\end{tabular} & \begin{tabular}[c]{@{}c@{}}\textbf{False}\\\textbf{positives}\end{tabular} & \begin{tabular}[c]{@{}c@{}}\textbf{True}\\\textbf{negatives}\end{tabular} & \begin{tabular}[c]{@{}c@{}}\textbf{False}\\\textbf{negatives}\end{tabular} & 
\begin{tabular}[c]{@{}c@{}}\textbf{Sensitivity}\\\textbf{(\%)}\end{tabular} &
\begin{tabular}[c]{@{}c@{}}\textbf{Specificity}\\\textbf{(\%)}\end{tabular} &
\begin{tabular}[c]{@{}c@{}}\textbf{Precision}\\\textbf{(\%)}\end{tabular} & 
\begin{tabular}[c]{@{}c@{}}\textbf{F-score}\\\textbf{(\%)}\end{tabular} & 
\begin{tabular}[c]{@{}c@{}}\textbf{MCC}\\\textbf{(\%)}\end{tabular}\\ \hline
\multirow{2}{*}{\textbf{0.05fmol}} & \textbf{DAPAR}    & 74 & 2989 & 8880  & 3  & 96.1 & 74.8 & 2.4  & 4.7  & 13   \\ \cline{2-11} 
                                   & \textbf{MI4P} & 74 & 2989 & 8880  & 3  & 96.1 & 74.8 & 2.4  & 4.7  & 13   \\ \hline
\multirow{2}{*}{\textbf{0.25fmol}} & \textbf{DAPAR}    & 76 & 2837 & 9032  & 1  & 98.7 & 76.1 & 2.6  & 5.1  & 13.9 \\ \cline{2-11} 
                                   & \textbf{MI4P} & 76 & 2837 & 9032  & 1  & 98.7 & 76.1 & 2.6  & 5.1  & 13.9 \\ \hline
\multirow{2}{*}{\textbf{0.5fmol}}  & \textbf{DAPAR}    & 76 & 1905 & 9964  & 1  & 98.7 & 83.9 & 3.8  & 7.4  & 17.8 \\ \cline{2-11} 
                                   & \textbf{MI4P} & 76 & 1905 & 9964  & 1  & 98.7 & 83.9 & 3.8  & 7.4  & 17.8 \\ \hline
\multirow{2}{*}{\textbf{1.25fmol}} & \textbf{DAPAR}    & 75 & 1411 & 10458 & 2  & 97.4 & 88.1 & 5    & 9.6  & 20.7 \\ \cline{2-11} 
                                   & \textbf{MI4P} & 75 & 1411 & 10458 & 2  & 97.4 & 88.1 & 5    & 9.6  & 20.7 \\ \hline
\multirow{2}{*}{\textbf{2.5fmol}}  & \textbf{DAPAR}    & 70 & 232  & 11637 & 7  & 90.9 & 98   & 23.2 & 36.9 & 45.3 \\ \cline{2-11} 
                                   & \textbf{MI4P} & 70 & 232  & 11637 & 7  & 90.9 & 98   & 23.2 & 36.9 & 45.3 \\ \hline
\multirow{2}{*}{\textbf{5fmol}}    & \textbf{DAPAR}    & 67 & 686  & 11183 & 10 & 87   & 94.2 & 8.9  & 16.1 & 26.7 \\ \cline{2-11} 
                                   & \textbf{MI4P} & 67 & 686  & 11183 & 10 & 87   & 94.2 & 8.9  & 16.1 & 26.7 \\ \hline
\end{tabular}
\caption{Performance evaluation on the \textit{Arabidopsis thaliana} + UPS1 dataset, filtered with at least 2 quantified values in each condition.}
\label{Table:A+UPS:2of3:impMLE:adjp}
\end{table}
\end{landscape}

\begin{landscape}
\begin{table}[ht]
\begin{tabular}{|c|c|c|c|c|c|c|c|c|c|c|}
\hline
\textbf{\begin{tabular}[c]{@{}c@{}}Condition\\ (vs 10fmol)\end{tabular}} & \textbf{Method}   & \begin{tabular}[c]{@{}c@{}}\textbf{True}\\\textbf{positives}\end{tabular} & \begin{tabular}[c]{@{}c@{}}\textbf{False}\\\textbf{positives}\end{tabular} & \begin{tabular}[c]{@{}c@{}}\textbf{True}\\\textbf{negatives}\end{tabular} & \begin{tabular}[c]{@{}c@{}}\textbf{False}\\\textbf{negatives}\end{tabular} & 
\begin{tabular}[c]{@{}c@{}}\textbf{Sensitivity}\\\textbf{(\%)}\end{tabular} &
\begin{tabular}[c]{@{}c@{}}\textbf{Specificity}\\\textbf{(\%)}\end{tabular} &
\begin{tabular}[c]{@{}c@{}}\textbf{Precision}\\\textbf{(\%)}\end{tabular} & 
\begin{tabular}[c]{@{}c@{}}\textbf{F-score}\\\textbf{(\%)}\end{tabular} & 
\begin{tabular}[c]{@{}c@{}}\textbf{MCC}\\\textbf{(\%)}\end{tabular}\\ \hline
\multirow{2}{*}{\textbf{0.05fmol}} & \textbf{DAPAR}    & 16 & 1567 & 6173 & 1 & 94.1 & 79.8 & 1   & 2    & 8.6  \\ \cline{2-11} 
                                   & \textbf{MI4P} & 16 & 1567 & 6173 & 1 & 94.1 & 79.8 & 1   & 2    & 8.6  \\ \hline
\multirow{2}{*}{\textbf{0.25fmol}} & \textbf{DAPAR}    & 16 & 1461 & 6279 & 1 & 94.1 & 81.1 & 1.1 & 2.1  & 9    \\ \cline{2-11} 
                                   & \textbf{MI4P} & 16 & 1461 & 6279 & 1 & 94.1 & 81.1 & 1.1 & 2.1  & 9    \\ \hline
\multirow{2}{*}{\textbf{0.5fmol}}  & \textbf{DAPAR}    & 15 & 895  & 6845 & 2 & 88.2 & 88.4 & 1.6 & 3.2  & 11.1 \\ \cline{2-11} 
                                   & \textbf{MI4P} & 15 & 895  & 6845 & 2 & 88.2 & 88.4 & 1.6 & 3.2  & 11.1 \\ \hline
\multirow{2}{*}{\textbf{1.25fmol}} & \textbf{DAPAR}    & 16 & 880  & 6860 & 1 & 94.1 & 88.6 & 1.8 & 3.5  & 12.1 \\ \cline{2-11} 
                                   & \textbf{MI4P} & 16 & 880  & 6860 & 1 & 94.1 & 88.6 & 1.8 & 3.5  & 12.1 \\ \hline
\multirow{2}{*}{\textbf{2.5fmol}}  & \textbf{DAPAR}    & 13 & 139  & 7601 & 4 & 76.5 & 98.2 & 8.6 & 15.4 & 25.2 \\ \cline{2-11} 
                                   & \textbf{MI4P} & 13 & 139  & 7601 & 4 & 76.5 & 98.2 & 8.6 & 15.4 & 25.2 \\ \hline
\multirow{2}{*}{\textbf{5fmol}}    & \textbf{DAPAR}    & 11 & 419  & 7321 & 6 & 64.7 & 94.6 & 2.6 & 4.9  & 12.1 \\ \cline{2-11} 
                                   & \textbf{MI4P} & 11 & 419  & 7321 & 6 & 64.7 & 94.6 & 2.6 & 4.9  & 12.1 \\ \hline
\end{tabular}
\caption{Performance evaluation on the \textit{Arabidopsis thaliana} + UPS1 dataset, extracted without Match Between Runs and filtered with at least 1 quantified value in each condition.}
\label{Table:A+UPS:noMBR:1of3:impMLE:adjp}
\end{table}
\end{landscape}

\begin{landscape}
\begin{table}[ht]
\begin{tabular}{|c|c|c|c|c|c|c|c|c|c|c|}
\hline
\textbf{\begin{tabular}[c]{@{}c@{}}Condition\\ (vs 10fmol)\end{tabular}} & \textbf{Method}   & \begin{tabular}[c]{@{}c@{}}\textbf{True}\\\textbf{positives}\end{tabular} & \begin{tabular}[c]{@{}c@{}}\textbf{False}\\\textbf{positives}\end{tabular} & \begin{tabular}[c]{@{}c@{}}\textbf{True}\\\textbf{negatives}\end{tabular} & \begin{tabular}[c]{@{}c@{}}\textbf{False}\\\textbf{negatives}\end{tabular} & 
\begin{tabular}[c]{@{}c@{}}\textbf{Sensitivity}\\\textbf{(\%)}\end{tabular} &
\begin{tabular}[c]{@{}c@{}}\textbf{Specificity}\\\textbf{(\%)}\end{tabular} &
\begin{tabular}[c]{@{}c@{}}\textbf{Precision}\\\textbf{(\%)}\end{tabular} & 
\begin{tabular}[c]{@{}c@{}}\textbf{F-score}\\\textbf{(\%)}\end{tabular} & 
\begin{tabular}[c]{@{}c@{}}\textbf{MCC}\\\textbf{(\%)}\end{tabular}\\ \hline
\multirow{2}{*}{\textbf{0.05fmol}} & \textbf{DAPAR}    & 8 & 1234 & 4119 & 1 & 88.9 & 76.9 & 0.6 & 1.3  & 6.4  \\ \cline{2-11} 
                                   & \textbf{MI4P} & 8 & 1234 & 4119 & 1 & 88.9 & 76.9 & 0.6 & 1.3  & 6.4  \\ \hline
\multirow{2}{*}{\textbf{0.25fmol}} & \textbf{DAPAR}    & 8 & 1150 & 4203 & 1 & 88.9 & 78.5 & 0.7 & 1.4  & 6.7  \\ \cline{2-11} 
                                   & \textbf{MI4P} & 8 & 1150 & 4203 & 1 & 88.9 & 78.5 & 0.7 & 1.4  & 6.7  \\ \hline
\multirow{2}{*}{\textbf{0.5fmol}}  & \textbf{DAPAR}    & 8 & 742  & 4611 & 1 & 88.9 & 86.1 & 1.1 & 2.1  & 8.9  \\ \cline{2-11} 
                                   & \textbf{MI4P} & 8 & 742  & 4611 & 1 & 88.9 & 86.1 & 1.1 & 2.1  & 8.9  \\ \hline
\multirow{2}{*}{\textbf{1.25fmol}} & \textbf{DAPAR}    & 8 & 536  & 4817 & 1 & 88.9 & 90   & 1.5 & 2.9  & 10.7 \\ \cline{2-11} 
                                   & \textbf{MI4P} & 8 & 536  & 4817 & 1 & 88.9 & 90   & 1.5 & 2.9  & 10.7 \\ \hline
\multirow{2}{*}{\textbf{2.5fmol}}  & \textbf{DAPAR}    & 6 & 83   & 5270 & 3 & 66.7 & 98.4 & 6.7 & 12.2 & 20.9 \\ \cline{2-11} 
                                   & \textbf{MI4P} & 6 & 83   & 5270 & 3 & 66.7 & 98.4 & 6.7 & 12.2 & 20.9 \\ \hline
\multirow{2}{*}{\textbf{5fmol}}    & \textbf{DAPAR}    & 6 & 274  & 5079 & 3 & 66.7 & 94.9 & 2.1 & 4.2  & 11.3 \\ \cline{2-11} 
                                   & \textbf{MI4P} & 6 & 274  & 5079 & 3 & 66.7 & 94.9 & 2.1 & 4.2  & 11.3 \\ \hline
\end{tabular}
\caption{Performance evaluation on the \textit{Arabidopsis thaliana} + UPS1 dataset, extracted without Match Between Runs and filtered with at least 2 quantified values in each condition.}
\label{Table:A+UPS:noMBR:2of3:impMLE:adjp}
\end{table}
\end{landscape}

\begin{landscape}
\begin{table}[ht]
\centering
\begin{tabular}{|c|c|c|c|c|c|c|c|c|c|c|}
\hline
\textbf{\begin{tabular}[c]{@{}c@{}}Condition\\ (vs 10fmol)\end{tabular}} & \textbf{Method}   & \begin{tabular}[c]{@{}c@{}}\textbf{True}\\\textbf{positives}\end{tabular} & \begin{tabular}[c]{@{}c@{}}\textbf{False}\\\textbf{positives}\end{tabular} & \begin{tabular}[c]{@{}c@{}}\textbf{True}\\\textbf{negatives}\end{tabular} & \begin{tabular}[c]{@{}c@{}}\textbf{False}\\\textbf{negatives}\end{tabular} & 
\begin{tabular}[c]{@{}c@{}}\textbf{Sensitivity}\\\textbf{(\%)}\end{tabular} &
\begin{tabular}[c]{@{}c@{}}\textbf{Specificity}\\\textbf{(\%)}\end{tabular} &
\begin{tabular}[c]{@{}c@{}}\textbf{Precision}\\\textbf{(\%)}\end{tabular} & 
\begin{tabular}[c]{@{}c@{}}\textbf{F-score}\\\textbf{(\%)}\end{tabular} & 
\begin{tabular}[c]{@{}c@{}}\textbf{MCC}\\\textbf{(\%)}\end{tabular}\\ \hline
\multirow{2}{*}{\textbf{0.05fmol}} & \textbf{DAPAR} & 41 & 1040 & 1557 & 0 & 100 & 60 & 3.8 & 7.3 & 15.1 \\ \cline{2-11} 
 & \textbf{MI4P} & 41 & 753 & 1844 & 0 & 100 & 71 & 5.2 & 9.8 & 19.1 \\ \hline
\multirow{2}{*}{\textbf{0.25fmol}} & \textbf{DAPAR} & 41 & 1072 & 1525 & 0 & 100 & 58.7 & 3.7 & 7.1 & 14.7 \\ \cline{2-11} 
 & \textbf{MI4P} & 41 & 797 & 1800 & 0 & 100 & 69.3 & 4.9 & 9.3 & 18.4 \\ \hline
\multirow{2}{*}{\textbf{0.5fmol}} & \textbf{DAPAR} & 40 & 848 & 1749 & 1 & 97.6 & 67.3 & 4.5 & 8.6 & 17 \\ \cline{2-11} 
 & \textbf{MI4P} & 40 & 585 & 2012 & 1 & 97.6 & 77.5 & 6.4 & 12 & 21.8 \\ \hline
\multirow{2}{*}{\textbf{1.25fmol}} & \textbf{DAPAR} & 41 & 409 & 2188 & 0 & 100 & 84.3 & 9.1 & 16.7 & 27.7 \\ \cline{2-11} 
 & \textbf{MI4P} & 41 & 142 & 2455 & 0 & 100 & 94.5 & 22.4 & 36.6 & 46 \\ \hline
\multirow{2}{*}{\textbf{2.5fmol}} & \textbf{DAPAR} & 41 & 208 & 2389 & 0 & 100 & 92 & 16.5 & 28.3 & 38.9 \\ \cline{2-11} 
 & \textbf{MI4P} & 40 & 69 & 2528 & 1 & 97.6 & 97.3 & 36.7 & 53.3 & 59 \\ \hline
\multirow{2}{*}{\textbf{5fmol}} & \textbf{DAPAR} & 41 & 475 & 2122 & 0 & 100 & 81.7 & 7.9 & 14.7 & 25.5 \\ \cline{2-11} 
 & \textbf{MI4P} & 37 & 203 & 2394 & 4 & 90.2 & 92.2 & 15.4 & 26.3 & 35.5 \\ \hline
\end{tabular}
\caption{Performance evaluation on the \textit{Arabidopsis thaliana} + UPS1 dataset at the protein-level, filtered with at least 1 quantified values in each condition.}
\label{Table:A+UPS:1of3:impMLE:Aggreg:adjp}
\end{table}
\end{landscape}

\begin{landscape}
\subsection{\textit{Saccharomyces cerevisiae} + UPS1 experiment}

\begin{table}[ht]
\begin{tabular}{|c|c|c|c|c|c|c|c|c|c|c|}
\hline
\textbf{\begin{tabular}[c]{@{}c@{}}Condition\\ (vs 25fmol)\end{tabular}} & \textbf{Method}   & \begin{tabular}[c]{@{}c@{}}\textbf{True}\\\textbf{positives}\end{tabular} & \begin{tabular}[c]{@{}c@{}}\textbf{False}\\\textbf{positives}\end{tabular} & \begin{tabular}[c]{@{}c@{}}\textbf{True}\\\textbf{negatives}\end{tabular} & \begin{tabular}[c]{@{}c@{}}\textbf{False}\\\textbf{negatives}\end{tabular} & 
\begin{tabular}[c]{@{}c@{}}\textbf{Sensitivity}\\\textbf{(\%)}\end{tabular} &
\begin{tabular}[c]{@{}c@{}}\textbf{Specificity}\\\textbf{(\%)}\end{tabular} &
\begin{tabular}[c]{@{}c@{}}\textbf{Precision}\\\textbf{(\%)}\end{tabular} & 
\begin{tabular}[c]{@{}c@{}}\textbf{F-score}\\\textbf{(\%)}\end{tabular} & 
\begin{tabular}[c]{@{}c@{}}\textbf{MCC}\\\textbf{(\%)}\end{tabular}\\ \hline
\multirow{2}{*}{\textbf{0.5fmol}} & \textbf{DAPAR}    & 188 & 439 & 18067 & 4   & 97.9 & 97.6 & 30   & 45.9 & 53.5 \\ \cline{2-11} 
                                  & \textbf{MI4P} & 183 & 144 & 18362 & 9   & 95.3 & 99.2 & 56   & 70.5 & 72.7 \\ \hline
\multirow{2}{*}{\textbf{1fmol}}   & \textbf{DAPAR}    & 186 & 246 & 18260 & 6   & 96.9 & 98.7 & 43.1 & 59.6 & 64.1 \\ \cline{2-11} 
                                  & \textbf{MI4P} & 183 & 71  & 18435 & 9   & 95.3 & 99.6 & 72   & 82.1 & 82.7 \\ \hline
\multirow{2}{*}{\textbf{2.5fmol}} & \textbf{DAPAR}    & 185 & 161 & 18345 & 7   & 96.4 & 99.1 & 53.5 & 68.8 & 71.4 \\ \cline{2-11} 
                                  & \textbf{MI4P} & 179 & 39  & 18467 & 13  & 93.2 & 99.8 & 82.1 & 87.3 & 87.4 \\ \hline
\multirow{2}{*}{\textbf{5fmol}}   & \textbf{DAPAR}    & 182 & 108 & 18398 & 10  & 94.8 & 99.4 & 62.8 & 75.5 & 76.9 \\ \cline{2-11} 
                                  & \textbf{MI4P} & 156 & 23  & 18483 & 36  & 81.2 & 99.9 & 87.2 & 84.1 & 84   \\ \hline
\multirow{2}{*}{\textbf{10fmol}}  & \textbf{DAPAR}    & 148 & 109 & 18397 & 44  & 77.1 & 99.4 & 57.6 & 65.9 & 66.2 \\ \cline{2-11} 
                                  & \textbf{MI4P} & 86  & 27  & 18479 & 106 & 44.8 & 99.9 & 76.1 & 56.4 & 58.1 \\ \hline
\end{tabular}
\caption{Performance evaluation on the \textit{Saccharomyces cerevisiae} + UPS1 dataset, filtered with at least 1 quantified value in each condition.}
\label{Table:Y+UPS:1of3:impMLE:adjp}
\end{table}
\end{landscape}

\begin{landscape}
\begin{table}[ht]
\begin{tabular}{|c|c|c|c|c|c|c|c|c|c|c|}
\hline
\textbf{\begin{tabular}[c]{@{}c@{}}Condition\\ (vs 25fmol)\end{tabular}} & \textbf{Method}   & \begin{tabular}[c]{@{}c@{}}\textbf{True}\\\textbf{positives}\end{tabular} & \begin{tabular}[c]{@{}c@{}}\textbf{False}\\\textbf{positives}\end{tabular} & \begin{tabular}[c]{@{}c@{}}\textbf{True}\\\textbf{negatives}\end{tabular} & \begin{tabular}[c]{@{}c@{}}\textbf{False}\\\textbf{negatives}\end{tabular} & 
\begin{tabular}[c]{@{}c@{}}\textbf{Sensitivity}\\\textbf{(\%)}\end{tabular} &
\begin{tabular}[c]{@{}c@{}}\textbf{Specificity}\\\textbf{(\%)}\end{tabular} &
\begin{tabular}[c]{@{}c@{}}\textbf{Precision}\\\textbf{(\%)}\end{tabular} & 
\begin{tabular}[c]{@{}c@{}}\textbf{F-score}\\\textbf{(\%)}\end{tabular} & 
\begin{tabular}[c]{@{}c@{}}\textbf{MCC}\\\textbf{(\%)}\end{tabular}\\ \hline
\multirow{2}{*}{\textbf{0.5fmol}} & \textbf{DAPAR}    & 131 & 146 & 16316 & 4  & 97   & 99.1 & 47.3 & 63.6 & 67.4 \\ \cline{2-11} 
                                  & \textbf{MI4P} & 131 & 146 & 16316 & 4  & 97   & 99.1 & 47.3 & 63.6 & 67.4 \\ \hline
\multirow{2}{*}{\textbf{1fmol}}   & \textbf{DAPAR}    & 130 & 59  & 16403 & 5  & 96.3 & 99.6 & 68.8 & 80.2 & 81.2 \\ \cline{2-11} 
                                  & \textbf{MI4P} & 130 & 59  & 16403 & 5  & 96.3 & 99.6 & 68.8 & 80.2 & 81.2 \\ \hline
\multirow{2}{*}{\textbf{2.5fmol}} & \textbf{DAPAR}    & 130 & 30  & 16432 & 5  & 96.3 & 99.8 & 81.2 & 88.1 & 88.4 \\ \cline{2-11} 
                                  & \textbf{MI4P} & 130 & 30  & 16432 & 5  & 96.3 & 99.8 & 81.2 & 88.1 & 88.4 \\ \hline
\multirow{2}{*}{\textbf{5fmol}}   & \textbf{DAPAR}    & 127 & 19  & 16443 & 8  & 94.1 & 99.9 & 87   & 90.4 & 90.4 \\ \cline{2-11} 
                                  & \textbf{MI4P} & 127 & 19  & 16443 & 8  & 94.1 & 99.9 & 87   & 90.4 & 90.4 \\ \hline
\multirow{2}{*}{\textbf{10fmol}}  & \textbf{DAPAR}    & 96  & 18  & 16444 & 39 & 71.1 & 99.9 & 84.2 & 77.1 & 77.2 \\ \cline{2-11} 
                                  & \textbf{MI4P} & 96  & 18  & 16444 & 39 & 71.1 & 99.9 & 84.2 & 77.1 & 77.2 \\ \hline
\end{tabular}
\caption{Performance evaluation on the \textit{Saccharomyces cerevisiae} + UPS1 dataset, filtered with at least 2 quantified values in each condition.}
\label{Table:Y+UPS:2of3:impMLE:adjp}
\end{table}
\end{landscape}

\begin{landscape}
\begin{table}[ht]
\centering
\begin{tabular}{|c|c|c|c|c|c|c|c|c|c|c|}
\hline
\textbf{\begin{tabular}[c]{@{}c@{}}Condition\\ (vs 25fmol)\end{tabular}} & \textbf{Method}   & \begin{tabular}[c]{@{}c@{}}\textbf{True}\\\textbf{positives}\end{tabular} & \begin{tabular}[c]{@{}c@{}}\textbf{False}\\\textbf{positives}\end{tabular} & \begin{tabular}[c]{@{}c@{}}\textbf{True}\\\textbf{negatives}\end{tabular} & \begin{tabular}[c]{@{}c@{}}\textbf{False}\\\textbf{negatives}\end{tabular} & 
\begin{tabular}[c]{@{}c@{}}\textbf{Sensitivity}\\\textbf{(\%)}\end{tabular} &
\begin{tabular}[c]{@{}c@{}}\textbf{Specificity}\\\textbf{(\%)}\end{tabular} &
\begin{tabular}[c]{@{}c@{}}\textbf{Precision}\\\textbf{(\%)}\end{tabular} & 
\begin{tabular}[c]{@{}c@{}}\textbf{F-score}\\\textbf{(\%)}\end{tabular} & 
\begin{tabular}[c]{@{}c@{}}\textbf{MCC}\\\textbf{(\%)}\end{tabular}\\ \hline
\multirow{2}{*}{\textbf{0.5fmol}} & \textbf{DAPAR} & 42 & 90 & 2285 & 0 & 100 & 96.2 & 31.8 & 48.3 & 55.3 \\ \cline{2-11} 
 & \textbf{MI4P} & 42 & 24 & 2351 & 0 & 100 & 99 & 63.6 & 77.8 & 79.4 \\ \hline
\multirow{2}{*}{\textbf{1fmol}} & \textbf{DAPAR} & 42 & 65 & 2310 & 0 & 100 & 97.3 & 39.3 & 56.4 & 61.8 \\ \cline{2-11} 
 & \textbf{MI4P} & 41 & 13 & 2362 & 1 & 97.6 & 99.5 & 75.9 & 85.4 & 85.8 \\ \hline
\multirow{2}{*}{\textbf{2.5fmol}} & \textbf{DAPAR} & 41 & 27 & 2348 & 1 & 97.6 & 98.9 & 60.3 & 74.5 & 76.2 \\ \cline{2-11} 
 & \textbf{MI4P} & 41 & 8 & 2367 & 1 & 97.6 & 99.7 & 83.7 & 90.1 & 90.2 \\ \hline
\multirow{2}{*}{\textbf{5fmol}} & \textbf{DAPAR} & 42 & 19 & 2356 & 0 & 100 & 99.2 & 68.9 & 81.6 & 82.6 \\ \cline{2-11} 
 & \textbf{MI4P} & 41 & 7 & 2368 & 1 & 97.6 & 99.7 & 85.4 & 91.1 & 91.2 \\ \hline
\multirow{2}{*}{\textbf{10fmol}} & \textbf{DAPAR} & 39 & 23 & 2352 & 3 & 92.9 & 99 & 62.9 & 75 & 75.9 \\ \cline{2-11} 
 & \textbf{MI4P} & 38 & 7 & 2368 & 4 & 90.5 & 99.7 & 84.4 & 87.4 & 87.2 \\ \hline
\end{tabular}
\caption{Performance evaluation on the \textit{Saccharomyces cerevisiae} + UPS1 dataset, at the protein-level and filtered with at least 1 quantified values in each condition.}
\label{Table:Y+UPS:1of3:impMLE:Aggreg:adjp}
\end{table}
\end{landscape}

\bibliography{References}